\documentclass{aa}  
\pdfoutput=1
\usepackage{graphicx}
\usepackage{comment}
\usepackage{mathrsfs}
\usepackage{orcidlink}

\usepackage{txfonts}
\usepackage{xcolor}
\usepackage{siunitx}
\usepackage{enumitem}
\definecolor{cornflowerblue}{RGB}{100,149,237}
\definecolor{forestgreen}{RGB}{34,139,34}
\definecolor{linkcolor}{rgb}{0,0,0.6} 
 
\newcommand{\e}{\text{e}}
\makeatletter
\renewcommand*\aa@pageof{, page \thepage{} of \pageref*{LastPage}}
\makeatother

\begin{document}

   \title{Magnetised Winds in Transition Discs I: 2.5\,D Global Simulations}

   %\subtitle{Subtitle}

   \author{\'Etienne Martel\,\orcidlink{0000-0002-0818-9034}
          \inst{1}
          \and
          Geoffroy Lesur\,\orcidlink{0000-0002-8896-9435}
          \inst{1}
          }

   \institute{ \inst{1}Univ. Grenoble Alpes, CNRS, IPAG, 38000 Grenoble, France\\
         \email{etienne.martel@univ-grenoble-alpes.fr}
             }

   \date{Accepted: apr. 2022}

% \abstract{}{}{}{}{} 
% 5 {} token are mandatory
 
  \abstract
   {Protoplanetary discs are cold, dense and weakly ionised environments that witness the planetary formation. 
   Among these discs, transition discs (TDs) are characterised by a wide cavity (up to tens of a.u.) in the dust and gas distribution.
   Despite this lack of material, a considerable fraction of TDs are still strongly accreting onto their central star, possibly indicating that a mechanism is driving fast accretion in TD cavities.}
   {The presence of radially extended `dead zones' in protoplanetary discs has recently revived the interest in magnetised disc winds (MDWs), where accretion is driven by a large magnetic field extracting angular momentum from the disc. We propose that transition discs could be subject to similar disc winds, and that these could naturally explain the fast-accreting and long-lived cavities inferred in TDs.
   }
   {We present the results of the first 2.5\,D global numerical simulations of transition discs harbouring MDWs using the PLUTO code. We impose a cavity in the gas distribution with various density contrasts, and consider a power law distribution for the large-scale magnetic field strength. We assume the disc is weakly ionised and is therefore subject to ambipolar diffusion, as expected in this range of densities and temperatures.}
   {We find that our simulated TDs always reach a steady state with an inner cavity and an outer `standard' disc. These models also maintain an approximately constant accretion rate through the entire structure, reaching $10^{-7}~M_\odot.\mathrm{yrs}^{-1}$ for typical surface density values. The MDW launched from the cavity is more magnetised and have a significantly larger lever arm (up to a few tens) than the MDW launched from the outer disc. The material in the cavity is accreted at sonic velocities, and the cavity itself is rotating at $70\%$ of the Keplerian velocity due to the efficient magnetic braking imposed by the MDW. Overall, our cavity matches the dynamical properties of an inner jet emitting disc (JED) and of magnetically arrested discs (MADs) in black hole physics. Finally, we observe that the cavity is subject to recurring accretion bursts that may be driven by a magnetic Rayleigh-Taylor instability of the cavity edge.}
   {Some strongly accreting TDs could be the result of magnetised wind sculpting protoplanetary discs. Kinematic diagnostics of the disc or the wind (orbital velocity, wind speeds, accretion velocities) could disentangle classical photo-evaporation from MDW models.}


   \maketitle
%
%-------------------------------------------------------------------
\section{Introduction}

Transition discs are protoplanetary discs exhibiting a deficit of near-infrared emission, indicating a significant drop in the abundance of small dust grains in the regions inside a few $10$s of a.u. \citep{espaillat_ppvi}.
These objects are believed to be the intermediate stage between `full' primordial T-Tauri discs and disc-less young stellar objects, hence their name. In this framework, TDs are the result of an inside-out dispersal process, which is usually believed to be a combination of viscous accretion, dust growth \citep{DullemondDominik05}, giant planets \citep{Marsh92} and photo-evaporation \citep{clarke01,Alexander-ppvii}.

Despite their cavities, a large fractions of TDs are accreting onto their protostars. While \cite{Najita07} quotes a median accretion rate reduced by one order of magnitude in Taurus compared to `primordial' discs, more recent studies find even stronger accretion rates. \cite{Fang13} shows that accreting TDs have a median accretion rate similar to normal optically thick discs. \cite{Manara14} finds that TDs accrete like classical T-tauri stars, and that there is no correlation between the accretion rate and the cavity size. The fact that TDs are accreting systems should not give the impression that their cavity is depleted only in dust grains: TDs also exhibit cavities in the gas distribution \citep{Zhang14}, with gas surface density increasing with radius \citep{carmona_constraining_2014}. Probing rotational emission of CO, \cite{Vdm15,Vdm16} find a drop in gas surface density by 2 to 4 orders of magnitude, while the drop in dust surface density goes up to 6 orders of magnitude. Similar results hold in ro-vibrational CO lines, probing the cavity further in, leading to a gas drop of 2 to 4 orders of magnitude in the inner ($<3$~a.u.) regions  \citep{Carmona17}.

Hence, the picture that emerges is that of discs with a drop in gas surface density by several orders of magnitude, which are accreting similarly to (or slightly less than) primordial discs. There can be only two explanation to this phenomenon: either accretion is due to a `hidden' mass reservoir localised close to the star, and what we observe is the transient accretion of this reservoir, or gas somehow manage to get through the cavity with a much larger velocity than the usual viscous accretion velocity. In that case, one typically needs an accretion velocity of the order of the sound speed to reconcile the accretion rate with the drop in surface density \citep{wang_wind-driven_2017}.

In the first category of models, we find scenarios involving photoevaporation combined to an inner dead zone \citep{Morishima12,garate_large_2021}. This inner dead zone, typically extending between 1 and 10 a.u., sets the radius of the mass reservoir, and therefore the cavity inner edge. While it predicts a fractions of TDs with accretion rates  $\dot{M}\sim 10^{-9}~M_\odot.\text{yrs}^{-1}$, it also predicts a large fraction of non-accreting TDs, which is not observed \citep{garate_large_2021}. In addition, these models rely on the Ohmic dead zone model of \cite{Gammie96}, while it is now understood that dead zones are much more extended radially because of ambipolar diffusion \citep{Simon13}, casting doubts on the applicability of inner dead zones models. The second type of models requires some mechanism to boost angular momentum transport in the cavity. The most studied candidate for this is planet-disc interaction with planets (typically more than 3) embedded in the inner cavity. This scenario however finds gaps which are not necessarily sufficiently `clean' \citep{Zhu11}, and predicts that multiple giant planet systems in resonance should be much more common than observed \citep{DongDawson16}.

It should be noted that all of these scenarios make the explicit assumption of viscous accretion, the viscosity being due to some kind of small scale turbulence, which could be of hydrodynamic (vertical shear instability or VSI, \citealt{nelson_linear_2013}) or  magnetic (magneto-rotational instability or MRI, \citealt{Bablus91}) origin. It is however becoming clear that accretion in the regions outside of 1~a.u. is probably partially driven by magnetic winds \citep{Bai13,lesur_thanatology_2014,bethune_global_2017}. While the accretion rate of viscous models is proportional to the gas surface density, the accretion rate of MHD wind-driven models is mostly controlled by the strength of the large scale magnetic field, and much less by the surface density (for instance, \citealt{lesur_systematic_2021} proposes $\dot{M}\propto \Sigma^{0.2}\,B^{1.6}$). Hence, if one carves a cavity in a disc without modifying significantly its magnetic field distribution, one could in principle create a population of accreting TDs not so different from classical T-Tauri discs in terms of accretion rates. This kind of scenario is found in secular evolution models that include a realistic dependence of the wind stress on the surface density (e.g. \citealt{Suzuki16}, see their $\Sigma$-dependent wind torque models). Hence, MHD winds could in principle generate and sustain a fast-accreting TD cavity.

The idea of having a magnetic wind-driven cavity was first proposed by \cite{Combet08}. In this work, the cavity (named jet emitting disc, or JED) is diluted, accreting at sonic velocities, and sustains accretion rates similar to that of the outer disc. The same angle of attack was more recently tackled by \cite{wang_wind-driven_2017}, who showed that the magnetic diffusion properties of TD cavities were reminiscent of the magnetic wind solutions of \cite{wardle_structure_1993}, indicating that all of the conditions required for efficient magnetic wind launching were met in TD cavities. While this picture is promising to explain accreting TDs, there exists no dynamical model connecting an outer `standard' disc to an inner cavity accreting thanks to magnetised winds.

In this work, we present the first self-consistent (under the standard MHD assumptions) numerical models of accreting TDs based on the MHD wind scenario.
The model we propose does not enforce accretion (for example with an $\alpha$ parameter that would be added by hand).
Accretion and the disc equilibrium are natural consequences of the first principles of MHD, in the sense that their origins lie within the magnetic stresses arising from the initial vertical magnetic field.
Our aim is to demonstrate that a fast accreting cavity can connect to a standard wind-emitting outer disc, subject to realistic magnetic diffusion, and that the resulting configuration can be long-lived. Given the richness of the dynamics, we first concentrate on 2.5\,D models in this first article, and we will discuss 3\,D models in a follow up paper. The paper is divided as follows: we first introduce the models equations, physical quantities and numerical setup. We then investigate in depth a fiducial model, which possesses a cavity with a drop of 4 order in magnitude in gas surface density. We finally explore alternative models, varying the cavity depth and size, and the diffusion coefficients before concluding. We stress that we focus here on a proof of concept that such a TD configuration is sufficiently stable to be observable, but we do not discuss `how' a primordial disc could have ended in such a configuration. This will be the subject of future work.

%--------------------------------------------------------------------

\section{Physical and numerical setups}

\subsection{Physical model}

\subsubsection{Governing non-ideal MHD equations}

In the following, we place ourselves in the non-relativistic, non-ideal MHD regime and consider a thin, locally isothermal disc to follow the evolution of the gas. The mass and momentum conservation equations and the induction equation respectively read
\begin{equation}
\partial_t\,\rho + \boldsymbol{\nabla}\cdot\left(\rho\,\boldsymbol{u}\right)=0,
\end{equation}
\begin{equation}
\partial_t\left(\rho\,\boldsymbol{u}\right)+\boldsymbol{\nabla}\cdot\left(\rho\,\boldsymbol{u}\otimes\boldsymbol{u}\right)=-\boldsymbol{\nabla}P-\rho\,\boldsymbol{\nabla}\Phi_* + \frac{\boldsymbol{J}\times\boldsymbol{B}}{c},
\end{equation}
\begin{equation}
\partial_t\,\boldsymbol{B}=-\nabla\times\boldsymbol{\mathcal{E}},
\label{induction}
\end{equation}
where $\rho$, $P$, $\boldsymbol{u}$ and $\boldsymbol{B}$ are respectively the density, the thermal pressure and the plasma velocity and magnetic field. $\Phi_*=-GM_*/r$ is the gravitational potential due to the central star of mass $M_*$, $G$ being the gravitational constant. To close this system of equations, we assume the plasma follows a non-ideal Ohm's law including ambipolar diffusion:
\begin{equation}
\label{ohm}
    \mathcal{E}=-\boldsymbol{u}\times\boldsymbol{B}-\frac{4\pi}{c}\,\eta_\text{A}\,\boldsymbol{J}\times\hat{\boldsymbol{b}}\times\hat{\boldsymbol{b}},
\end{equation}
where $\hat{\boldsymbol{b}}$ is a unit vector parallel to $\boldsymbol{B}$ and $\boldsymbol{J}$ the electric current, $c$ is the speed of light and $\eta_\text{A}$ is the ambipolar diffusivity. No turbulence is added in this model whatsoever. In addition to these equations, the plasma follows the Maxwell's equations
\begin{equation}
\boldsymbol{\nabla}\cdot\boldsymbol{B}=0
\end{equation}
and
\begin{equation}
\boldsymbol{J} = \frac{c}{4\pi}\,\boldsymbol{\nabla}\times\boldsymbol{B}.
\label{divB}
\end{equation}
We place ourselves in a spherical coordinate system $(r,\theta,\phi)$ centred on the star. For convenience, we also introduce the cylindrical coordinates $R=r\,\sin\theta$, $\vartheta=\varphi$ and $z=r\,\cos\theta$.

Since we work in a thin disc, the azimuthal angular velocity $\Omega$ is expected to be close to the Keplerian angular velocity $\Omega_\text{K}(r)=(G\,M_\star/r^3)^{1/2}$. It is therefore useful to introduce a deviation from the Keplerian velocity $\boldsymbol{v}$ defined as
\begin{equation}
\boldsymbol{v} = \boldsymbol{u} - r\,\sin\theta\,\tilde{\Omega}(r)\,\boldsymbol{e}_\varphi,
\end{equation}
with $\tilde{\Omega}(r)\equiv \Omega_\text{K}(r)/\sin^2\theta$. We note that the latitudinal dependence of $\tilde{\Omega}$ is somewhat arbitrary, and need not be a particular equilibrium state. Here, our choice of $\tilde{\Omega}(r)$ ensures that our reference Keplerian velocity has constant specific angular momentum on spherical shells and eliminates surface terms which are otherwise present in angular momentum conservation equations (e.g. the last term of equation (16) in \citealt{zhu_global_2018}). This will simplify the interpretation of angular momentum budgets later.

\subsubsection{Equation of state and cooling function}

As a simplification, we assume the flow follows an ideal equation of state, and is approximately locally isothermal, i.e. $T\approx T_\mathrm{eff.}(R)$ where $T_\mathrm{eff.}$ is a prescribed radial temperature profile. This is achieved solving the energy equation
\begin{equation}
\partial_t\,P + \boldsymbol{u}\cdot\boldsymbol{\nabla}P + \Gamma\,P\,\boldsymbol{\nabla}\cdot\boldsymbol{u} = \Lambda,
\end{equation}
where we have defined a heating/cooling function
\begin{equation}
\Lambda = \frac{P}{T}\frac{T-T_\mathrm{eff.}}{\tau},
\end{equation}
where $\tau$ is the cooling time that equals $0.1$ time code unit (see below) and $\Gamma=1.0001$ is the polytropic index of the gas. The target temperature profile is
\begin{equation}
T_\mathrm{eff.}(R) = T_0\,\left(\frac{R}{R_\text{int}}\right)^{-1},
\label{temperature}
\end{equation}
where $T_0$ is the midplane temperature at the inner radius $R_\text{int}$. This choice of cooling function allows us to enforce a chosen temperature profile which mimics the real radiative equilibrium, and avoid the development of the vertical shear instability \citep[VSI, ][]{nelson_linear_2013}, which would appear in a strictly locally isothermal approximation. 

Since the gas is ideal, we can define an isothermal sound speed $c_\text{s}^{\,2}=P/\rho$. It can be shown that as a result of the vertical hydrostatic equilibrium, $c_\text{s}$ and $\Omega_\text{K}$ are related to the vertical disc thickness $h(R)$ through
\begin{equation}
h(R) = c_\text{s}(R)/\Omega_\text{K}(R).
\end{equation}
Assuming the disc is at thermal equilibrium ($T=T_\mathrm{eff.}(R)$), we have $c_\text{s}\propto R^{-1/2}$ and hence the disc aspect ratio $\varepsilon\equiv h/R$ is constant.
In the following, we choose $T_0$ in (\ref{temperature}) so that $\varepsilon=0.1\,$.

\subsection{Numerical method and parameters}

\subsubsection{Integration scheme}

The simulations are performed using the PLUTO code \citep{mignone_pluto_2007} that solves the MHD equations with a conservative Godunov type scheme and a second order Runge-Kutta time stepping. We use a HLLD type Riemann solver to compute the intercell fluxes.
In order to ensure the solenoidal constraint (\ref{divB}), we use the constrained transport approach \citep{kane_yee_numerical_1966,evans_simulation_1988}. The implementation of ambipolar diffusion in the PLUTO code follows that of \cite{lesur_thanatology_2014} and \cite{bethune_global_2017}.

\subsubsection{Code units and notations}

The internal radius is $R_\text{int}=1$, which sets the length code unit, and is chosen to be $1~$a.u. while $R_\text{ext}=50$. The time code unit is $\Omega_0^{\,-1}\equiv\Omega_\text{K}(R_\text{int})^{-1} = 1$ which is set to~$1 / 2\pi~$years so that $G\,M_*=1$ with $M_*=1~M_\odot$, $M_\odot$ being $1$~solar mass. Therefore $\Omega_\text{K}(R) = \Omega_0\,\left(R/R_\text{int}\right)^{-3/2}=R^{-3/2}$.
We choose as a unit for the surface density $300\,\si{\gram\,\centi\metre^{-2}}$ and express the accretion rate in $M_\odot.\text{yrs}^{-1}$.
We denote by `c.u.' the use of code units.
We use the subscript $X_0$ to indicate that the quantity $X$ is considered on the midplane ($\theta=\pi/2$) and the subscript $X_\text{p}$ when $X$ is a poloidal quantity.

\subsubsection{Dimensionless numbers and definitions}

We use the plasma parameter $\beta$ to quantify the disc magnetisation, defined from the midplane properties of the disc as
\begin{equation}
\beta = \frac{8\pi\,P_0}{\boldsymbol{B}_{\text{p},\,0}^{\,2}}.
\end{equation}
When considering the initial state of a given simulation, we refer to the initial magnetisation inside the cavity as $\beta_\text{in}$ and to the initial magnetisation in the external part of the disc as $\beta_\mathrm{out}$.
The second key parameter of this study is the strength of ambipolar diffusion, quantified with the Elsasser number
\begin{equation}
\Lambda_\text{A}\equiv\frac{v_\text{A}^{\;2}}{\Omega_\text{K}\,\eta_\text{A}},
\label{lambdaA}
\end{equation}
where $v_\text{A}=B/(4\pi\,\rho)^{1/2}$ is the Alfv\'en speed.
We refer to the appendix~\ref{app_lambda} for detailed information on the justifications of the model we adopt for $\Lambda_\text{A}$ and how we model its spatial dependencies in our simulations.
These two dimensionless numbers are the main control parameters of our study. 

The disc refers to the whole part of the simulation that covers $r \in \left[1; 50\right]$ 
and $z/R \in \left[-0.3;0.3\right]$. The cavity is the region where the surface density is reduced by a given factor in the innermost part of the disc (i.e. from $r=1$ to $r=10$ in most of the models).
The external part of the disc or so called `outer disc' refers to the region where the disc is full and described by a standard protoplanetary disc (without a drop in the density profile) and which extends from $r\approx 10$ to $r=50$.

Finally, we call `seed' the region defined by $r\leq 1$ of our disc, which is at play in our simulations through the inner radial boundary condition.

\subsubsection{Computational domain}

 The radial direction is divided into~$320$ cells that expand from the inner radius $r\equiv R_\text{int}$ to the external one $r\equiv R_\text{ext}$ that are uniformly meshed on a logarithmically shaped grid. The colatitude domain is mapped on a stretched grid near the poles (from $\theta=0$ to $\theta=1.279$ and from $\theta=1.862$ to $\theta=\pi$, with~$72$ cells in each zone) while the grid is chosen to be uniform around the midplane (from $\theta=1.279$ to $\theta=1.862$ with~$96$ cells) for a total of~$240$ that increases the precision in the region of interest. The disc scale height $h$ is then covered by 16~points in the case where $\varepsilon$ is fixed constant and equal to $0.1$.

\subsubsection{Boundary conditions}

Outflow boundary conditions are used in the radial direction so that no matter can come from the inner radius.
In addition, we add a wave absorbing zone for radii $r<1.5$ which damps poloidal motions on an orbital timescale. We detail the impact of this procedure in appendix~\ref{relaxation}.

In these~2.5\,D simulations, axisymmetric conditions with respect to the polar axis are enough to handle the boundaries for the colatitude direction. With the aim of reducing the impact of the outer boundary conditions, we will focus on radii lower than~$30$.

\subsubsection{Initial condition, wind and cavity}

The initial temperature profile is the effective temperature profile given in (\ref{temperature}). The initial states for the density and the azimuthal velocity $v_\varphi=R\,\Omega_\text{K}$ mimic \cite{nelson_linear_2013} to account for the hydrostatic equilibrium, while $v_r=v_\theta=0$ initially. These profiles read, without taking into account the cavity yet

\begin{align}
\rho(R,z) &= \rho_0\,\left(\frac{R}{R_\text{int}}\right)^p\,\exp{\left[\left(\frac{\Omega_\text{K}(R)\,R^3}{c_\text{s}(R)}\right)^2\left(\frac{1}{\sqrt{R^2 + z^2}} - \frac{1}{R}\right)\right]}
\label{rho_eq}\\
v(R,z) &= v_\text{K}(R)\left[(p+q)\left(\frac{c_\text{s}(R)}{\Omega_\text{K}(R)\,R^2}\right)^2 + (1+q) - \frac{q\,R}{\sqrt{R^2+z^2}}\right]^{1/2},
\label{vphi_eq}
\end{align}
with $\rho_0$ being the density at the internal radius. We choose $q=-1$ and $p=-3/2$ for the equations (\ref{rho_eq}) and (\ref{vphi_eq}) which is consistent with self-similar stationary disc solutions \citep{jacquemin-ide_magnetic_2020}. 

The initial vertical magnetic field follows a power law $B_z\propto R^{\;(p+q)/2}$ so that the plasma $\beta$ parameter in the unperturbed disc is constant. To ensure that $\boldsymbol{\nabla}\cdot\boldsymbol{B}=0$, we initialise the magnetic field using its vector potential $\boldsymbol{A}$ defined so that $\boldsymbol{B}=\boldsymbol{\nabla}\times\boldsymbol{A}$. Following \cite{zhu_global_2018}, we choose
\begin{equation}
  A_\varphi = \left \{
  \begin{aligned}
 & \frac{1}{2}\, B_0\,R  ~~~	&\text{if}~	R\leq R_\text{int} \\
  &B_0\frac{R_\text{int}^{\;2}}{R}\,\left(\frac{1}{2}-\frac{1}{m+2}\right)+R\,\left(\frac{R}{R_\text{int}}\right)^{\;m}\frac{1}{(m+2)}	~~~&\text{if}~R>R_\text{int}
  \end{aligned} \right.,
\end{equation} 
where $m=(p+q)/2=-5/4$. This results in a poloidal magnetic field which depends on the radius only
\begin{align}
\boldsymbol{B} = B_0\,\left(\frac{R}{R_\text{int}}\right)^{\,m}\,\boldsymbol{e}_z.
\label{Magnetic_field}
\end{align}
The initial strength of the magnetic field is controlled by $\beta_\mathrm{out}$, so that $B_0\propto \beta_\mathrm{out}^{\,-1/2}$.

To add a cavity and mimic a transition disc, we multiply the density profile by a function $f$ that depends on the radius only so that
\begin{equation}
\Sigma(R) = f(R) \times \Sigma_0(R),
\label{sigmaf}
\end{equation}
with
\begin{equation}
f(R) = a\,\left(1-c\,\tanh\left[b\,\left(1-\frac{R}{R_0}\right)\right]\right),
\end{equation}
where $\Sigma_0(R)\propto R^{p+1}$ is a standard surface density profile for protoplanetary disc.
The $a$, $b$, $c$ coefficients are defined as
\begin{equation*}
  \left \{
  \begin{aligned}
    b &= \frac{2}{n}\,\left(\frac{\delta R}{R_0}\right)^{-1} \\
a &= \frac{\beta_\text{in}/\beta_\mathrm{out}+\tanh(b)}{1+\tanh(b)} \\
c &= \frac{1 - \beta_\text{in}/\beta_\mathrm{out}}{\beta_\text{in}/\beta_\mathrm{out} + \tanh(b)}
  \end{aligned} \right.,
\end{equation*} 
where $R_0$ is the radius of the cavity (in code units), $n$ the number of cells on which the transition spans and $n\,\delta R$ the corresponding length in code units. Note that while the density profile exhibit an inner `hole', the magnetic field distribution is kept as a power law (\ref{Magnetic_field}). As a consequence, the initial magnetisation $\beta(R)$ also exhibits a jump in the cavity since $P\propto\Sigma(R)$. 

\begin{figure*}
\begin{center}
\includegraphics[width=0.85\linewidth]{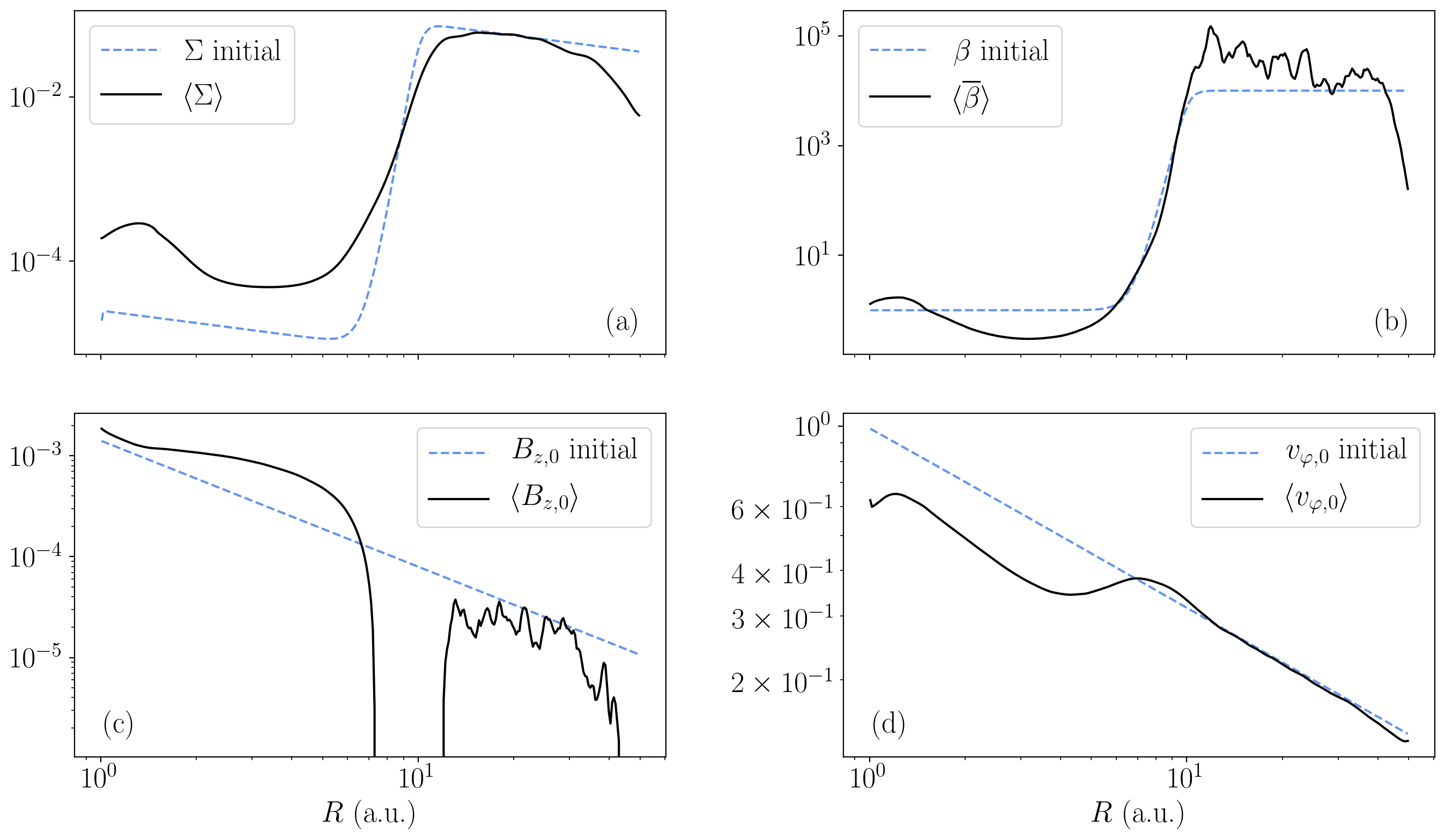}
\caption{\label{initial_avg_inifid2D}Initial and time-averaged profile of $\Sigma$ (top left panel), $\overline{\beta}$ (top right), $B_{z,\,0}$ (bottom left) and $v_{\varphi,\,0}$ (bottom right) with respect to $R$.}
\end{center}
\end{figure*}

Therefore, $\beta_\mathrm{out}/\beta_\text{in}$ is equal to the contrast in the gas surface density.
In short, the function $f$ creates a cavity in $\Sigma$ but does not affect $B_\text{p}$.
As a result, we simulate a transition disc with a strongly magnetised cavity ($\beta_\text{in}=1$). A typical radial profile of the quantities discussed above are shown in Fig.~\ref{initial_avg_inifid2D}.

\noindent

\subsection{Integration and averages}
\begin{figure}
\begin{center}
\includegraphics[scale=0.75]{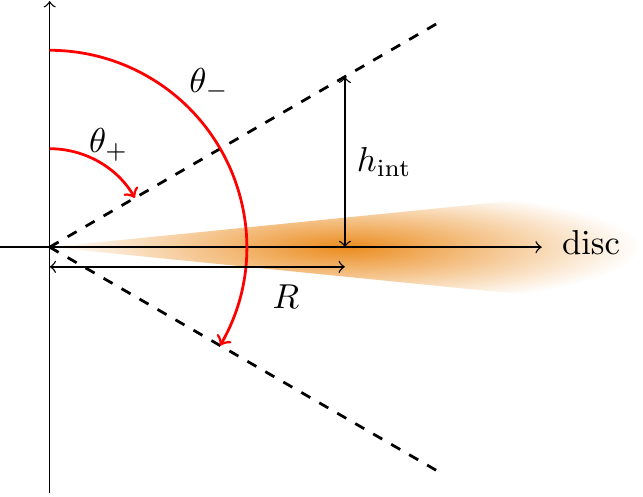}
\caption{\label{schema_disc}Schematic view of the disc which is represented in orange. $\theta_\pm$ define the vertical integration surface and $h_\text{int}$ is the integration scale height at a given radius $R$.}
\end{center}
\end{figure}

Several integrations and averages are used throughout the text.
In this manuscript, we use the following proxy for the vertical integration along $\theta$
\begin{equation}
\overline{X}(r, t) = r \, \int_{\theta_+}^{\,\theta_-} X(\boldsymbol{r}, t) \sin\theta \, \text{d} \theta.
\label{procedure}
\end{equation}
$\theta_\pm$ quantify the integration height as shown in Fig. \ref{schema_disc} so that
\begin{equation}
\frac{\theta_- - \theta_+}{2} = \arctan\left(\frac{h_\mathrm{int}}{R}\right) = \arctan\varepsilon_\mathrm{int},
\end{equation}
with $h_\mathrm{int}$ the integration height at radius $R$ given by an integration effective aspect ratio $\varepsilon_\mathrm{int}\equiv h_\mathrm{int}/R$.
Note that this integration `height' is not necessarily the disc thickness $h$.
We introduce $\overline{\beta}$ as
\begin{equation}
    \overline{\beta} \equiv \frac{8\pi\,\Sigma\,\overline{c}_{\text{s},0}^{\,2}}{\sqrt{2\pi}\,R\,\varepsilon\,\left(\overline{B_r}^{\,2}+\overline{B_\theta}^{\,2}\right)},
\end{equation}
which corresponds to a theta-averaged `effective' midplane $\beta$ plasma parameter. It is defined so that it matches the midplane $\beta$ parameter in a hydrostatic isothermal disc. This more general definition is needed when the disc midplane is displaced vertically such as inside the cavity (see section \ref{Fast_variability}).
Finally, we add the time-average defined by
\begin{equation}
\langle X(\boldsymbol{r}) \rangle = \frac{1}{T}\int_{t_0}^{\,t_0 + T}X(\boldsymbol{r}, t)\, \text{d}t.
\end{equation}
We run the 2.5\,D simulations so that we reach $1000$ orbits at $R=10$ which means $\approx 31000$ orbits at $R_\text{int}$.
If not specified, time-averages are calculated taking into account the whole simulation without the first $4000$ orbits at $R_\text{int}$ to suppress the transient state.
Otherwise, we indicate our choice of notation when needed $\langle X\rangle_{1000}$ being the time-averaged value of $X$ during the last $1000$ orbits at $R_\text{int}$ for example.

\subsection{Simulations table}

All the simulations performed are listed in table \ref{SimuList}. The characteristic parameters are the external initial plasma parameter $\beta_\mathrm{out}$, the internal initial plasma parameter $\beta_\text{in}$ and the initial ambipolar Elsasser number $\Lambda_{\text{A},\,0}$.
Additionally, we perform a convergence test by running a high resolution $(640\times480)$ simulation similar to the fiducial one that exhibits profiles that differ by less than $8~$\% in the cavity and by less than $1~$\% when considering the entire domain.

 \begin{table}
         \label{SimuList}
     $$ 
         \begin{array}{p{0.3\linewidth}lcccc}
            \hline
            \noalign{\smallskip}
            Name      &  \beta_{\text{out}} & \beta_{\text{in}} & \Lambda_{\text{A},\,0} &
            R_0~\text{(a.u.)}\\
            \noalign{\smallskip}
            \hline
            \noalign{\smallskip}
            \textbf{B4Bin0Am0} & 10^4 & 1 & 1 & 10 \\
            B3Bin0Am0        & 10^3 & 1 & 1 & 10 \\
            B5Bin0Am0        & 10^5 & 1 & 1 & 10 \\
            B4Bin0Am1       & 10^4 & 1 & 10 & 10 \\
            B4Bin1Am0        & 10^4 & 10 & 1 & 10 \\
            B4Bin2Am0        & 10^4 & 10^2 & 1 & 10 \\
            B4Bin3Am0        & 10^4 & 10^3 & 1 & 10 \\
            B5Bin1Am0        & 10^5 & 10 & 1 & 10 \\
            B5Bin2Am0			& 10^5 & 10^2 & 1 & 10 \\ 
            B5Bin3Am0			& 10^5 & 10^3 & 1 & 10 \\
            B5Bin4Am0			& 10^5 & 10^4 & 1 & 10 \\
            B3Bin1Am0			& 10^3 & 10^1 & 1 & 10 \\ 
            B3Bin2Am0			& 10^3 & 10^2 & 1 & 10 \\
            R20FID	& 10^4 & 1 & 1 & 20 \\ 
            \noalign{\smallskip}
            \hline
         \end{array}
    $$
    \caption[Set of simulations.]{Simulations information. B4Bin0Am0 is the fiducial simulation. B4Bin0Am1 quantifies the influence of $\Lambda_{\text{A},\,0}$ while B5Bin0Am0 and S2DB3Bin0Am0 are the reference runs for $\beta_\mathrm{out} = 10^5$ and $\beta_\mathrm{out}=10^3$. All the runs with $\text{Bin}\neq 0$ in their label explore the role of the initial value of $\beta$ at $R_\text{int}$. R20FID is the same simulation as the fiducial one, with a cavity twice larger.}
   \end{table} 

\section{Fiducial simulation}

We start by describing in details our fiducial simulation ($\beta_\mathrm{out}=10^4$, $\beta_\text{in}=1$, $\Lambda_\text{A}=1$ and $R_0=10$), before turning to an exploration of the parameter space.
\subsection{Evolution of surface density and plasma magnetisation}
\label{evolution}
\begin{figure*}
\begin{center}

\includegraphics[width=1.\linewidth]{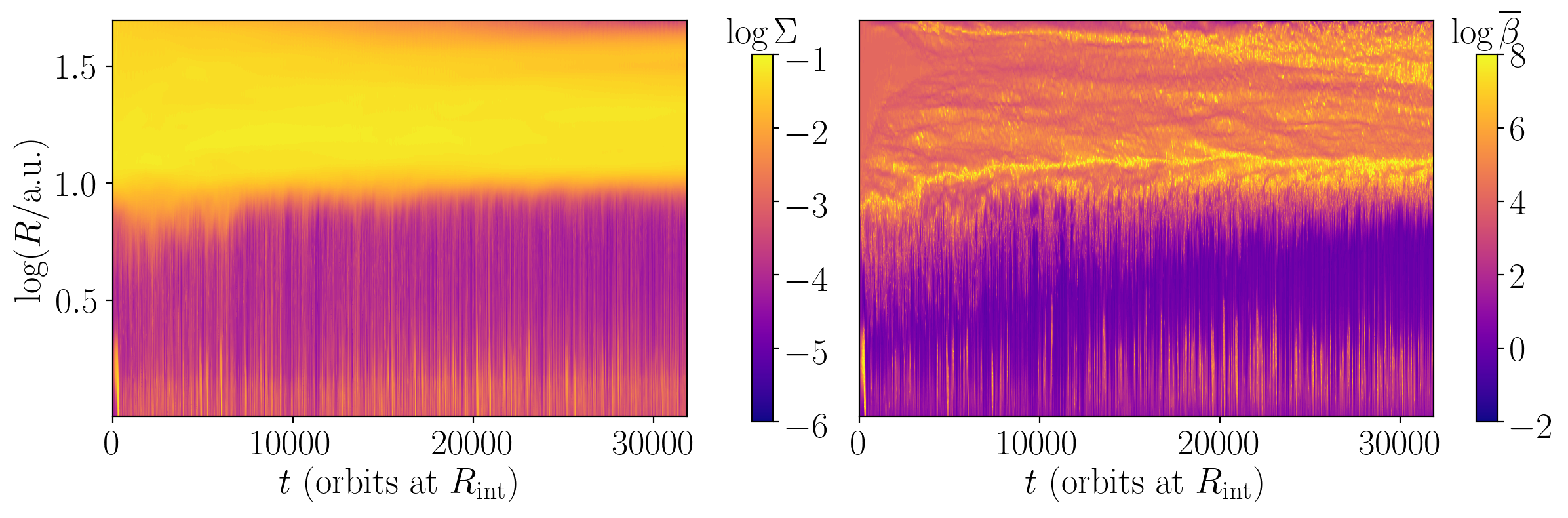}

\caption{\label{sig_beta_avg_fid2D}Surface density (left panel) and plasma beta parameter $\overline{\beta}$ (right panel) as a function of $R$ at midplane and time for the fiducial simulation. The cavity remains during the entire simulation and keeps a relatively strong magnetisation with $\overline{\beta}\sim 1$. }
\end{center}
\end{figure*}

We first look at the temporal evolution of the surface density (Fig.~\ref{sig_beta_avg_fid2D}). 
We find that the cavity stands during the whole simulation as its radius remains close to its initial value.
As it will be shown in section \ref{slow_evolution}, the cavity tends to expand slightly. The cavity location, defined as the radius where the surface density equals half of its maximum value, is subject to a small variation $\Delta R / R = 10.3\%$ over the duration of the simulation.
While the external disc is relatively smooth with respect to time, the cavity is striped by temporal variations of $\Sigma$ that may suggest that matter is moving inside the cavity at relatively fast speeds. We study in depth these stripes in section \ref{Fast_variability}. A small accumulation of material is seen close to the inner radius at $R\leq 1.5$. We refer to the appendix~\ref{relaxation} for a quantitative discussion on this accumulation.

Figure \ref{sig_beta_avg_fid2D} also pictures the evolution of $\overline{\beta}$ whose results are similar to the ones for $\Sigma$. Inside the cavity, $\overline{\beta}$ exhibits a striped-like pattern with an accumulation close to the internal radius. The edge of the cavity is not smooth at all but varies around its initial value of~$10$. Though $\overline{\beta}$ stays on average around~$1$ in the cavity, some low values around~$10^{-2}$ are reached from time to time.
After approximately~$4000$ orbits at the internal radius, both $\Sigma$ and $\overline{\beta}$ reach a quasi-stationary state.

\begin{center}
\begin{figure}
\includegraphics[width=1.\linewidth]{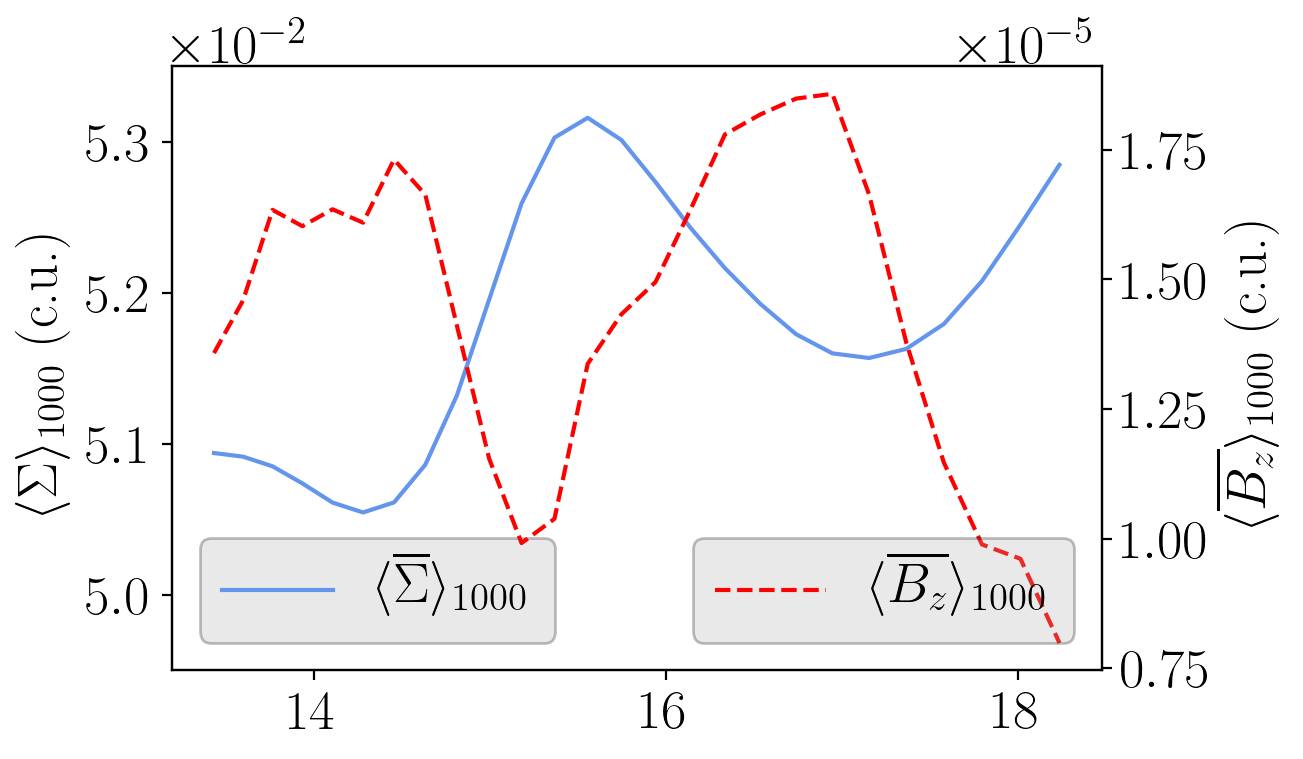}
\caption{\label{gaps_fid2D}Time-averaged on the last $1000$~orbits at $R_\mathrm{int}$ of the surface density and vertical magnetic field in the midplane. The vertical magnetic field is vertically averaged and both profiles are given in arbitrary units.}
\end{figure}
\end{center}

Gaps and rings are detected in the outer part of the disc, in the spatio-temporal diagram of both $\Sigma$ and $\beta$ (Fig.~\ref{sig_beta_avg_fid2D}).
We also emphasise that these structures are observed in all of our simulations (see Fig.~\ref{sig_beta_avg_S2DB4Bin0Am1}, \ref{sig_beta_avg_2D_beta} and \ref{sig_beta_avg_R20fid2D}).
Regarding the fiducial simulation, we detect two main gaps after the cavity edge and before $R=30$.
For better visibility, we show the surface density and the vertical magnetic field, time-averaged on the last $1000$~inner orbits and a focus in the region $R=12$---$18$~a.u. (where the gaps are detected) in Fig.~\ref{gaps_fid2D}.
Gaps are characterised by a drop of $\sim 5\,\%$ of the local surface density and their location is correlated with a sharp increase of the vertical magnetic field, which matches the secular wind instability described by \cite{riols_ring_2020}.
These structures are enhanced in the simulation with a higher ambipolar Elsasser number as it can be seen in Fig.~\ref{stream_lines_mag_zoom_S2DB4Bin0Am1}. In addition, we observe the merging of gaps on longer timescales (Fig.~\ref{sig_beta_avg_S2DB4Bin0Am1}) similarly to \cite{cui_global_2021}. While of interest for the dynamics of the outer disc, we do not address the evolution of these rings and gaps any further and instead focus on the dynamics of the cavity.

\subsection{Disc structure}

\subsubsection{Magnetic structure}

\begin{figure}
\begin{center}

\includegraphics[width=0.5\textwidth]{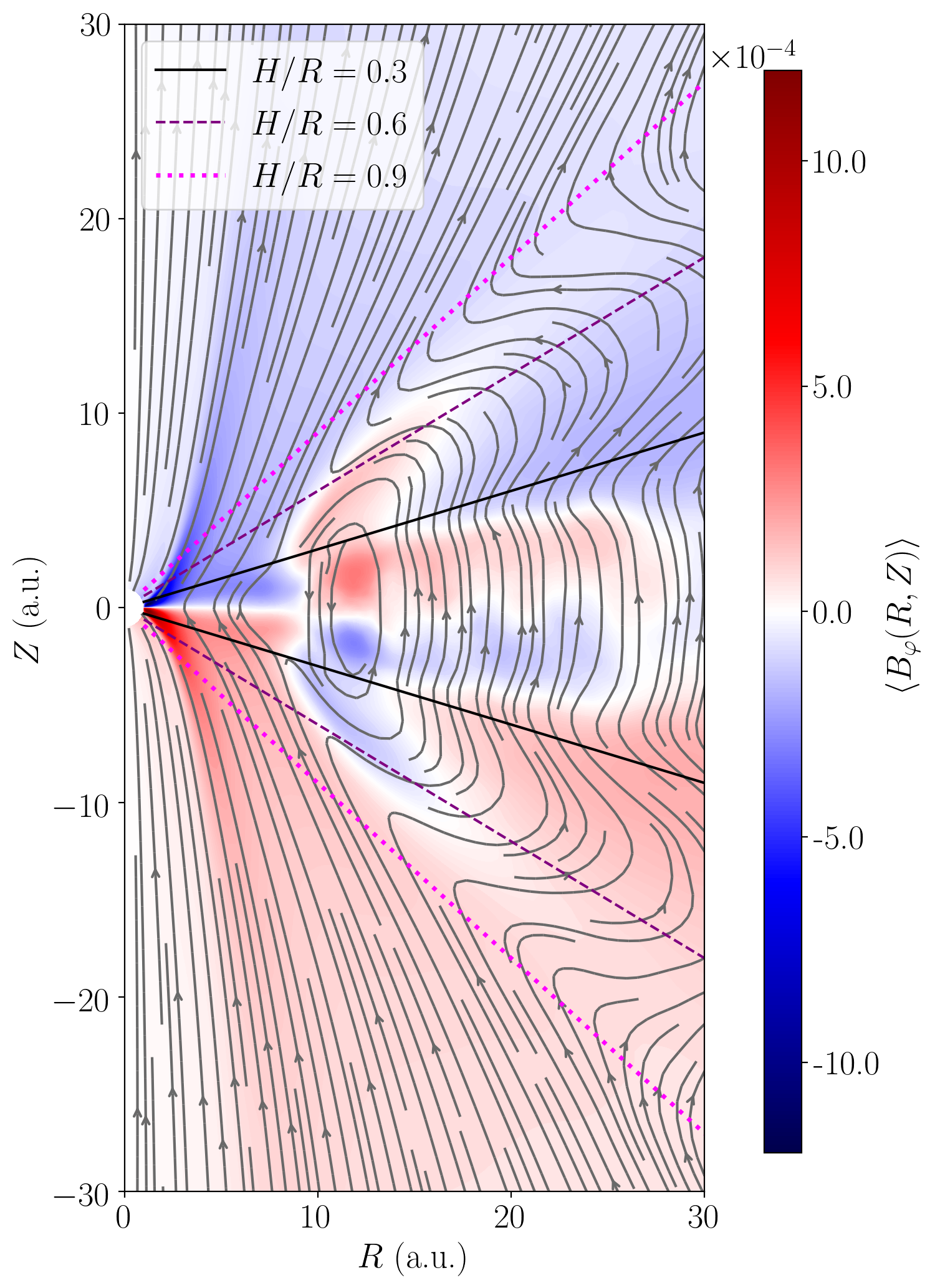}
\caption{\label{mag_field_struct_fid2D}Time-averaged poloidal magnetic field lines and toroidal field component $\langle B_\varphi\rangle$ for the fiducial simulation. Note the peculiar field topology close to the truncation radius.}

\end{center}
\end{figure}

We show the time-averaged magnetic field in Fig.~\ref{mag_field_struct_fid2D}.
In the cavity, the poloidal magnetic field lines are pinched at the midplane but they remain vertical in the outer disc.
These two regions are separated by a transition zone located at the cavity edge which exhibits a magnetic loop.
Inside this loop, the polarity of the azimuthal component is reversed, with $B_\varphi>0$ in the upper hemisphere close to the disc.
The poloidal field lines present an elbow-shaped structure above and below the transition with significant changes of direction at $h_\text{int}/R\approx \pm 0.3$, $\pm 0.6$ and $\pm 0,9$.

\subsubsection{Velocity stream lines}

\begin{figure}
\begin{center}

\includegraphics[width=0.5\textwidth]{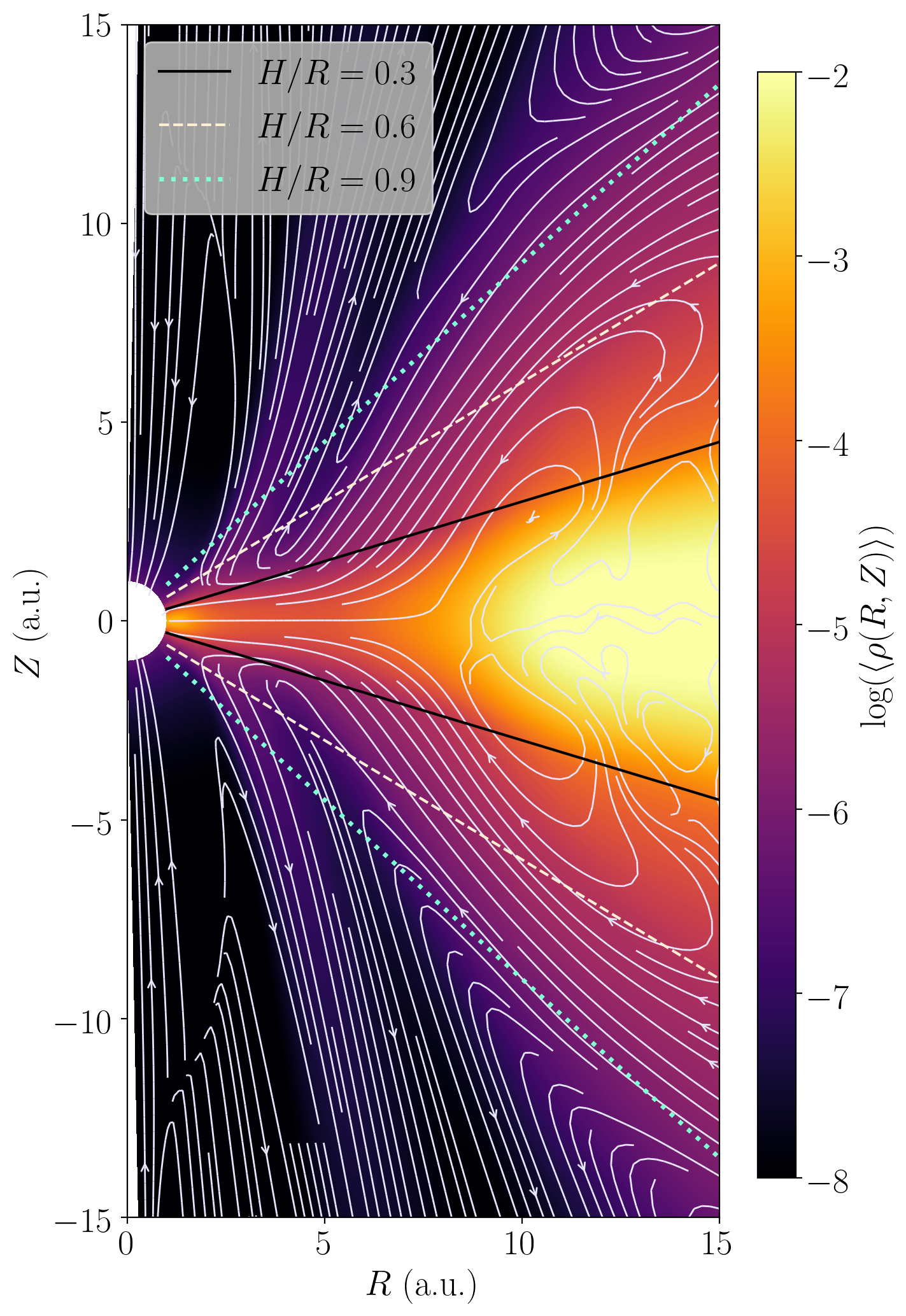}
\caption{\label{stream_lines_fid2D}Time-averaged stream lines and density for the fiducial simulation. Note the peculiar shape of the streamlines around the transition radius.}

\end{center}
\end{figure}

We show the time-averaged density and streamlines in Fig.~\ref{stream_lines_fid2D}.
The disc clearly appears around the midplane for $R\gtrsim 10$ while the depleted profile in $\rho$ indicates the cavity for $R<10$. We find that a wind is emitted from  the cavity, with poloidal streamlines approximately parallel to  magnetic field lines, as expected from ideal MHD. A closer inspection of the streamlines however shows that in the regions close to the transition radius $R\gtrsim 8$, matter is falling into the cavity.  Figure \ref{stream_lines_fid2D} shows that this material is actually coming from the outer disc. It is originally ejected from this disc, before being deflected and accreted into the cavity, generating an elbow-like shape similar to the one found for magnetic field lines (Fig.~\ref{mag_field_struct_fid2D}). This accretion stream then stays localised close to the cavity midplane down to the inner radius of the simulation. In the outer disc, the motion of the gas is not as well organised though it is approximately symmetric with respect to the midplane.

\subsubsection{Angular Momentum stream lines}

\begin{figure}
\begin{center}

\includegraphics[width=0.5\textwidth]{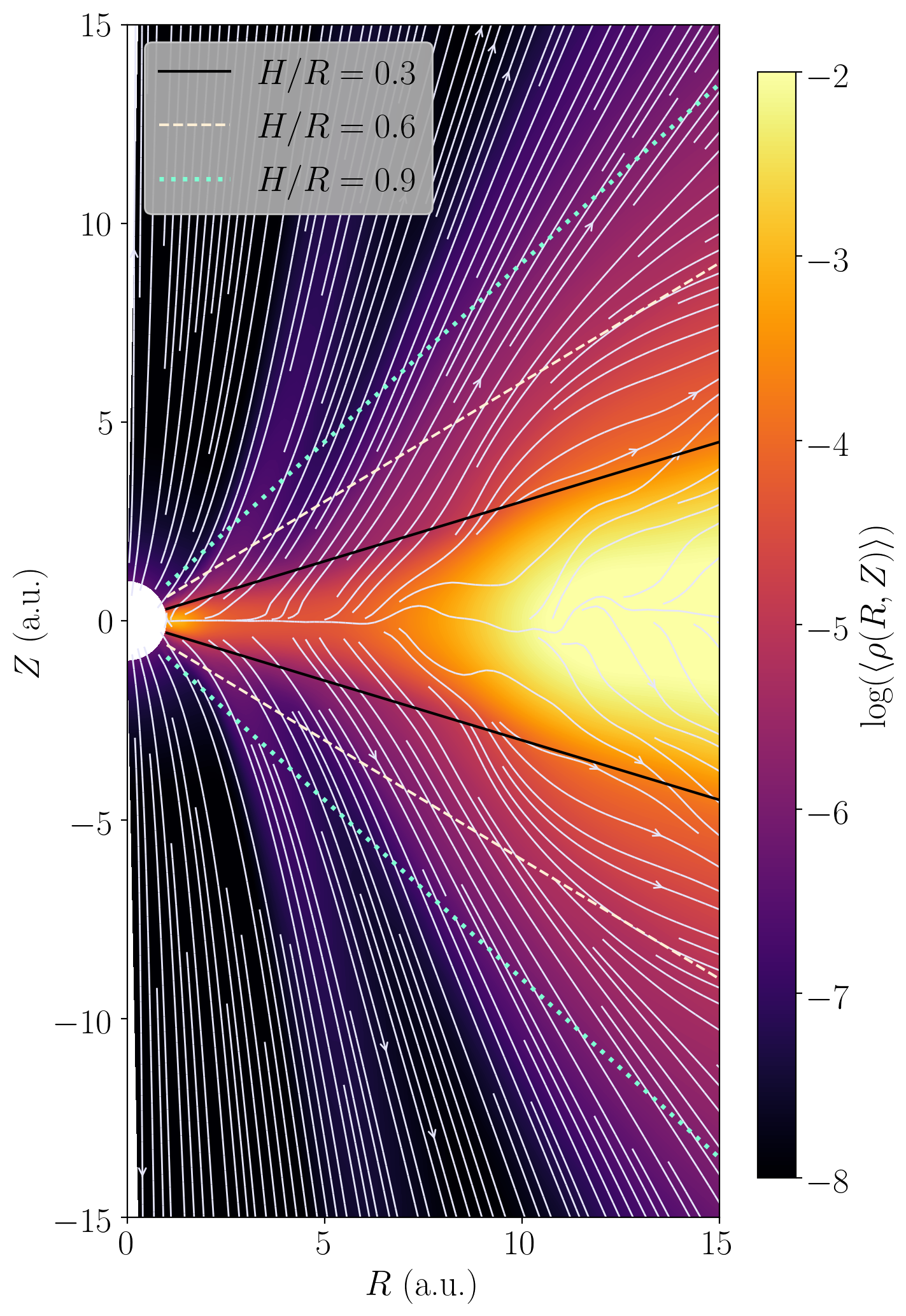}
\caption{\label{momentum_stream_lines_fid2D}Time-averaged angular momentum flux stream lines over time-averaged density for the fiducial simulation. Angular momentum leaves the disc midplane because of the wind.}

\end{center}
\end{figure}

In order to deeper the analysis of the role of the magnetic structure, we concentrate on the time-averaged angular momentum flux, defined by
\begin{equation}
\boldsymbol{\mathcal{L}}_\text{p} = r\,\sin\theta\,\langle\rho\,\boldsymbol{u}_\text{p}\,u_\varphi\rangle - r\,\sin\theta\,\langle \boldsymbol{B}_\text{p}\,B_\varphi\rangle.
\end{equation}
The poloidal flux lines associated to this angular momentum flux are shown in Fig.~\ref{momentum_stream_lines_fid2D}. It appears that angular momentum is extracted from the disc midplane and carried both radially and vertically in a relatively homogeneous manner. In particular, we note that there is no elbow-like shape for the angular momentum flux, in contrast to the magnetic and velocity streamlines, indicating that the cavity$\,+\,$outer disc system has adapted its magnetic topology to transport angular momentum homogeneously.

\subsection{Accretion theory}
\subsubsection{Accretion rate}

The first step to study the accretion in the disc is to investigate the accretion rate $\dot{M}$ defined as
\begin{equation}
\dot{M}(R, t) = - 2\pi\, R\,\overline{\rho v_r}.
\end{equation}

\begin{figure}
\begin{center}
\includegraphics[width=1.0\linewidth]{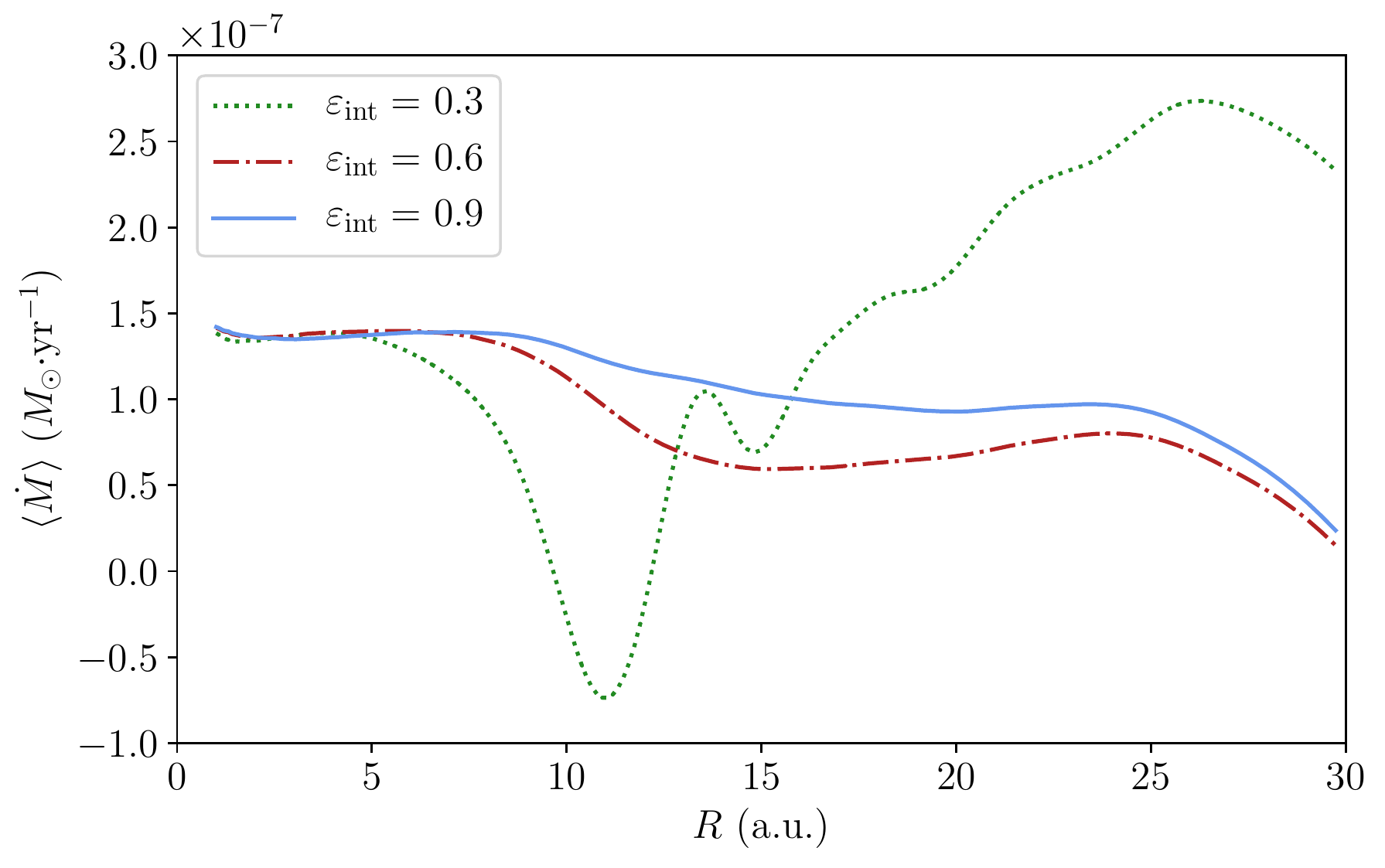}
\caption{\label{mdot_fid2D}Accretion rate for different integration height scales with respect to the radius inside the disc. The higher $\rho v_r$ is integrated the closer to a constant value $\dot{M}$ is in the cavity. The average value inside the cavity (from $R=1$ to $R=10$) is $\dot{M} = 1.4\pm 0.2\times 10^{-7}~M_\odot.\text{yrs}^{-1}$.}
\end{center}
\end{figure}

\noindent The height over which $\rho v_r$ is integrated has a direct influence on $\dot{M}$ mostly because of the elbow-shaped stream lines.
It is then useful to change the thickness of the integration domain which is controlled by the parameter $\varepsilon_\text{int}\equiv \tan\left[\left(\theta_+-\theta_-\right)/2\right]$.
Results are presented in Fig. \ref{mdot_fid2D} for~$3$ values of $\varepsilon_\text{int}$.
For $\varepsilon_\text{int}=0.3$ and around $R=10$, the accretion rate is close to zero indicating that the gas does not plunge directly in the cavity from the disc midplane. This radius corresponds to the location of the basis of the elbow-shaped loop along which the gas is moving.
Averaging higher above the disc allows us to cancel out this effect. Moving to $\varepsilon_\text{int}=0.6$ and $0.9$, the accretion rates in the disc and in the cavity eventually match by less than $50\%$, despite a jump of more than two orders of magnitude in $\Sigma$. This clearly indicates that the accreted material effectively `jumps' above the transition radius, and that a steady state is reached with the whole system (cavity$\,+\,$outer disc) accreting at a constant rate.

\begin{figure}
\begin{center}

\includegraphics[width=1.0\linewidth]{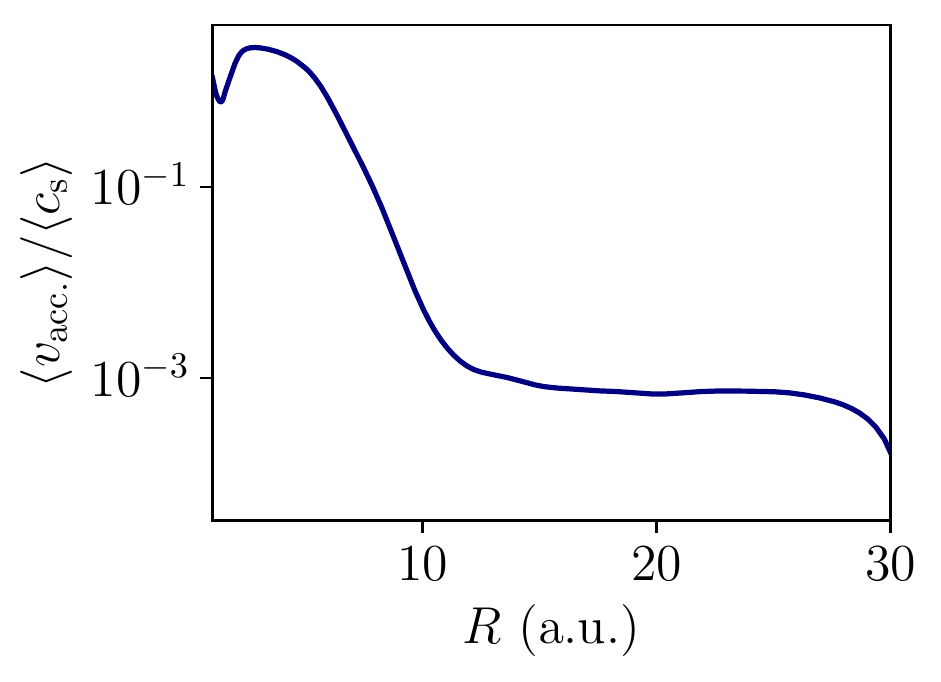}
\caption{\label{vacc_fid2D}Accretion speed for $\varepsilon_\text{int}=0.9$ in units of local sound speed $c_\text{s}$. The profile exhibits a clear transition between subsonic and transsonic accretion that occurs where the edge of the cavity is located.}

\end{center}
\end{figure}

The fact that the accretion rate is approximately constant while the surface density decreases by two orders of magnitude implies that the accretion speed should increase dramatically. This is clearly visible in Fig.~\ref{vacc_fid2D} which shows the radial profile of the accretion speed $v_\text{acc.}$ for $\varepsilon_\text{int}=0.9$, defined by
\begin{equation}
\langle v_\text{acc.} \rangle \equiv \frac{\langle\dot{M}\rangle}{2\pi\,R\,\langle \Sigma\rangle}.
\end{equation}
This velocity profile exhibits a well-defined transition between subsonic accretion outside the cavity with $\langle v_\text{acc.} \rangle \sim 10^{-3}\,\langle c_\text{s} \rangle$ and transsonic accretion inside with $\langle v_\text{acc.} \rangle \sim \langle c_\text{s} \rangle$.

\subsubsection{Governing equations for accretion}

Accretion theory can be understood as the secular evolution of $\dot{M}$ and $\Sigma$.
In systems driven by MHD processes, these two quantities are usually supplemented by the magnetic field $B_z$ threading the disc. 
We apply the vertical integration procedure to the mass and angular momentum conservation equations which become
\begin{align}
\partial_t\Sigma - \frac{1}{2\pi\,r}\partial_r\dot{M} &= -\left[\sin\theta\,\rho v_\theta\right]_{\theta_+}^{\theta_-}\label{mass_conservation} \\
\partial_t\left(\overline{r\,\sin\theta\,\rho v_\varphi}\right) -\frac{1}{2\pi\,r}\dot{M}\,\partial_r\left(r^2\,\sin^2\theta\,\tilde{\Omega}(r)\right) &=-\frac{1}{r}\partial_r\left(r^2\,W_{r\varphi}\right) - W_{\theta\varphi} \label{momentum_conservation}
\end{align}
where we have defined $W_{r\varphi}$ and $W_{\theta\varphi}$ respectively the radial and surface stresses by
\begin{equation}
\label{stresses}
  \left \{
  \begin{aligned}
    W_{r\varphi} &\equiv \overline{\rho\,\sin\theta\,v_r\,v_\varphi} - \overline{\sin\theta\,\frac{B_r\,B_\varphi}{4\pi}} \\
W_{\theta\varphi} &\equiv \left[r\,\sin^2\theta\,\left(\rho\, v_\theta\,v_\varphi-\frac{B_\theta\,B_\varphi}{4\pi}\right)\right]_{\theta_+}^{\theta_-}
  \end{aligned} \right..
\end{equation} 
We recall that we use a peculiar definition of the velocity deviation $\boldsymbol{v}$ so that no additional surface terms appear in Eq.~\ref{momentum_conservation}.
In order to take into consideration the role of the magnetic wind, we complete this set of equations by the vertical magnetic flux conservation 
\begin{equation}
\partial_t B_{\theta,\,0} = \frac{1}{r}\,\partial_r\left(r\,\mathcal{E}_{\varphi,\,0}\right).
\label{flux_conservation}
\end{equation}

\subsubsection{Mass conservation and mass loss rate parameter}

\begin{figure}
\begin{center}

\includegraphics[width=1.0\linewidth]{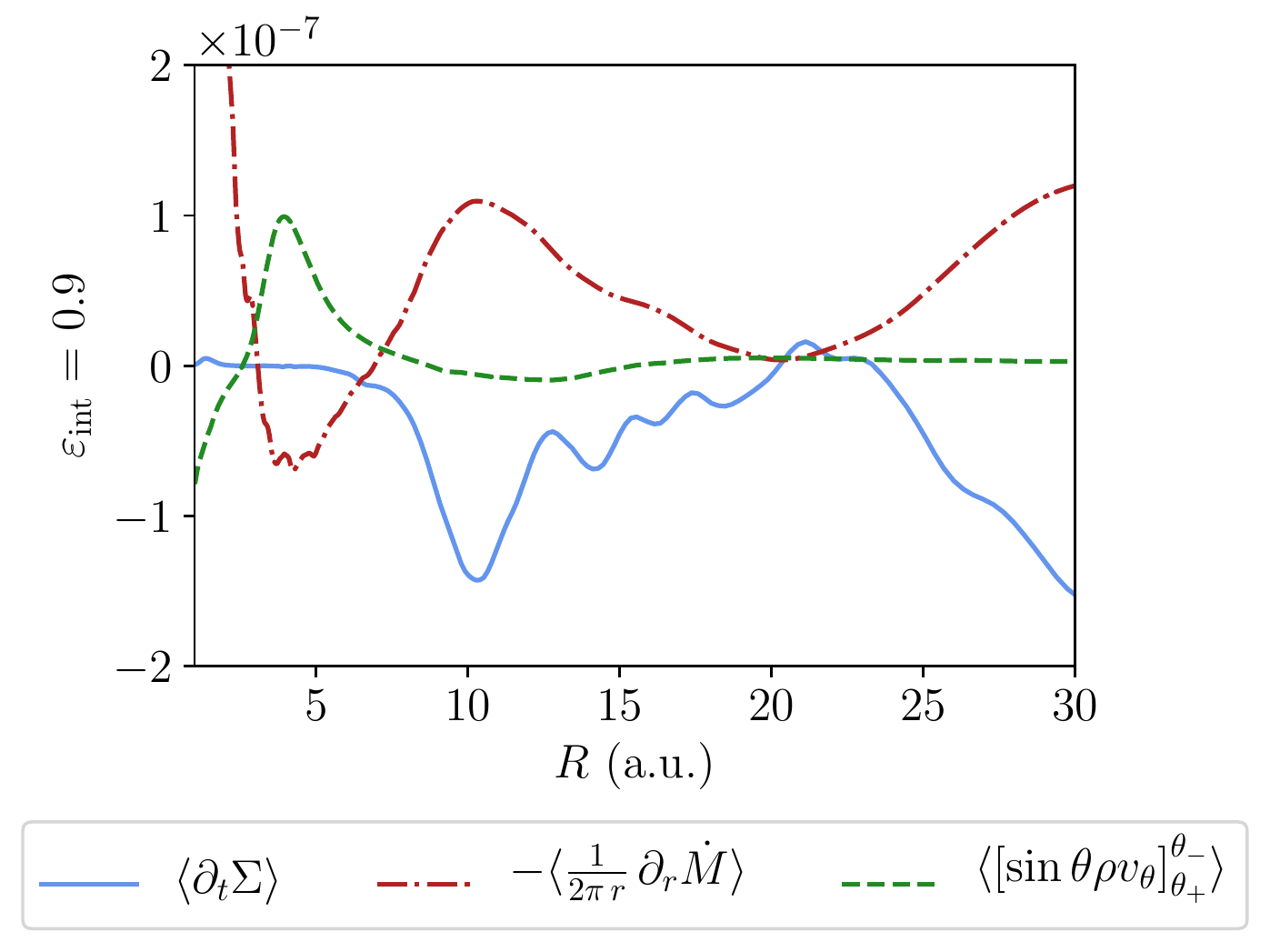}
\caption{\label{mass_hr_fid}Mass conservation for  $\varepsilon_\text{int}=0.9$. The time derivative of $\Sigma$ remains perfectly constant and equal to zero inside the cavity and only gets moderately negative at the cavity edge $R=10$. This suggests that the simulation indeed reaches a steady state for radii up to $\approx 8$. The mass conservation is also correctly recovered. The three lines do not add up to zero because we use a moving average for better visibility and the quantities are time-averaged on a sample selection of output files that do not contain all the timesteps computed by the code.}
\end{center}
\end{figure}

The mass conservation equation is given by Eq.~\ref{mass_conservation}.
Figure \ref{mass_hr_fid} shows the mass conservation for $\varepsilon_\text{int}=0.9$ with time-averaged quantities.
The first information is that inside the cavity, the time derivative of $\Sigma$ is close to zero, meaning the simulation reaches a steady state up to $R\approx 8$.
Closer to the cavity edge, we note that this same term is negative which is linked to the slow expansion of the cavity, as it will be discussed later in section~\ref{slow_evolution}.

The main contribution of the wind mass loss is located in the cavity at $R<5$ and is completely compensated by the radial derivative of the accretion rate.
Additionally, the `wind' mass flux turns negative around the cavity edge, which is due to  matter being accreted from the outer disc atmosphere (see the `elbow-shaped structure' in the poloidal streamlines).

\begin{figure}
\begin{center}

\includegraphics[width=1.0\linewidth]{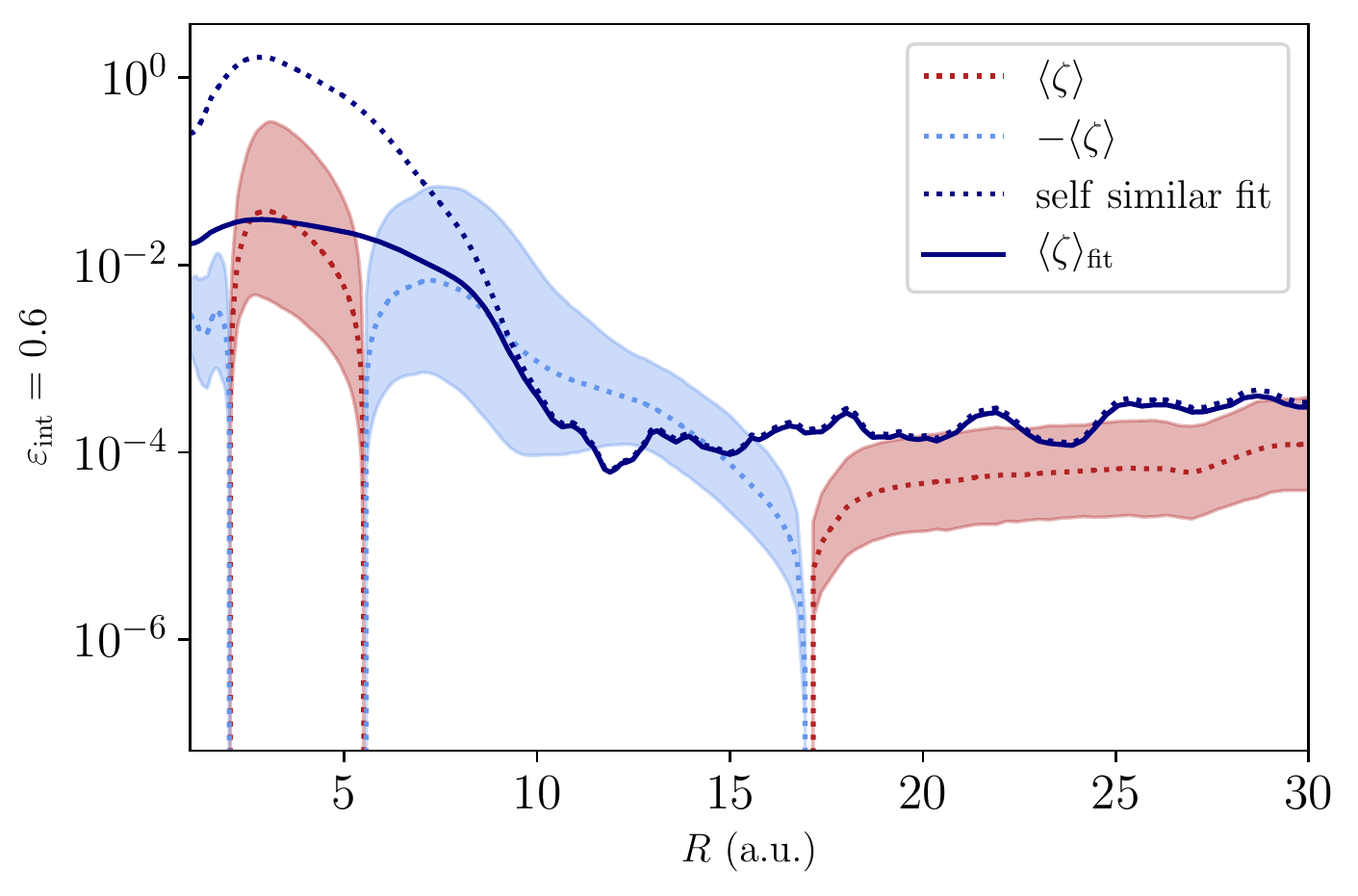}
\caption{\label{zeta_fit_fid2D}$\langle\zeta\rangle$ parameter for $\varepsilon_\text{int} = 0.6$. The self similar fit shown here is for comparison only and was obtained also for $\varepsilon_\text{int}=0.6$. It appears that it is coherent for the external disc while it predicts a wind way too intense in the internal part, therefore, another model is used to describe $\langle\zeta\rangle$ in the whole disc.}

\end{center}
\end{figure}

In order to quantitatively account for the role of the wind, we construct the mass loss rate parameter $\zeta=\zeta_++\zeta_-$ \citep{lesur_systematic_2021}, where $\zeta_+$ and $\zeta_-$ are defined by
\begin{equation}
\langle\zeta_\pm\rangle\equiv\pm \frac{\langle\rho v_z\rangle(\theta_\pm)}{\langle\Sigma\rangle\,\Omega_\text{K}} = \pm \frac{\langle\rho v_r\,\cos\theta\rangle - \langle\rho v_\theta\,\sin\theta\rangle}{\langle\Sigma\rangle\,\Omega_\text{K}},
\end{equation}
where the corresponding quantities are time-averaged.
The signs of $\zeta_\pm$ are chosen accordingly so that a positive value of $\zeta_\pm$ corresponds to matter leaving the surface at $\theta_\pm$.
Since $\zeta_+$ and $\zeta_-$ are pretty much symmetric with respect to the midplane, we focus on $\zeta$ only.
The results are illustrated in Fig.~\ref{zeta_fit_fid2D} where both $\langle\zeta\rangle$ and $-\langle\zeta\rangle$ are shown.
In order to compare with self-similar models \citep{lesur_systematic_2021}, we study the values of $\langle\zeta\rangle$ at $z_0 = 6\,h$ which corresponds to $\varepsilon_\text{int}=0.6$.
The mass loss rate parameter is approximately constant in the external part of the disc around $6.2\times 10^{-5}$,  while it peaks at $2.9\times 10^{-2}$ in the inner part.
We find two zones where $\langle\zeta\rangle<0$.
One is close to the inner boundary and probably a boundary condition artefact, while the other extends from $R\approx 5$ to $R\approx 17$ a.u. and is related to the material falling down on the disc around the transition zone,
such a contribution being notably stronger for $\varepsilon_\text{int}=0.6$.

To compare to self-similar solutions, we show the self-similar scaling of the mass loss rate parameter with respect to $\langle\beta\rangle$ derived by \citep{lesur_systematic_2021} which reads $\langle\zeta_\text{self}\rangle = 0.24\,\langle\beta\rangle^{-0.69}$.
It comes as no surprise that this fit does not account for negative values of $\langle\zeta\rangle$ since these are due to the transition radius, which is not self-similar by essence. 

The wind mass loss rate parameter is smaller than the self-similar scaling in the outer disc by a factor of a few. This discrepancy is probably due to the influence of the cavity magnetosphere that compresses the disk magnetosphere, resulting in a deviation of $\zeta$ from the self-similar result. Moreover, it seems that the further we move outward, the closer we get to the self-similar values, indicating that we recover self-similar scalings far `enough' from the cavity, as expected.

In the cavity, $\zeta$ is significantly weaker than expected from a naive extrapolation of self-similar scaling laws. This indicates that the mass loss rate saturates at $\beta\sim 1$, a regime which has not been explored by \cite{lesur_systematic_2021}.

An alternative model to the self similar one is used to describe $\langle\zeta\rangle$ with greater accuracy. The self similar fit is kept for the external parts of the disc $\zeta_\text{ext} = \zeta_{0,\,\text{ext}}\,\langle\beta\rangle^{a_
 \text{ext}}$, with $a_\text{ext}=-0.69$ and $\zeta_{0,\,\text{ext}}=0.24$. Another one is then calculated for the inner part only, $\zeta_\text{int} = \zeta_{0,\,\text{int}}\,\langle\beta\rangle^{a_
 \text{int}}$, with $a_\text{int}<0$, so that the final profile is given by
 \begin{equation}
 \langle \zeta \rangle_\text{fit} = \frac{ \zeta_{0,\,\text{ext}}\,\langle\beta\rangle^{a_
 \text{ext}}}{1 + \frac{\zeta_{0,\,\text{ext}}}{\zeta_{0,\,\text{int}}}\,\langle\beta\rangle^{a_\text{ext} - a_\text{int}}}.
 \end{equation}
We get $a_\text{int} = - 0.20$ and $\zeta_{0\,\text{int}} = 0.018$.
Such a model, with $a_\text{ext} - a_\text{int}<0$ allows to recover both the~2 previous regimes with a reasonably accurate depiction of the disc.
The final profile exhibits a transition occurring at $\beta_\text{t}\approx 5$ which is close to the lowest value of the ones used to build the self similar fit in \citep{lesur_systematic_2021}.
The final curves are rendered in Fig.~\ref{zeta_fit_fid2D}.
The fit does not account for the negative values, but properly catches both the inner and external parts of the disc.

\subsubsection{Angular momentum conservation}

\begin{figure}
\begin{center}

\includegraphics[width=1.0\linewidth]{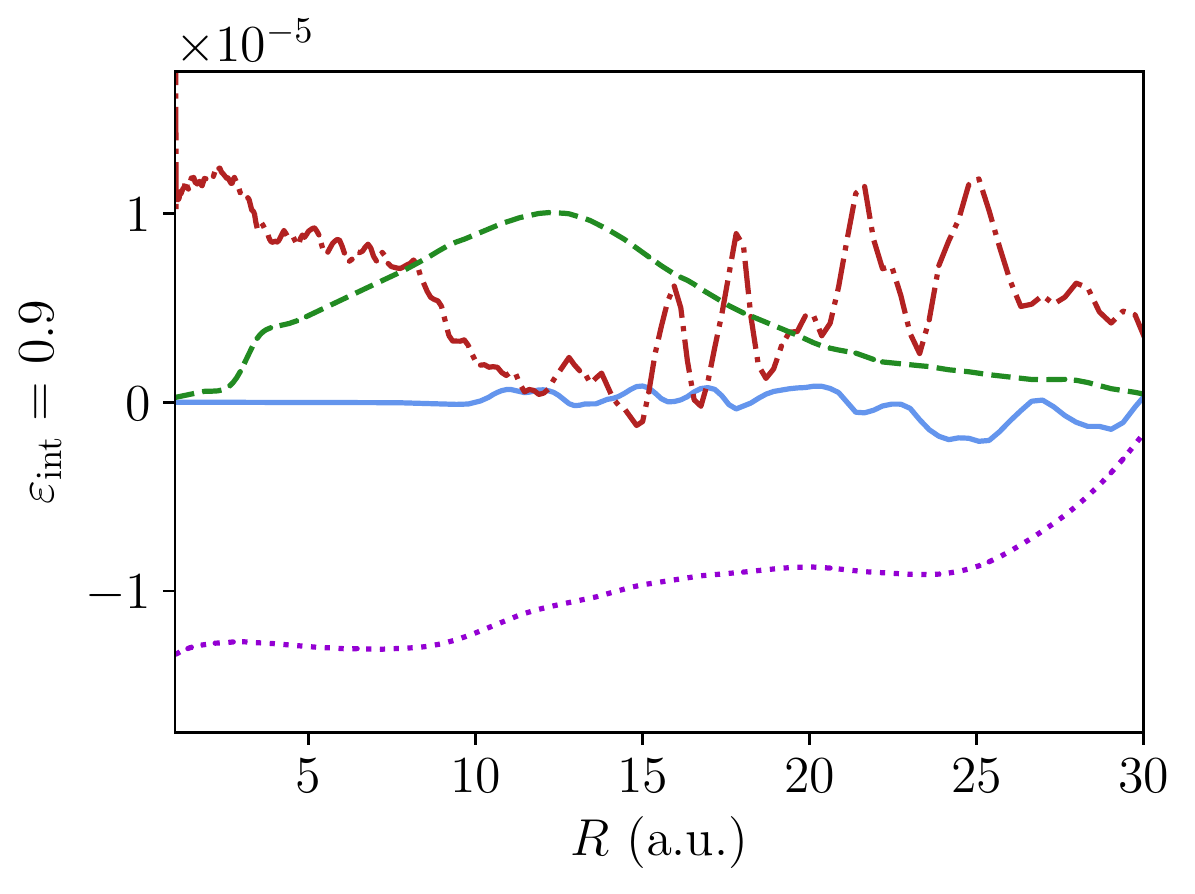}
\caption{\label{momentum_hr_fid2D}Angular momentum conservation multiplied by $r^{-3/2}$ and time-averaged. Full blue line is $\langle\partial_t(\overline{r\,\sin\theta\,\rho v_\varphi})\rangle\,r^{-3/2}$, red dot-dashed line is $\langle \partial_r(r^2\,W_{r\varphi})\rangle\,r^{-3/2}$, green dashed line is $\langle W_{\theta\varphi}\rangle\,r^{-3/2}$ and purple dotted line is $- \langle\frac{1}{2\pi\,r}\dot{M}\,\partial_r(r^2\,\tilde{\Omega})\rangle\,r^{-3/2}$.}%, where $\tilde{\Omega}=\Omega_\text{K}/\sin^2\theta$.}

\end{center}
\end{figure}

We show in Fig.~\ref{momentum_hr_fid2D} the terms involved in the angular momentum conservation equation (\ref{momentum_conservation}), time-averaged and multiplied by $r^{-3/2}$ for better readability.

The integration height is $\varepsilon_\text{int}=0.9$ and chosen so that the influence of  the cavity edge is diminished.
In contrast to the mass conservation equation, the time derivative is negligible.
The surface stress (`wind') removes angular momentum from the whole disc with a major contribution right after the cavity at $R\approx13$. We also observe that the radial stress is always positive except at the cavity edge.

\begin{figure}
\begin{center}

\includegraphics[width=1.0\linewidth]{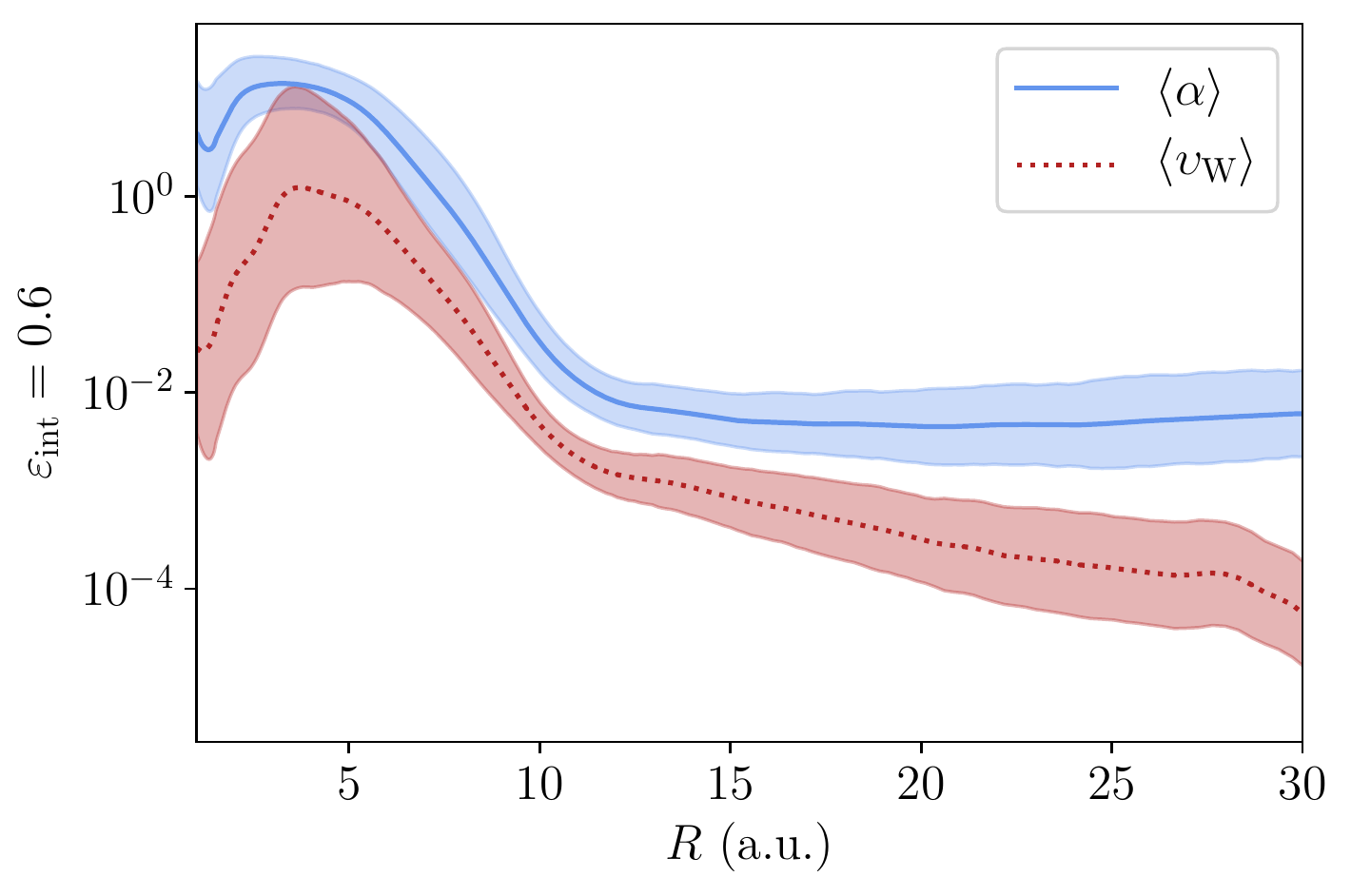}
\caption{\label{upsilon_alpha_fid2D}Time-averaged transport coefficients $\langle\alpha\rangle$ and $\langle\upsilon_\mathrm{W}\rangle$ for $\varepsilon_\text{int} = 0.9$.}

\end{center}
\end{figure}

Such a cancellation suggests that 2~accretion regimes are observed in the disc, which echoes the radial profile of both the accretion rate and speed.
To characterise the radial stress term, we introduce \cite{shakura_black_1973} $\alpha$ parameter.
It must be noted that the origin of this stress is in no way solely linked to turbulence and considerably driven by the laminar structure of the magnetic wind.
The appendix~\ref{Laminar_coef} details the origin of the stress and sheds light on the turbulent vs. laminar origin of $\alpha$.
Nevertheless, the $\alpha$ parameter can still be used in this wind model whose definition when time-averaged is
\begin{equation}
\langle\alpha\rangle \equiv \frac{\langle W_{r\varphi}\rangle}{\langle\overline{P}\rangle}.
\end{equation}
The corresponding profile is given in Fig.~\ref{upsilon_alpha_fid2D}, where $\varepsilon_\text{int}=0.9$.
In the external part of the disc, $\langle\alpha\rangle = 49\pm 5 \times 10^{-4}$ while it reaches a maximum value inside the cavity $\langle\alpha\rangle = 13\pm5$.

Following a similar procedure as the one for $\alpha$, we define a dimensionless number associated to the surface stress component, $\upsilon_\mathrm{W}$. 
As for $\zeta$, we define $\upsilon_{\mathrm{W},\,\pm}$ which are chosen to be positive for angular momentum leaving the disc on both sides:
\begin{equation}
\langle \upsilon_\mathrm{W} \rangle = \langle\upsilon_{\mathrm{W},\,+}\rangle + \langle\upsilon_{\mathrm{W},\,-}\rangle = \frac{\langle W_{\theta\varphi\rangle}}{r\,\langle P_0\rangle}.
\end{equation}
We show the dependence of $\upsilon_\mathrm{W}$ on $R$ in Fig.~\ref{upsilon_alpha_fid2D}.
In the external disc, $\langle\upsilon_\mathrm{W}\rangle = 2.3\pm1.1\times 10^{-4}$ while it rises up to $\langle \upsilon_\mathrm{W} \rangle = 1.0\pm 0.1 $ inside the cavity.
The same observations as for $\langle\zeta\rangle$ are drawn for both $\langle\alpha\rangle$ and $\langle\upsilon_\mathrm{W}\rangle$.
Therefore, two separated regimes are at stake in the disc.
The outer disc regime is typical of wind-emitting protoplanetary discs,  with transport coefficients close to the ones found in self-similar wind models for $\beta\sim 10^4$, indicating that the dynamical properties of the outer disc are not perturbed by the presence of the cavity. 
On the contrary, the second regime describes the inner part of the disc with fast accretion and high values for  $\alpha$ and $\upsilon_\mathrm{W}$, which are both of the order of unity.
Table~\ref{transport_table} displays the transport coefficients values for all the simulations.

\subsection{MHD wind}

It is well known that steady-state MHD winds in ideal MHD can be characterised by a set of MHD invariants \citep{blandford_hydromagnetic_1982} which are conserved quantities along each poloidal field lines (Fig.~\ref{mag_field_struct_fid2D}).
In our axisymmetric simulations, a steady-state is approximately achieved above the disc, in the ideal MHD region.
Hence, we can measure these invariants on fields lines attached in the cavity and in the outer disc.

In the following, we select a field line anchored in the disc midplane at $R\equiv R_\text{w}$.
The corresponding Keplerian angular velocity is $\Omega_\text{w}$ while $B_\text{w}$ is the poloidal magnetic field at the midplane.
We then consider the following invariants, built on time-averaged quantities and listed in table~\ref{wind_table}

\begin{itemize}
\item The mass loading parameter which accounts for the quantity of matter that escapes the disc with the wind
\begin{equation}
\kappa \equiv 4\pi\,\frac{\rho\,v_\text{p}\,\Omega_\text{w}\,R_\text{w}}{B_\text{p}\,B_\text{w}}.
\end{equation}
\item The rotation parameter
\begin{equation}
\omega \equiv \frac{\Omega}{\Omega_\text{w}} - \frac{\kappa\,B_\text{w}\,B_\text{p}}{4\pi\,\rho\,R\,R_\text{w}\,{\Omega_\text{w}}^2}.
\end{equation}
\item The magnetic lever arm that accounts for the angular momentum that is removed from the disc by the wind
\begin{equation}
\lambda \equiv \frac{\Omega\,R^2}{\Omega_\text{w}\,{R_\text{w}}^2} - \frac{R\,B_\varphi}{R_\text{w}\,B_\text{w}\,\kappa}.
\end{equation}
\end{itemize}
Of course, these invariants echoes the transport coefficients definitions previously used to describe the disc and one expects $\kappa\approx \beta\,\zeta /4\,\varepsilon$ and $\lambda\approx 1+\varepsilon\,\upsilon_\mathrm{W}/\zeta$ \citep{lesur_magnetohydrodynamics_2021}.

\begin{figure*}
\begin{center}
\includegraphics[width=1.0\linewidth]{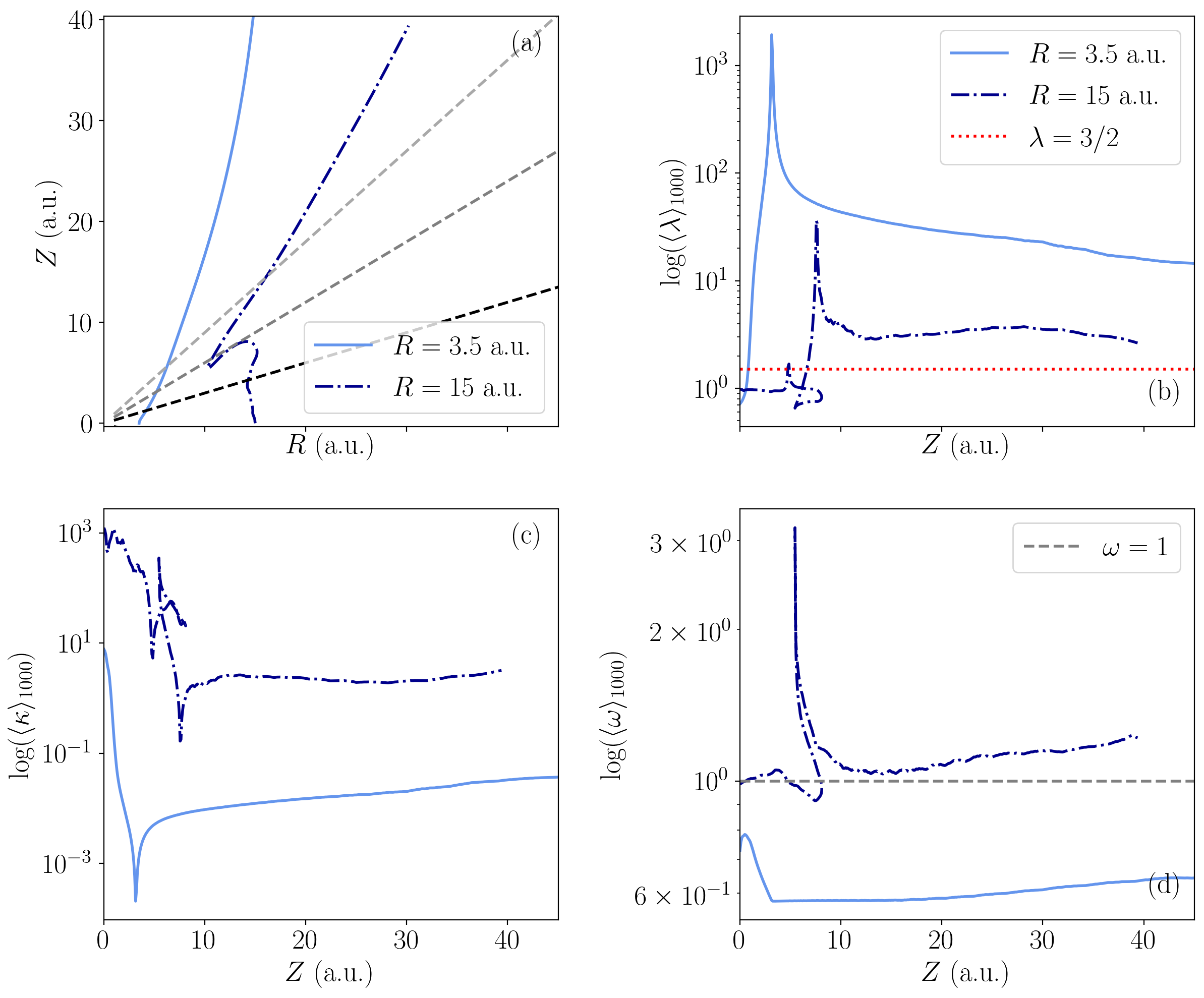}
\caption{\label{wind_fid2D}The first panel shows the field lines in the internal and external disc. The grey dashed lines represent the surfaces at $h/R=0.3;\,0.6$ and $0.9$. The 3 other plots display the MHD invariants for the internal field line (full light blue) and the external one (semi dashed dark blue). The invariants are time-averaged on the last 1000 orbits.}
\end{center}
\end{figure*}

To compute these invariants, we arbitrarily choose one field line in the cavity (referred to as `in') leaving the midplane at $R_\text{in}=5~$a.u. and one in the external disc (referred to as `ext') leaving the midplane at $R_\text{ext}=15~$a.u (see the first panel of Fig.~\ref{wind_fid2D}). Note that the disc thickness affects the MHD invariants since the physical foot points of the field lines are not located at the midplane but slightly above.
Such limitation especially concerns the field lines in the external disc which are subject to a large scale oscillation close to the transition radius. Therefore, the calculated MHD invariants are subject to caution and we only draw general conclusions regarding the nature of the wind.

We show the invariants along the chosen field lines in Fig.~\ref{wind_fid2D}. We find that all of the invariants remain reasonably constant once high enough above the disc, as expected from a steady-state ideal MHD flow. The wind launched from the cavity is different from the disc one. The cavity wind has a much weaker mass loading parameter and a much larger lever arm (by almost a factor $10$). We also find that its rotation parameter differs significantly from 1, indicating that field lines are rotating at 80\% of $\Omega_\text{K}$ in the cavity. This point is probably related to the fact that the disc itself is sub-Keplerian in this region (Fig.~\ref{initial_avg_inifid2D}).
Quantitatively, we find $\kappa_\text{in} = 2.2\times 10^{-2}$, $\kappa_\text{ext} = 2.5$, $\lambda_\text{in} = 23$ and $\lambda_\text{ext}=3.2$.
These values are coherent with the transport coefficients computed in previous sections. We also note that the values of $\kappa$ and $\lambda$ in the cavity match some of the historical solutions of \cite{blandford_hydromagnetic_1982} (see their figure~2), which correspond to super-Alfv\'enic and collimated outflows. These values are also consistent with the magnetic outflow solutions of \cite{Ferreira97} (see figure 3). Hence, the cavity we find quantitatively matches the inner JED proposed by \cite{Combet08}.

\subsection{Temporal evolution}

We observe two kinds of time variability in the fiducial simulation: a secular variability responsible for the slow expansion of the cavity, and a short timescale variability, responsible for the striped patterns observed in space-time diagrams (Fig.~\ref{sig_beta_avg_fid2D}). We start here our exploration of time variability by focusing on the secular evolution, beginning with a discussion of the cavity expansion.

\subsubsection{Slow cavity edge expansion}
\label{slow_evolution}
As previously mentioned, the cavity edge moves slowly outwards during the simulation.
Neglecting the impact of the wind in terms of mass loss rate at the cavity edge location, which is coherent with Fig.~\ref{mass_hr_fid}, and assuming piecewise constant accretion rates and surface densities across the cavity edge, one gets
\begin{equation}
\dot{R}_0 = -\frac{1}{2\pi\,R_0}\,\frac{\delta\dot{M}}{\delta\Sigma},
\label{cavity_edge}
\end{equation}
where $\dot{R}_0$ is the cavity edge `velocity' and  $\delta\dot{M}$ and $\delta\Sigma$ are the jump in accretion rate and surface density at the cavity edge. By calculating $\dot{M}$ and $\Sigma$ around $R_0$, we find $\dot{R}_0 = 1.8\times 10^{-5}$ while evaluating directly the cavity edge motion $\dot{R}_0$ yields $\dot{R}_0 = 1.4\times 10^{-5}$ (both in c.u.).
Therefore, the cavity is expanding because of the slight mismatch in accretion rate observed in Fig.~\ref{mdot_fid2D}.

\subsubsection{Magnetic field transport}

\begin{figure}
\begin{center}
\includegraphics[width=1.0\linewidth]{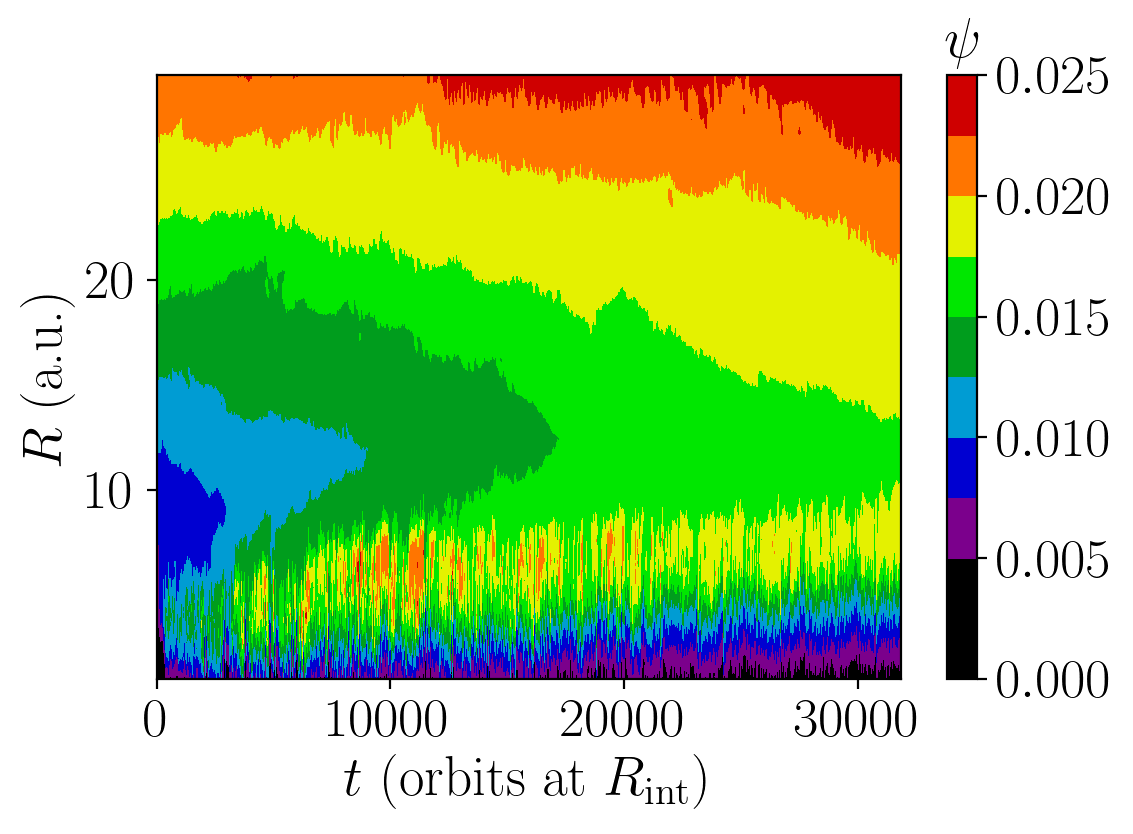}
\caption{\label{psi_field_fid}Flux function $\psi(R, t)$ for the magnetic field, taking into account the flux at the surface of the seed and the radial flux. For radii larger than the one of the cavity, the field lines are advected towards the centre during the whole simulation. At $R\approx 13$, the flux accumulates and exhibits a striped structure for smaller radii.}
\end{center}
\end{figure}
To interpret the time evolution of the magnetic field, we study the transport of magnetic flux inside the disc and define a flux function $\psi$ such that
\begin{equation}
\psi(r, t) = R_\text{int}^{\;2}\,\int_{0}^{\,\pi/2}B_{r}(R_{\text{int}},\theta, t) \,\sin\theta\,\text{d}\theta - \int_{r=R_\text{int}}^{\,r} r\,B_{\theta}(r, \pi/2, t)\,\text{d}r.
\label{psi}
\end{equation}

Assuming the total flux is constant with respect to time, the iso-contours of $\psi$ describe the motion of the magnetic field lines in the disc plane.
The spatio-temporal diagram for $\psi$ is shown in Fig.~\ref{psi_field_fid}.
The magnetic flux is advected slowly towards the star in the external disc while it tends to diffuse outwards from the inner part of the disc to the cavity edge. The poloidal magnetic field lines on Fig. \ref{mag_field_struct_fid2D} show that $\langle B_{z,0}\rangle<0$ in the transition region ($8 \lesssim R\lesssim 12$) and $\langle B_{z,0}\rangle>0$ otherwise. This transition region is recovered in Fig.~\ref{psi_field_fid} as a region where $\partial_r\psi<0$.

Overall, we observe that the negative field of the transition region is diffusing outwards, while the positive field of the outer disc is advected inwards. We therefore observe a reconnection of the large scale field around $R\approx 12$, which progressively `eats' the negative field of the transition region. In addition to this, we observe that field lines deep in the cavity also diffuse outwards. 

To get a quantitative estimate of the field line advection speed, we first note that the evolution equations for $\psi$ read
\begin{equation}
  \left \{
  \begin{aligned}
    \partial_t\psi(R,t) &= -R\,\mathcal{E}_{\varphi,0}(R,t)\\
    \partial_R\psi(R,t) &= -R\,B_{\theta,0}(R,t)
  \end{aligned} \right..
\end{equation} 
Following \cite{guilet_global_2014}, we rewrite these evolution equations as an advection equation for $\psi$
\begin{equation}
\partial_t\psi +v_\psi\,\partial_R\psi = 0,
\label{advection_equation}
\end{equation}
where we have defined the `field advection velocity'
\begin{equation}
v_\psi = -\frac{\mathcal{E}_{\varphi,0}(R,t)}{B_{\theta,0}(R,t)}.
\end{equation}
Eventually, we define a dimensionless advection parameter $\nu_\text{B} = v_\psi/v_\text{K}$ which quantifies the advection speed \citep{bai_hall-effect_2017}. In this framework,  positive values of $\nu_\text{B}$ implies an outward transport field while negative values trace inward field transport.

\begin{figure}
\begin{center}

\includegraphics[width=1.0\linewidth]{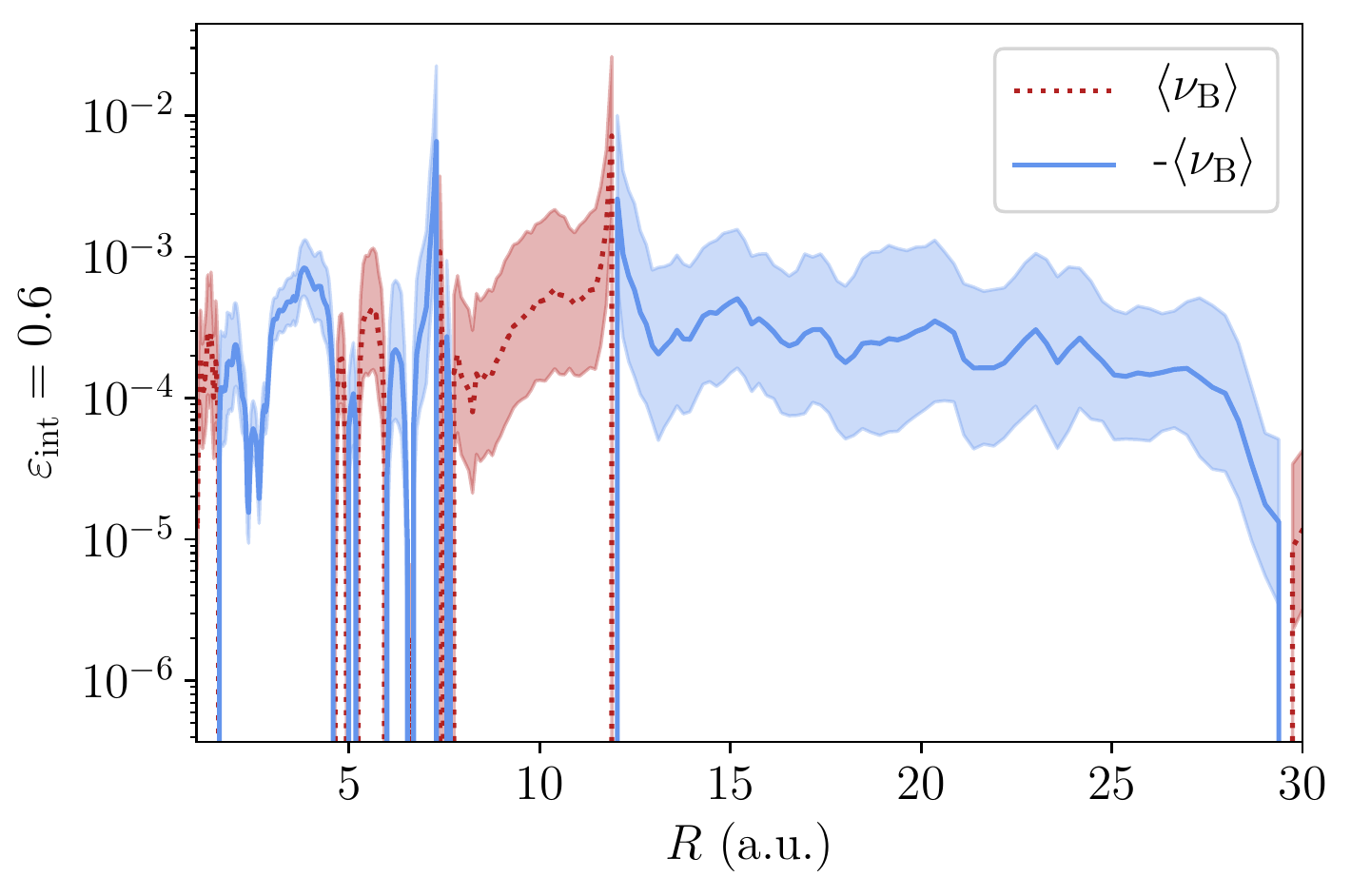}
\caption{\label{nu_B_fit_fid2D}The magnetic field transport parameter $\nu_\text{B}$ as a function of radius. Note that the outer disc is transporting magnetic field lines inwards ($\nu_\text{B}<0$).}

\end{center}
\end{figure}

We show the radial dependence of $\langle\nu_\text{B}\rangle$ in Fig.~\ref{nu_B_fit_fid2D}.
In the external disc we find that the magnetic field is advected inwards with a velocity $v_\psi = -2.6\times 10^{-3}\,v_\text{K}$.
$\nu_\text{B}$ changes its sign multiple times in the cavity, but remains negative close to the cavity edge, between $R\approx 7$ and $R\approx 11~$a.u. where $v_\psi=+3.2\times 10^{-3}\,v_\text{K}$.
Such a result is in accordance with Fig.~\ref{psi_field_fid} and indicates that field lines are converging at the transition radius with opposite vertical polarity.
In the external parts, $\nu_\text{B}$ is negative and $v_\psi = -2.6\times 10^{-3}\,v_\text{K}$ so that vertical magnetic field pointing upwards is advected. We note that this inwards advection of the outer disc field lines is in sharp contrast to other work which focused on `full' discs \citep{bai_hall-effect_2017,lesur_systematic_2021}. We will come back to this discrepancy in the discussion.

\subsubsection{Fast variability of the cavity}
\label{Fast_variability}

Up to this point, we have mostly considered time-averaged quantities, and ignored fast variability. While our numerical solution are quasi-steady if one looks at averages on 100s of orbits, they also exhibit a fast time variability (see the temporal stripes in Fig.~\ref{sig_beta_avg_fid2D}) whose origin ought to be clarified.

\begin{figure*}
\begin{center}
\includegraphics[width=1.0\linewidth]{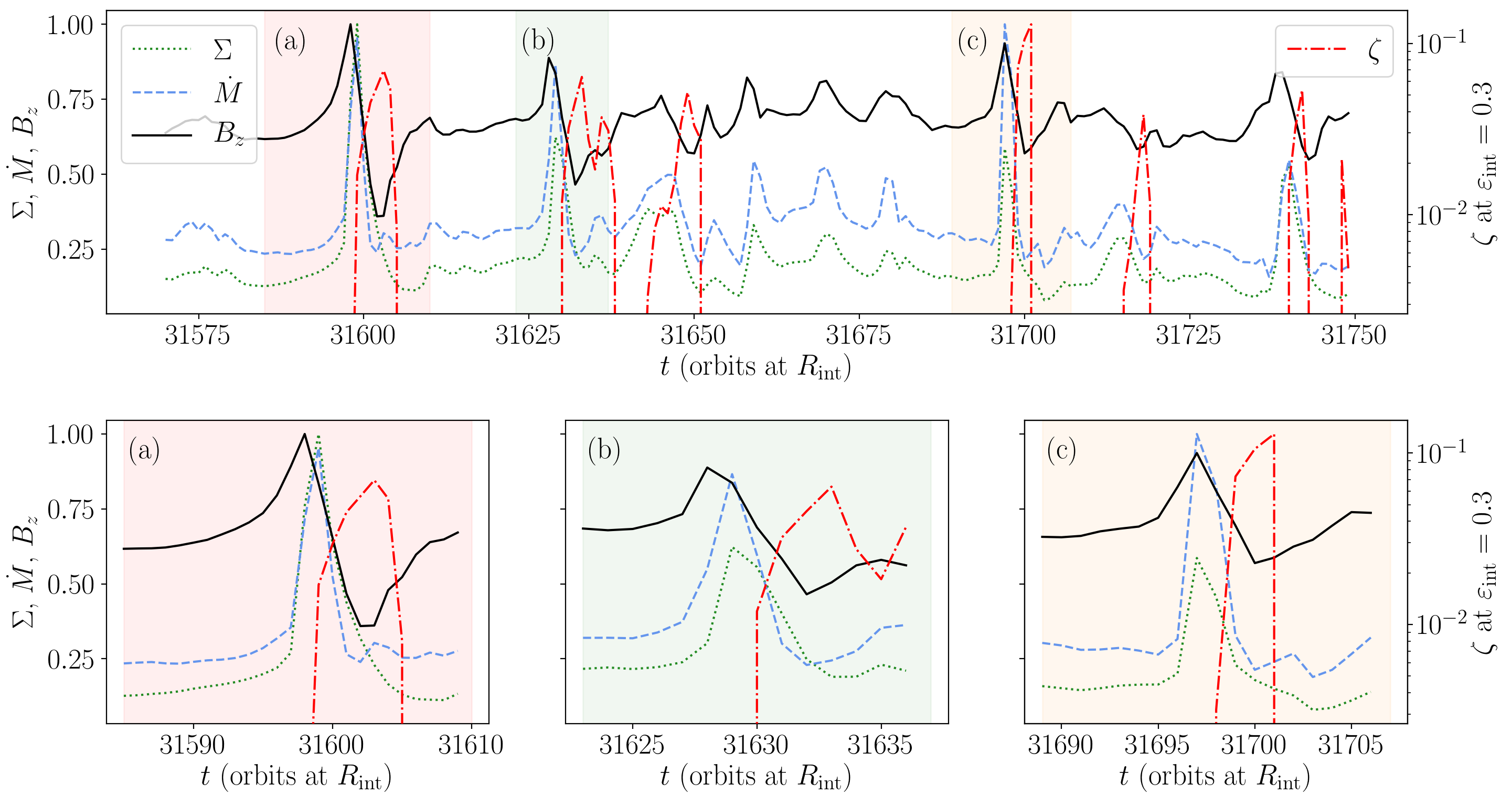}
\caption{\label{temporal_analysis_fid2D} Temporal evolution of $\Sigma$ in dotted green, $\dot{M}$ in dashed blue and $B_z$ (vertically averaged) in black full line at $R=3$~a.u. for the fiducial simulation. $\zeta$ is calculated at $\varepsilon_\text{int}=0.3$ and shown in semi-dashed red line with a logarithmic scale on the right of the panels. Apart from $\zeta$, all the profiles are given in arbitrary units and divided by their maximum value reached during the timescale of the top panel. 
}
\end{center}
\end{figure*}

Figure \ref{temporal_analysis_fid2D} shows such a temporal evolution of $\Sigma$, $\dot{M}$ and $B_z$ at $R=3$.
These profiles encounter sharp fluctuations over time, chaotically distributed.
Therefore, the cavity is subject to bursts of matter that quickly falls onto the star (the typical width of a peak is $\sim 5\,$ orbits at $R_\text{int}$, which is still far larger than our temporal resolution).
This variability explains the stripes seen in the spatio-temporal diagram (Fig.~\ref{sig_beta_avg_fid2D}).

\begin{figure*}
\begin{center}
\includegraphics[width=1.0\linewidth]{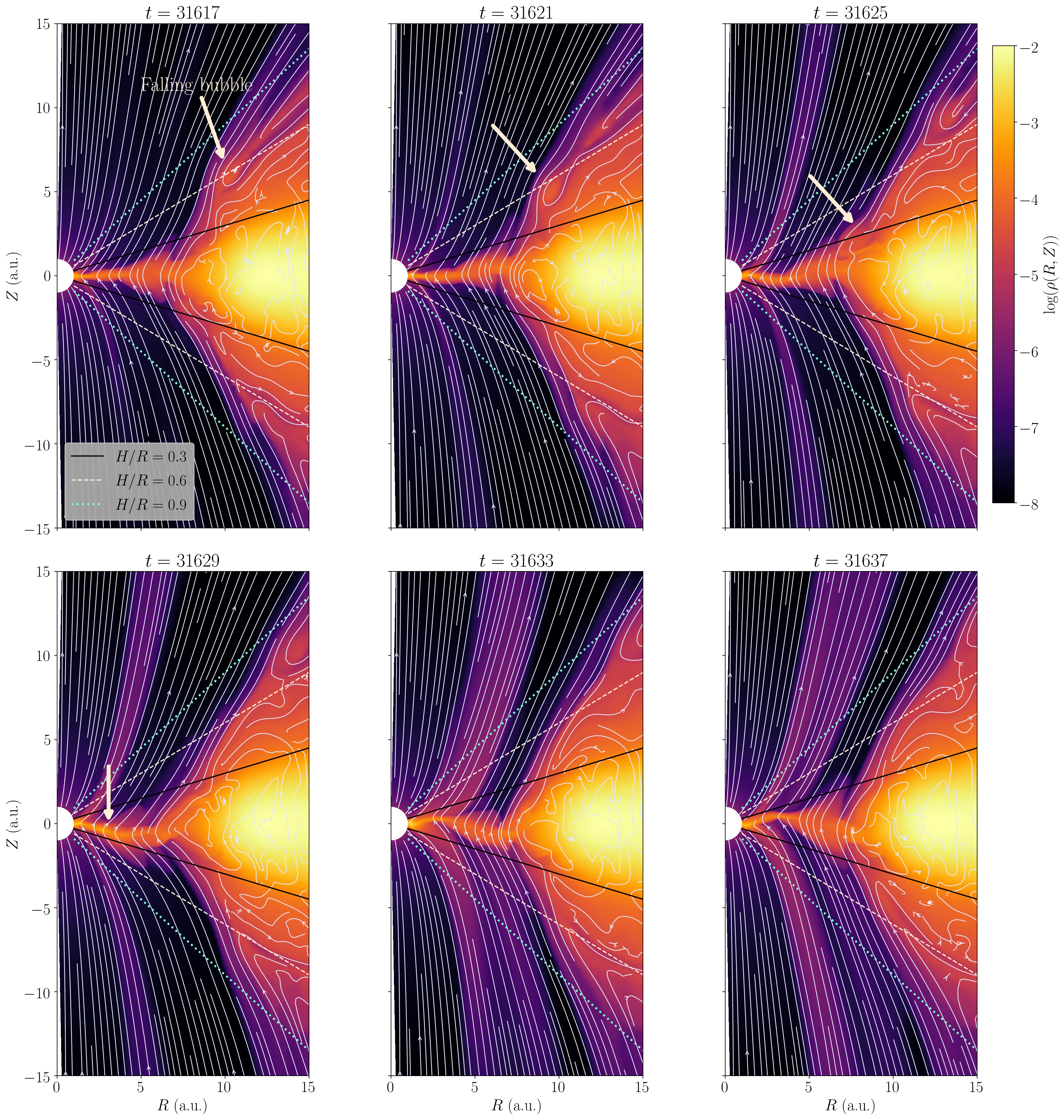}
\caption{\label{blob_panel} Density and magnetic field lines at different times showing the advection of a bubble of material (arrow) from the outer disc through the cavity.}
\end{center}
\end{figure*}

We focus on a few of these bursts in the bottom panels of Fig.~\ref{temporal_analysis_fid2D}, while instantaneous pictures of the density corresponding to the (b) panel are given in Fig.~\ref{blob_panel}.
For these bursts, we see that the local maximum values of $B_z$, $\Sigma$ and $\dot{M}$ are correlated.
When an inflow of matter crosses the cavity, $\Sigma$ peaks as well as $\dot{M}$ which increases $\zeta$.
In terms of temporal sequence, it seems that $B_z$ increases slightly before $\Sigma$ and $\dot{M}$, which would indicate that $B_z$ is the driver of these bursts, but we cannot be definitive on this sequence because of the lack of temporal resolution. Finally, we observe that $\zeta$ is always clearly delayed compared to the other quantities, indicating that the wind inside the cavity ejects more material once the bubble of material has passed.

For a more precise insight on accretion and temporal variability, we refer to Fig. \ref{blob_panel} that shows the density and poloidal magnetic field lines at different times.
On the first panel, we see a filament of matter located above the disc that extends from $(R=10,~Z=5)$ to $(R=15,~Z=10)$.
This structure is cut in two on the second panel, revealing two bubbles of matter, one being about to fall while the other is about to be ejected and to leave the disc in the wind.
Concerning the filament as well as the bubbles formation, we detect a current sheet localised at the location of the filaments, where the total magnetic field cancels ($B_\varphi=0$ at the edge of the magnetic loop and $B_\text{p}=0$ because two antiparallel poloidal field lines meet at the elbow shape structure location). It is therefore a possibility that these structures form due to magnetic reconnection.
Focusing on the falling material, we see it reaching the edge of the cavity on the third panel before crossing it on the next one.
When the gas crosses the cavity, the disc oscillates locally above and below the midplane and is therefore highly dynamical.
With a slight delay (last three panels), we see an outflow emerging from the cavity and the wind density increases.
Such an observation exhibits the link between wind and accretion (see Fig. \ref{mass_hr_fid}).
The ejection of gas from the cavity is not constant with respect to time and occurs occasionally with burst events for which $\zeta$ eventually peaks at $0.1$.
This explain why the effective value of $\langle\zeta\rangle$ is lower than the one predicted by self-similar models for which the ejection is continuous with a higher mass loss rate parameter.

Combining \ref{stream_lines_fid2D} and \ref{blob_panel}, we unveil a general scheme for feeding the cavity.
First, the gas located inside the outer disc elevates from the midplane up to approximately $2$~local disc height and organises itself in a filamentary way.
Then, bubbles of matter fall and cross the cavity, forming the elbow-shaped structure on the time-averaged profile.

\subsubsection{Magnetic Rayleigh Taylor instability}
\label{RTI}
To account for the formation and stability of the bubbles of matter at the cavity edge, we explore the possibility of having a magnetic Rayleigh Taylor instability (RTI) (or interchange instability) in the cavity.

The disc is geometrically thin inside the cavity and the density is relatively continuous radially. Under these conditions, we refer to the analysis of \cite{spruit_mass_1990, spruit_interchange_1995, stehle_stability_2001}, which assume an infinitely thin disc.
We reformulate the instability criterion of \cite{spruit_interchange_1995} (see their equation 59) in terms of the plasma parameter in appendix~\ref{Interchange_appendix}.
The resulting criterion \ref{crit2} states that a necessary condition for the occurrence of the RTI is $\overline{\beta}<\beta_\text{crit.}\simeq 0.0355$.
Figure \ref{sig_beta_avg_fid2D} shows that $\overline{\beta}$ is of the order of $0.1$ in the cavity and rarely go beyond this value, except for very short periods of time, for instance during the accretion `bursts'.

We conclude that the cavity $\beta$ plasma parameter is too large to sustain the RTI on average, but we cannot exclude that it could be triggered in the rare excursions where the cavity reach $\overline{\beta}<0.1$, as during some of the bursts.

\section{Parameters space exploration}

\subsection{Ambipolar Diffusion}

We check the influence of $\Lambda_{\text{A},\,0}$ in the simulation B4Bin0Am1, which is the same as the fiducial one except for the initial value of $\Lambda_\text{A}$ which is set to $10$.

\subsubsection{General structure of the disc and gaps}

\begin{figure*}
\begin{center}
\includegraphics[width=1.0\linewidth]{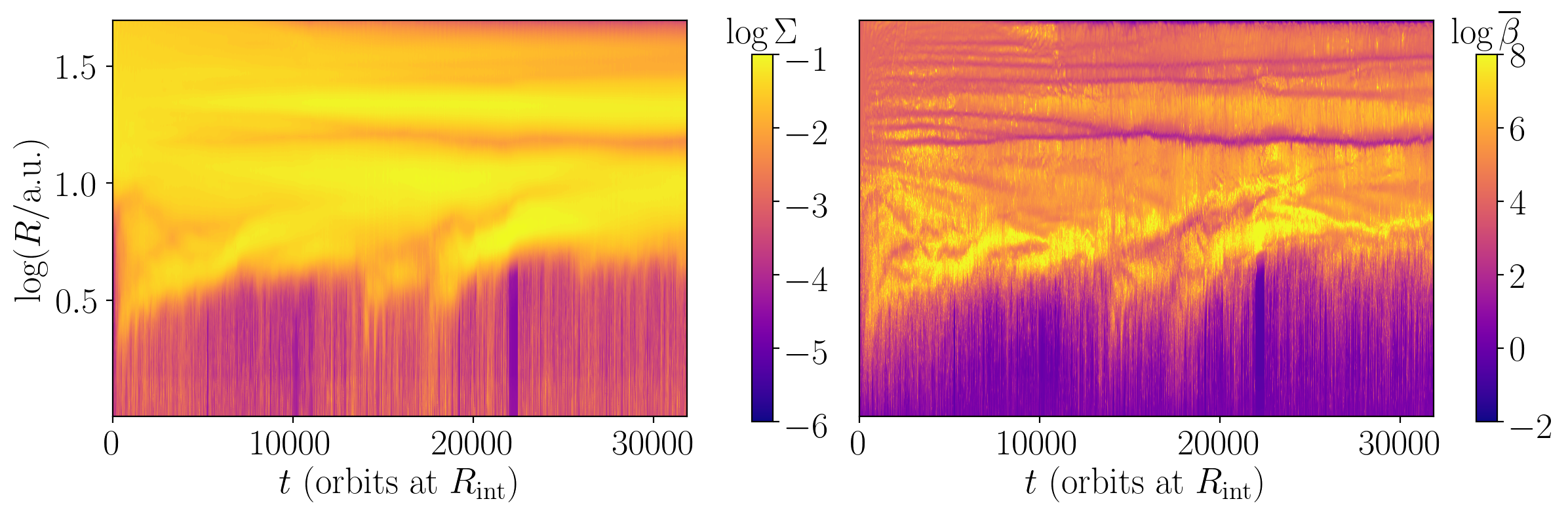}
\caption{\label{sig_beta_avg_S2DB4Bin0Am1} $\Sigma(R,\,t)$ and $\overline{\beta}(R,\,t)$ of B4Bin0Am1. The cavity stands during the whole simulation though its edge falls down to $R\approx 2$~a.u. during the transient, before broadening up to $R\approx 4$~a.u. in a few thousands of orbits at $R_\text{int}$. The profile of $\overline{\beta}$ is characterised by the presence of gaps in the external disc.}
\end{center}
\end{figure*}

The spatio-temporal evolution of $\Sigma$ and $\beta$ are shown in Fig.~\ref{sig_beta_avg_S2DB4Bin0Am1} for $\Lambda_\text{A}=10$.
During the transient state, the cavity edge falls down to $R\lesssim 2~\text{a.u.}$ before expanding back up to $R\gtrsim 4~\text{a.u.}$ in a few thousands of orbits at $R_\text{int}$.
Overall, the transient state lasts for a shorter period of time than in the fiducial run and the cavity extension is smaller.

We observe the apparition of gaps in both the profiles of $\Sigma$ and $\overline{\beta}$ (Fig. \ref{sig_beta_avg_S2DB4Bin0Am1}) located in the external disc and broadening with time.
Such structures are observed in numerous occasions in protoplanetary discs simulations either with ideal \citep{jacquemin-ide_magnetic_2020} or non-ideal MHD \citep{bethune_global_2017, suriano_formation_2019, riols_ring_2020, cui_global_2021}. We observe that gaps are associated with low $\beta$ regions and are localised relatively far from the disc inner boundary. Some gaps merge  with one another, so that only 3 of them remain after $15000$ orbits at $R_\text{int}$, similarly to \cite{cui_global_2021}. We deserve the study of the interaction between these gaps and the cavity to a future paper.

\begin{figure*}
    \centering
    \includegraphics[width=1.0\linewidth]{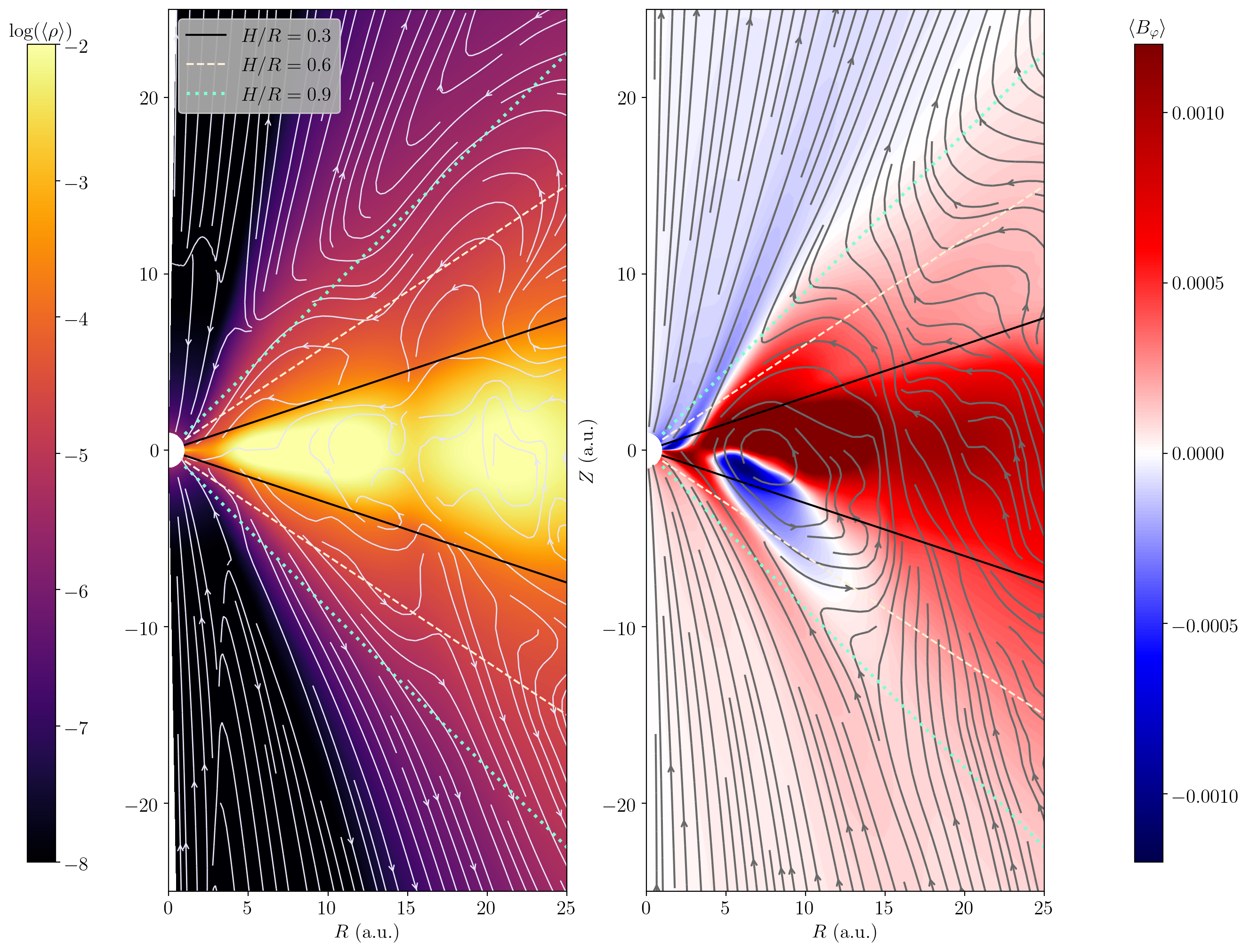}
    \caption{Time-averaged structure of the disc for B4Bin0Am1. Left panel: poloidal stream lines and density. Right panel: magnetic structure of the disc with magnetic poloidal field lines and $\langle B_\varphi\rangle$.}
    \label{stream_lines_mag_zoom_S2DB4Bin0Am1}
\end{figure*}

Figure \ref{stream_lines_mag_zoom_S2DB4Bin0Am1} shows the flow and field topology for $\Lambda_\text{A}=10$ as well as the time-averaged magnetic structure of the disc.
The main features of the fiducial simulation are recovered, namely the elbow-shaped structure and the associated magnetic loop. These are however located closer to the star, the cavity radius being smaller in this simulation.

In contrast to the fiducial simulation, the outer disc is this time top/down asymmetric, which has an impact on the shape of the elbow above and below the disc plane. The elbow is prominent above the disc but almost disappear below, except for a small set of stream lines close to the cavity.
The magnetic field lines exhibit a local slanted symmetry in the external disc at the gaps location.
This is similar to the topology observed in ambipolar dominated discs \citep{riols_dust_2018, riols_spontaneous_2019}.
The gaps seem to be characterised by small vortices in the ($r,\theta$) plane, located at the disc surface at the corresponding radii, indicating a meridional circulation.

\subsubsection{Transport coefficients and wind invariants}

The accretion rate remains constant in the whole disc with a value close to $1.2\times 10^{-7}\,M_\odot.\text{yrs}^{-1}$ and an accretion velocity that is still subsonic in the outer disc and peaks up to $2\,c_\text{s}$ at the internal radius.
Therefore, the accretion picture is identical to the one for the fiducial run with an internal transsonic regime connecting through the cavity edge to a weakly magnetised wind.

Regarding the wind, we obtain a highly mass loaded field line in the external disc that removes little angular momentum ($\lambda_\text{ext}=1.5$ and $\kappa_\text{ext}=5.0$) and a lighter one in the internal disc that carries a massive load of angular momentum ($\lambda_\text{in}=4.9$ and $\kappa_\text{in}=0.24$).
We note that the disc wind is overall less magnetised and more massive, while the general picture of the fiducial run remains.
The rotational invariant contrast is higher than in the fiducial simulation, its internal value being 3 times lower and the external 3 times higher.

 \begin{table*}
     $$ 
         \begin{array}{p{0.15\linewidth}lccccccc}
            \hline
            \noalign{\smallskip}
            Name      & \dot{M}~(10^{-7}\,M_\odot. \text{yrs}^{-1}) & \zeta_\text{in}~(10^{-2}) & \zeta_\text{ext}~(10^{-5}) & \alpha_\text{in} & \alpha_\text{ext}~(10^{-3}) & \upsilon_{\mathrm{W},\,\text{in}} & \upsilon_{\mathrm{W},\,\text{ext}}~(10^{-4})\\
            \noalign{\smallskip}
            \hline
            \noalign{\smallskip}
            \textbf{B4Bin0Am0}& 1.4 & 2.9 & 6.2 & 13 & 4.9 & 1.0 & 2.3\\
            B3Bin0Am0        & 5.1 & 3.8 & 14 & 19 & 23 & 1.4 & 31\\
            B5Bin0Am0        & 0.27 & 2.1 & 4.5 & 6.6 & 1.0 & 0.18 & 0.19\\
            B4Bin0Am1       & 1.2 & -1.8 & 10 & 2.8 & 16 & 0.16 & 2.7\\
            R20FID & 1.1 & 4.5 & 5.8 & 15 & 2.4 & 1.2 & 10\\
            \noalign{\smallskip}
            \hline
         \end{array}
    $$
    \caption{Transport coefficients for a subset of simulations. The accretion rate is calculated inside the cavity.}
    \label{transport_table}
   \end{table*} 
   
   \begin{table*}
     $$ 
         \begin{array}{p{0.15\linewidth}lcccccc}
            \hline
            \noalign{\smallskip}
            Name      & \lambda_\text{in} & \lambda_\text{ext} &\kappa_\text{in} & \kappa_\text{ext} & \omega_\text{in} & \omega_\text{ext} \\
            \noalign{\smallskip}
            \hline
            \noalign{\smallskip}
            \textbf{B4Bin0Am0}& 23 & 3.2 & 2.2\times 10^{-2} & 2.5 & 0.67 & 1.2\\
            B3Bin0Am0        & 185 & 1.3 & 3.1\times10^{-3}  & 9.9 & 0.64 & 0.93\\
            B5Bin0Am0        & 4.4 & 1.2 & 1.3 & 18 & 0.69 & 1.1\\
            B4Bin0Am1       & 4.9 & 1.5 & 0.24 & 5.0 & 0.23 & 3.6 \\
            R20FID	&	26 & 2.1	&	1.8\times 10^{-2} & 2.8 &	0.52 & 1.4 \\
            \noalign{\smallskip}
            \hline
         \end{array}
    $$
    \caption{MHD invariants for a subset of simulations, computed with time-averaged quantities on the last $1000$ orbits at $R_\text{int}$.}
    \label{wind_table}
   \end{table*}

\subsection{Influence of the initial plasma parameter}

We study the impact of the plasma parameter varying both its internal $\beta_\text{in}$ and external $\beta_\mathrm{out}$ initial value.

\subsubsection{Role of the external initial plasma parameter}
\label{beta_ext}
We explore here how the outer disc magnetisation impacts the general properties of the system. We vary the initial value of $\beta$ between $\beta_\mathrm{out}=10^3$ (run B3Bin0Am0) and $\beta_\mathrm{out}=10^5$ (run B5Bin0Am0).

\begin{figure*}
\begin{center}
\includegraphics[width=1.0\linewidth]{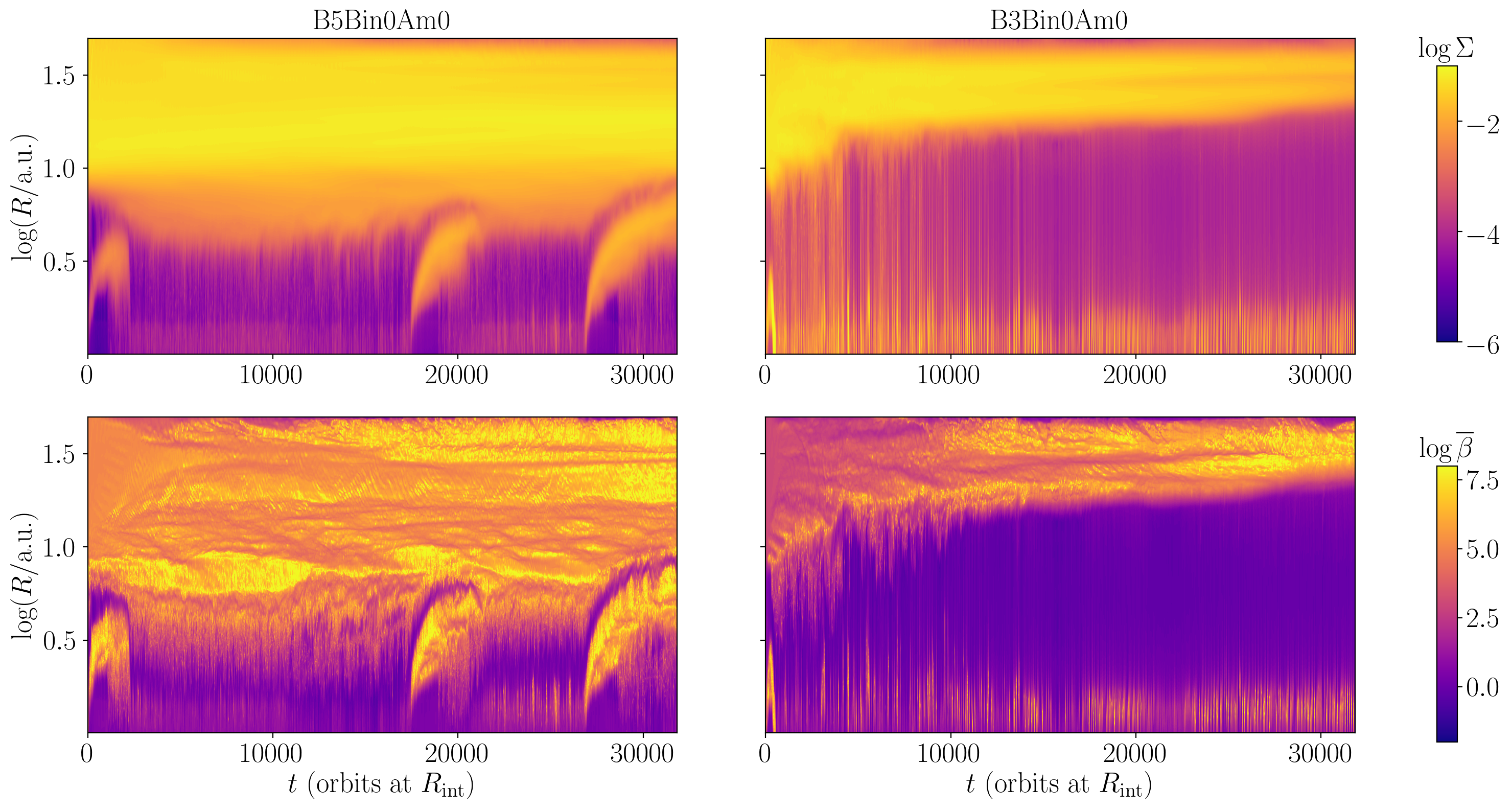}
\caption{\label{sig_beta_avg_2D_beta} Spatio-temporal diagrams for $\Sigma$ and $\overline{\beta}$ for B5Bin0Am0 and B3Bin0Am0. The cavity expands more in B3Bin0Am0 but shrinks in B5Bin0Am0. Bursts of matter occur in B5Bin0Am0, we refer to the text for more explanations.}
\end{center}
\end{figure*}

\textbf{General observations for B5Bin0Am0:} 
The spatio-temporal evolutions of $\Sigma$ and $\overline{\beta}$ are shown in the left panels of Fig.~\ref{sig_beta_avg_2D_beta}.
Right at the beginning of the simulation, a burst of matter appears in the cavity which is subsequently refilled.
Its radius then remains fixed at $\sim 4$~a.u. until other bursts happen at $\sim \num{17400}$ and $\sim \num{27000}$ orbits at $R_\text{int}$.
Such local events do not dramatically change the general properties of the disc which is overall similar to the fiducial one.

The bursts of matter (at $\sim \num{17400}$ and $\sim \num{27000}$ orbits at $R_\text{int}$, assuming the first one is due to the initial transient) give the illusion that some gas might be created inside the cavity, challenging mass conservation.
These bursts are actually due to gas accumulating at the boundary of an accretion `barrier'.
We refer to the appendix~\ref{relaxation} for a more detailed description of these bursts.
For now, we point out that these bursts highlight a limitation of our model regarding the implementation of the inner boundary conditions, but only occur in the weakly magnetised ($\beta_\text{out}=10^5$) simulations.

Lastly, we add that estimating $\dot{R_0}$ for this simulation is too difficult, since the cavity edge barely move during the entire simulation.

\textbf{General observations for B3Bin0Am0:}
In contrast to the run B5Bin0Am0, the cavity quickly expands up to $R\approx 15$ a.u. and keeps growing during all the simulation, faster than in the fiducial run (see the right panels of Fig. \ref{sig_beta_avg_2D_beta}).
We estimate its velocity as $\dot{R}_0 \approx 3.0\times 10^{-5}$ c.u., which is about 3 times faster than the fiducial run.
We get $\delta\dot{M}\approx 2.3\times 10^{-4}$ and $\delta\,\Sigma\approx 4.2\times 10^{-2}$ (both in code units) so that Eq.~\ref{cavity_edge} gives $\dot{R_0}\approx 4.2\times 10^{-5}$ c.u., where we choose $R_0\approx 20$.
The simple model we use seems to overestimate the widening velocity of the cavity but still gives a correct order of magnitude.

\begin{figure}
\begin{center}
\includegraphics[width=1.0\linewidth]{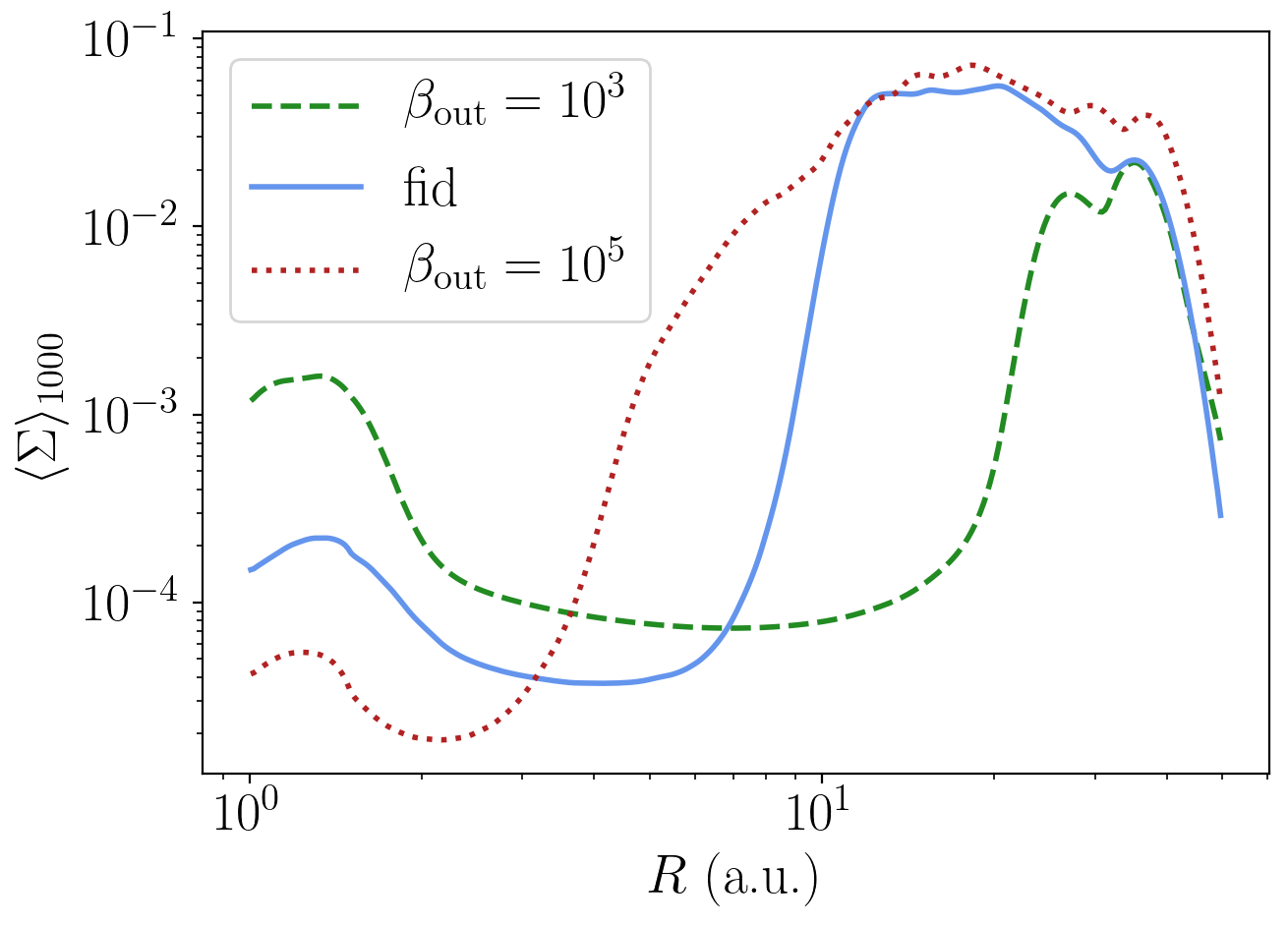}
\caption{\label{sigma_ext_last1000}Impact of the initial external magnetisation on the surface density. We average the profile on the last~1000 orbits at $R_\text{int}$. For B5Bin0Am0, we average on 1000 orbits at $R_\text{int}$ occurring between the 2~burst events seen in Fig.~\ref{sig_beta_avg_2D_beta}.}
\end{center}
\end{figure}

The time-averaged surface density from the fiducial run, B3Bin0Am0 and B5Bin0Am0 are shown in Fig.~\ref{sigma_ext_last1000}, which shows that the size of the cavity is ruled by the initial external plasma parameter.
The lower $\beta_\text{out}$ is, the wider the cavity gets when the disc reaches a steady state.
On the contrary, the plasma parameter inside the cavity does not depend on its external structure and converges to $\beta_\text{in}\lesssim 1$ in all of these simulations (the subsection~\ref{beta_int_sec} tackles this observation in depth).
Once the transient state is gone, we note that the cavity expands faster for lower $\beta_\text{out}$. This can be understood using Eq.~\ref{cavity_edge}, which can be recast as
\begin{align}
    \dot{R}_0= v_{\mathrm{acc.,\,in}}\,\left(\frac{\Sigma_\mathrm{in}}{\Sigma_\mathrm{out}}-\frac{v_\mathrm{acc.,\,out}}{v_\mathrm{acc.,\,in}}\right),
\end{align}
where we have defined the accretion velocities $v_{\mathrm{acc.}}\equiv\dot{M}/2\pi\,R_0\,\Sigma$, and we have assumed $\Sigma_\mathrm{out}\gg\Sigma_\mathrm{in}$. The expansion speed is then controlled by the term in parenthesis, since the accretion velocity in the cavity is always sonic (see \ref{beta_int_sec}). It is well known that the accretion velocity in the outer `standard' disc is a decreasing function of $\beta$. Writing $v_\mathrm{acc.}\propto \beta^{-\sigma}$ with $\sigma >0$, \cite{lesur_systematic_2021} proposes $\sigma = 0.78$ and \cite{Bai13} $\sigma=0.66$, which indicates that $0<\sigma<1$. Assuming that there exists a value $\tilde{\beta}$ for which $\dot{R}_0=0$, we get the scaling
\begin{align}
\label{eq:R0final}
\dot{R}_0=v_{\mathrm{acc.,\,in}}\,\frac{1}{\beta_\mathrm{out}}\,\left[1-\left(\frac{\beta_\mathrm{out}}{\tilde{\beta}}\right)^{1-\sigma}\right],
\end{align}
where we have used the fact that $\Sigma_\mathrm{in}/\Sigma_\mathrm{out}=\beta_\mathrm{in}/\beta_\mathrm{out}=\beta_\mathrm{out}^{\,-1}$ in our setup. The relation (\ref{eq:R0final}) shows that for $\beta_\mathrm{out}\ll \tilde{\beta}$, we have approximately $\dot{R}_0\approx  v_{\mathrm{acc.,\,in}} \,\beta_\mathrm{out}^{\,-1}$, indicating that the cavity expansion speed should increase as $\beta_\mathrm{out}$ gets lower, which is precisely what we observe for B3Bin0Am0. For $\beta_\mathrm{out}\gg\tilde{\beta}$ we get on the contrary $\dot{R}_0\approx - v_{\mathrm{acc.,\,in}}\, \tilde{\beta}^{\,\sigma-1}\,\beta_\mathrm{out}^{\,-\sigma}$, showing a change of sign (hence a contraction of the cavity), albeit with a reduced speed. This regime might correspond to B5Bin0Am0, indicating that $\tilde{\beta}\simeq 10^4$.

\subsubsection{Role of the internal initial plasma parameter}
\label{beta_int_sec}

\begin{figure}
\begin{center}
\includegraphics[width=1.0\linewidth]{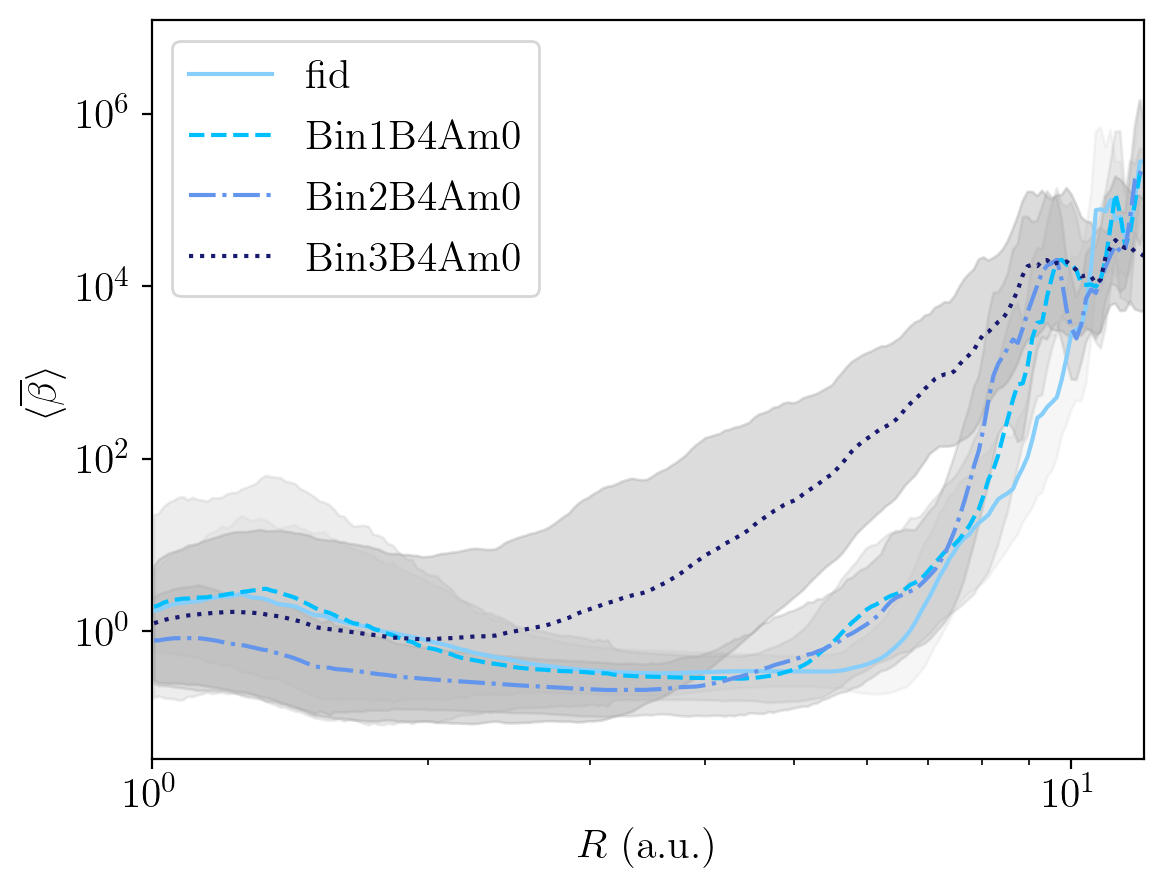}
\caption{\label{beta_int}Impact of the internal initial magnetisation on the plasma parameter for $\beta_\mathrm{out}=10^4$. }
\end{center}
\end{figure}

\begin{figure*}
\begin{center}
\includegraphics[width=1.\linewidth]{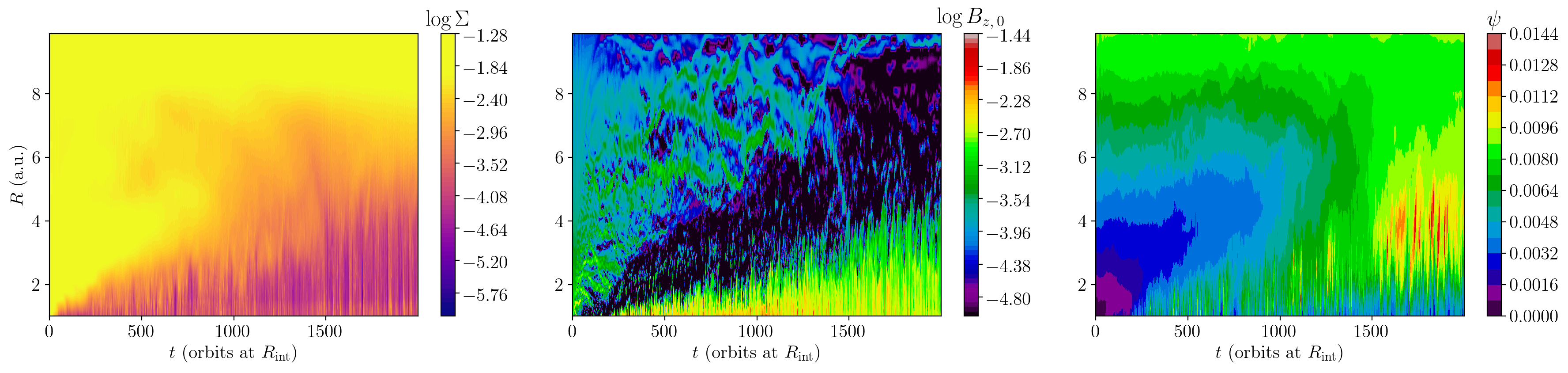}
\caption{\label{burst_S2DB4Bin3Am0}Spatio-temporal diagrams of $\Sigma$ (first panel), $B_{z,\,0}$ the vertical magnetic field at the midplane and $\psi$ the flux function defined in Eq.~\ref{psi}. These profiles focus on the first orbits of the run B4Bin3Am0.}
\end{center}
\end{figure*}

To study the impact of $\beta_\text{in}$, we run a set of simulations that covers all the possible initial gaps $\beta_\text{in}/\beta_\text{out}$ where $\log\beta_\text{out}\in\{3,\,4,\,5\}$ and $\log\beta_\text{in}\in\ \llbracket 0;\,\log\beta_\text{out}\llbracket$.
We compare each result to the one obtained with $\beta_\text{in}=1$ and the corresponding value of $\beta_\text{out}$.
A striking result is the fact that the disc inner structure does not depend on $\beta_\text{in}$.
No matter which $\beta_\text{in}$ we initially choose, a transition occurs in the cavity in order to impose $\beta_\text{in}\approx 1$.
Interestingly, this threshold value is the one required to get transsonic accretion as it is mentioned in \cite{wang_wind-driven_2017}.
We illustrate this statement with Fig.~\ref{beta_int} for the particular case of $\beta_\text{out}=10^4$.
We focus on the transient state of B4Bin3Am0 in Fig.~\ref{burst_S2DB4Bin3Am0}.
The transition is due to matter leaving the cavity because of the fast accretion at stake after a sharp increase of the magnetic field (and therefore a decrease of $\beta$).
This reorganisation of the cavity is a consequence of a rapid advection of magnetic flux from the cavity onto the seed which initially has a low magnetisation because of our initial setup.
Due to the total magnetic flux conservation, there is a shortage of magnetic flux inside the cavity, up until the inner seed reaches a state where its magnetisation is almost constant.
The magnetic field then accumulates at the inner boundary and $\beta$ decreases accordingly so that accretion is enhanced.
At this point, matter leaves the cavity as it is accreted onto the star.
It is then clear that the cavity converges towards the same overall structure as the fiducial simulation one.

We note that taking $\beta_\mathrm{in}$ equal to $\beta_\mathrm{out}$ would simulate a full disc with no cavity. 
Hence there should exist a threshold regarding the value of $\beta_\mathrm{in}$ above which no cavity is able to form.
Considering Fig.~\ref{beta_int}, it seems that this threshold is $\gtrsim 10^3$.

From these observations, we deduce that the cavity is regulated by the value of the plasma parameter which must take a value close to~$1$. 
The reason for this regulation is not entirely clear and we add a word of caution regarding the role of the inner radial boundary condition, especially with respect to the magnetic field transport at $R_\text{int}$.
We discuss this influence in appendix~\ref{relaxation}.

Referring to the sections \ref{Fast_variability} and \ref{RTI}, we suggest that the RTI may be responsible for this regulation, but a dedicated study would be required to ascertain this claim.

\subsection{Zoom with a larger cavity radius}

\begin{figure*}
\begin{center}
\includegraphics[width=1.0\linewidth]{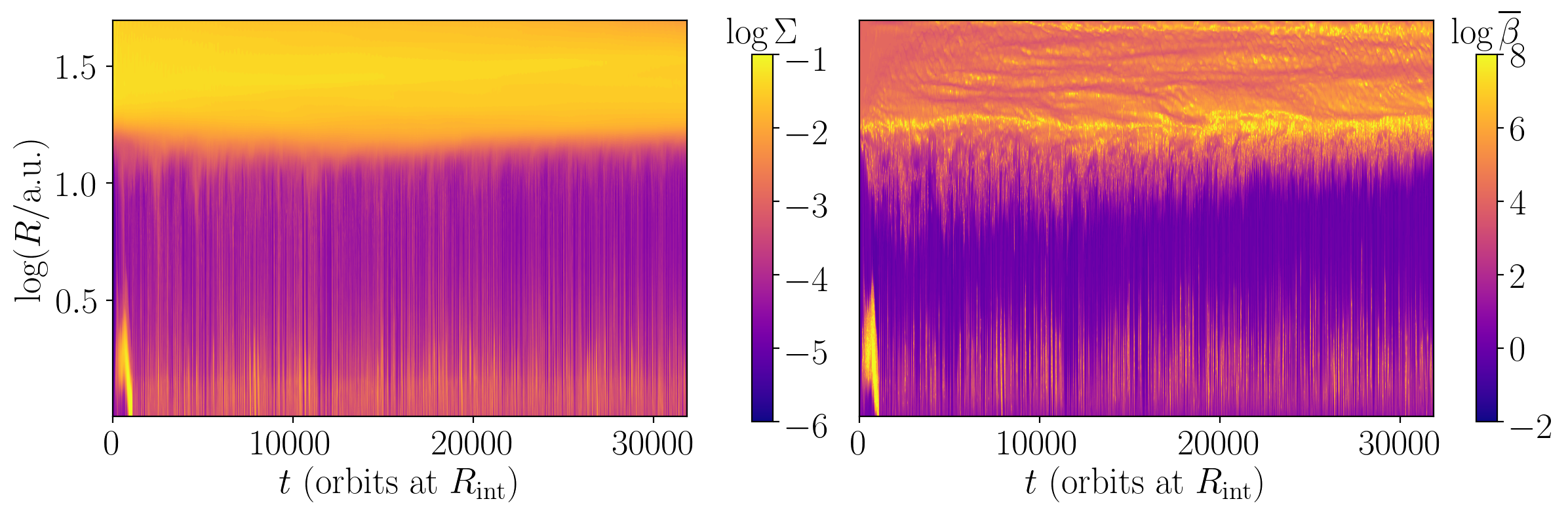}
\caption{\label{sig_beta_avg_R20fid2D} Spatio-temporal diagrams for $\langle\Sigma\rangle$ and $\langle\beta\rangle$ for R20FID.}
\end{center}
\end{figure*}

We perform a simulation with a double-sized cavity ($R_0=20$~a.u.) in order to check the impact of the cavity size.
The simulation was integrated for $1000$~orbits at $R=10$~a.u. so that it reaches $355$~orbits at $R=20$~a.u.
The general observations are confirmed such as the elbow-shaped structure, the magnetic loop, the magnetic field advection in the outer disc as well as the conclusions regarding the accretion.
While the cavity size is identical to B3Bin0Am0, the behaviour of the disc is exactly the same as the fiducial one (\ref{sig_beta_avg_R20fid2D}), indicating that $\beta_\mathrm{out}$ is the main parameter regulating the cavity expansion.
This means that the global picture where two types of discs are connected is robust and not linked to limitations in the cavity size or artefacts due to the inner boundary condition.

\section{Discussion and comparison with previous work}

We model transition discs sustained by MHD winds by performing 2.5\,D global simulations.
This model acts as a proof of concept, showing that steady-state discs with both a cavity and a wind can be obtained.
The resulting simulated discs are characterised by two different zones with contrasted dynamics.

First, our  `outer disc' behaves like a standard weakly-magnetised ambipolar-dominated protoplanetary discs \citep{lesur_systematic_2021,cui_global_2021}.
In particular we find  mass and angular momentum transport coefficients, wind properties and accretion rates comparable to those found in the literature for `full' discs.
We also find weak gaps which are characteristic of non-ideal MHD discs \citep{riols_spontaneous_2019, riols_ring_2020}.
However, the magnetic field transport in the outer disc differs from previous studies: we find that magnetic field lines are advected inwards in the outer disc, in contrast to measurements in full discs which always show outwards transport \citep{bai_hall-effect_2017,gressel_global_2020,lesur_systematic_2021}. This discrepancy is likely due to the fact that the field lines in the cavity are more collimated (i.e. less opened), which results in a lower pressure on the magnetic surfaces in the outer disc, but possibly also connected to the peculiar elbow shaped magnetic surfaces at the transition radius. In any case, it points to the fact that magnetic field transport is a non-local phenomenon: it depends on the global disc structure.

In contrast to the outer disc, the cavity (or inner disc) is strongly magnetised ($\beta\approx 1$) because of its low surface density. We emphasise here that the absolute magnetic field strength in the cavity is not stronger than standard protoplanetary disc models.
In practice, and given our set of units, we have $B_0 \approx 0.13~$G (see Eq.~\ref{Magnetic_field}, with $\beta_\mathrm{out}=10^4$) so that initially, $B_z\approx 1.25~\text{mG}$ at $R=42~$a.u. in our simulations, which is of the same order of magnitude as the upper limit of $B_z(R=42~\text{a.u.})=0.8~\text{mG}$ found in~\cite{vlemmings_stringent_2019} for example. Hence, while the cavity is strongly magnetised, its field strength is compatible with observational constraints.

Compared to the outer disc, the mass and angular momentum transport coefficients in the cavity are all of the order of unity, resulting in transsonic accretion velocities and faster wind with large lever arms ($\lambda\gtrsim10$). Overall, this picture matches quantitatively the inner jet emitting disc proposed by \cite{Combet08}. Interestingly, in all of our models, the cavity manages to reach an accretion rate close to the outer disc one by self-regulating the magnetic stresses. 
We find that most of the angular momentum transport is due to the laminar stress (appendix ~\ref{Laminar_coef}) indicating that turbulent transport (possibly MRI-driven) is unimportant in the cavity.
This is not surprising since our discs are dominated by ambipolar diffusion which mostly suppresses MRI turbulence \citep{bai_magnetorotational_2011}. 

We find a significant deviation of the rotation profile in the cavity as a result of the strong magnetic stress due to the wind and typical rotation velocities of the order of 70-80\% of the Keplerian velocity. This fact, combined with the transsonic accretion, implies that the kinematics of these cavities have singular observational signatures. Fast accretion kinematics have been observed in some transition discs \citep{Rosenfeld14} but we note that these signatures might also be due to a warped circumbinary disc \citep{Casassus_2015}.

As a result of the stress balance mentioned above, we obtain accreting cavities that survive thousands or orbits and which are slowly expanding or contracting, depending on the outer disc magnetisation.
This result suggests that a cavity could be carved spontaneously if the magnetisation of the outer disc is high enough. There are already hints of such a process in global simulations: for instance \cite{cui_global_2021} show a gas-depleted cavity forming in the inner profile of $\Sigma$ (see their figure~$5$, first row and first column panel). While this is by no mean a proof since the boundary conditions are probably unrealistic, it shows that the secular evolution of wind-driven discs should be investigated systematically to check whether or not cavities could spontaneously form in these models.

The temporal analysis of the disc reveals the appearance of dynamical structures.
In particular, we highlight the formation of gas filaments above the disc surface that end up forming $2$ bubbles of gas each, one being ejected while the other one falls down onto the cavity before crossing it.
At some point, the falling matter has to cross the poloidal magnetic field lines at the magnetic field loop location, recalling to some extent the magnetospheric accretion observed in young stars \citep{bouvier_magnetospheric_2007, pouilly_magnetospheric_2020, bouvier_probing_2020, bouvier_investigating_2020} and magnetospheric ejection events \citep{zanni_mhd_2013, cemeljic_magnetospheric_2013}.
However, there is no magnetosphere in our simulations so the magnetic topology is quite different from that of magnetospheric interaction.

By analogy with magnetospheric accretion, we have checked whether the time variability seen in our simulations could be due to a magnetic RTI. We have studied $2$~criteria for the RTI, in the form of a radial interchange of poloidal field lines (see section~\ref{RTI} and appendix~\ref{Interchange_appendix}). We found however that the RTI requires magnetisations stronger than the ones found in our simulations, ruling out the RTI in the form we have assumed. It is however still possible that another branch of this instability is present. It is also possible that the non-axisymmetric version of the RTI could be triggered in 3D simulations. We therefore defer this study to a future publication.

On longer timescales, averaging out the fast variability, the magnetic field strength appears to be self-regulated with $0.1\lesssim \beta\lesssim 1$ in the cavity, independently on the initial field strength. As a result, the cavity is strongly magnetised and rotates at sub-Keplerian velocities, indicating a substantial magnetic support against gravity in this region. In essence, the regime of our cavity is similar to the magnetically arrested disc (MAD) proposed by \cite{narayan_magnetically_2003} in the context of black hole accretion discs. \cite{mckinney_general_2012} shown that MADs could be regulated by magnetic RTI leading to magnetically chocked accretion flows (MCAF).
The MAD model is also associated to the formation of plasmoids by reconnection events  \citep{ripperda_black_2021}. These features are recovered in our models of transition discs, despite the fact that we have used Newtonian dynamics (MADs are usually found in GRMHD simulations) and the presence of a strong ambipolar diffusivity in our models. Hence, our models could be interpreted as non-ideal non-relativistic models of MADs.

The time variability of the cavity is likely to be related to the axisymmetric approximation used in this work since it suppresses non-axisymmetric instabilities which seem to play a key role in MADs simulations \citep[e.g.][]{mckinney_general_2012,liska2022formation}.
Additionally, we note that the question of non-axisymmetric hydrodynamical instabilities such as the Rossby Wave Instability (RWI) \citep{lovelace_rossby_1999, li_rossby_2000} at the cavity edge is still open to debate in a magnetised environment \citep{Bajer13}.
We will address these points using full 3\,D simulations in a follow up paper.

Regarding the caveats of our simulations, we remark that the inner radial boundary is probably the most stringent caveat of our numerical model. 
In particular, we found that this inner boundary condition is sometime expelling some poloidal magnetic flux, resulting in the bursts seen in ﬁg.~\ref{sig_beta_avg_2D_beta}.
However, the weakly magnetised simulations (such as B5Bin0Am0) are the only ones exhibiting these events, and once the transient state is over, all the simulations reach comparable steady states.
So the inner boundary condition is likely not affecting the long term evolution of our models.
Future models should nevertheless try to either include an inner turbulent disc, or possibly the magnetospheric interaction with the central star.

A possible limitation of our model one could raise concerns the role of the MRI.
Our simulated discs are dominated by ambipolar diffusion, and as such, subject to MRI quenching by the non-linearity embedded in the ambipolar diffusivity ($\eta\propto B^2$).
This saturation is different from the saturation by 3D turbulence observed in the ideal MHD regime.
It is suggested that the MRI saturates in very similar ways in 3D and 2D under strong ambipolar diffusion (see e.g. \citealt{bethune_global_2017,cui_global_2021}). This is also confirmed by our own 3D simulations which will be published in a forthcoming paper.
Hence, the fact that our simulations are 2.5D have a very limited impact on the turbulent transport one may observe.

Note that our simulations used a simplified treatment of thermodynamics and ionisation chemistry. More numerically involved models, such as \cite{wang_wind-driven_2017}, use a refined computation of the ionisation fraction and $\Lambda_\mathrm{A}$ inside the cavity of a TD, including several chemical species.
This work highlights in particular the influence of the X-ray luminosity of the star $L_\mathrm{X}$ (see their Fig.~2, panels~2 and 3) as well as the role of the temperature $T_0$ at 16~a.u. (Fig.~2, panels 6 and 7).
Regarding our profile of $\Lambda_\mathrm{A}\approx 1-10$, our work is similar to their models 2 (with $L_\mathrm{X}=10^{29}~\mathrm{erg}\,\mathrm{s}^{-1}$) and 6 (where $T_0=30~\mathrm{K}$).
Therefore, we anticipate that an increase of 2 orders of magnitude for $L_\mathrm{X}$ would lead to $\Lambda_\mathrm{A}>10^2$ in most of our cavity.
Such a change would greatly alter the dynamical regime of the cavity since MRI would then play a significant role \citep[see the appendix~\ref{app_lambda} and][]{blaes_local_1994, bai_magnetorotational_2011}.
However, the role of the temperature is less straightforward and seems to have a little impact on $\Lambda_\mathrm{A}$.

Additionally, dust plays a significant role in \citet{wang_wind-driven_2017} regarding the ionisation of the disc.
As a matter of fact, only their models with dust reach low values of $\Lambda_\mathrm{A}$.
The effect of dust in transition discs is a major subject that is not addressed in our work.
Dust can modify the ionisation fraction but also create peculiar structures at the interface between the disc and cavity.
We mention in particular the interplay between dust and the radiation pressure, that is known to create non-axisymmetric structures at the cavity edge \citep{bi_2022} or an inner rim with an accumulation of matter due to photophoresis \citep{cuello_2016}.

\section{Conclusions}

We performed 2.5\,D global numerical simulations of transition discs in the context of non-ideal MHD with MHD wind launching.
Our simulation design is initialised with a cavity in the gas surface density profile, and a power law distribution for the vertical magnetic field strength, resulting in a strongly magnetised cavity surrounded by a standard weakly magnetised disc.

The main results are summarised in the following points:

   \begin{enumerate}
      \item We have modelled strongly accreting transition discs that reach a quasi steady state that last for at least tens of kyrs. The accretion rate inside the cavity connects smoothly to the accretion rate in the external part of the disc
      \item The cavity itself is characterised by a strong sub-Keplerian rotation and a transsonic accretion velocity. These kinematic signatures could potentially be verified observationally.
      \item The magnetic field is advected inwards in the outer disc, in contrast to full disc simulations. This points to the possible non-locality of large-scale field transport. 
      \item The cavity structure (density and field strength) is self-regulated. In particular, it is insensitive to a change in the initial internal magnetisation and is characterised by $
      0.1 \lesssim\beta_\text{int}\lesssim 1$.
      \item The temporal analysis of the cavity dynamics highlights the formation and accretion of bubbles of gas above the disc which cross the cavity at sonic speeds. The magnetic Rayleigh-Taylor instability might be responsible for this unsteadiness.
      \item The physics of the cavity (accretion speed, wind lever arm and mass loading) match previously published jet emitting disc solutions \citep{Ferreira97,Combet08}. The presence of a strong radial magnetic support and possible regulation by the RTI is also reminiscent of MADs in black hole physics \citep{narayan_magnetically_2003,mckinney_general_2012}. These resemblances suggest that transition discs could be an instance of MADs applied to protoplanetary discs.
   \end{enumerate}

\begin{acknowledgements}
The authors would like to thank the anonymous referee for constructive comments that have greatly improved the quality of this work.
They wish to thank Jonatan Jacquemin-Ide, Andr\'es Carmona, Antoine Riols, Ileyk El Mellah and Jonathan Ferreira for fruitful discussions and comments.
This work is supported by the European Research Council (ERC) European Union Horizon 2020 research and innovation programme (Grant agreement No. 815559 (MHDiscs)). This work was granted access to the HPC resources of TGCC under the allocation 2021-A0100402231 made by GENCI. A part of the computations presented in this paper were performed using the GRICAD infrastructure (https://gricad.univ-grenoble-alpes.fr), which is supported by Grenoble research communities.
This work makes use of matplotlib \citep{Matplotlib} for graphics, NumPy \citep{Numpy}, SciPy \citep{Scipy} and Pickle \citep{Pickle}.
This article has been typeset from a \TeX / \LaTeX file prepared by the authors.
\end{acknowledgements}

\section*{Data availability}

The data underlying this article will be shared on reasonable request to the corresponding author.

\bibliographystyle{aa}
\bibliography{biblio_td}

\begin{thebibliography}{84}
\expandafter\ifx\csname natexlab\endcsname\relax\def\natexlab#1{#1}\fi

\bibitem[{{Alexander} {et~al.}(2014){Alexander}, {Pascucci}, {Andrews},
  {Armitage}, \& {Cieza}}]{Alexander-ppvii}
{Alexander}, R., {Pascucci}, I., {Andrews}, S., {Armitage}, P., \& {Cieza}, L.
  2014, in Protostars and Planets VI, ed. H.~{Beuther}, R.~S. {Klessen}, C.~P.
  {Dullemond}, \& T.~{Henning}, 475

\bibitem[{Bai(2011)}]{bai_magnetorotational_2011}
Bai, X.-N. 2011, ApJ, 739, 50

\bibitem[{Bai \& Goodman(2009)}]{bai_heat_2009}
Bai, X.-N. \& Goodman, J. 2009, ApJ, 701, 737

\bibitem[{{Bai} \& {Stone}(2013)}]{Bai13}
{Bai}, X.-N. \& {Stone}, J.~M. 2013, \apj, 769, 76

\bibitem[{Bai \& Stone(2017)}]{bai_hall-effect_2017}
Bai, X.-N. \& Stone, J.~M. 2017, ApJ, 836, 46

\bibitem[{{Bajer} \& {Mizerski}(2013)}]{Bajer13}
{Bajer}, K. \& {Mizerski}, K. 2013, \prl, 110, 104503

\bibitem[{{Balbus} \& {Hawley}(1991)}]{Bablus91}
{Balbus}, S.~A. \& {Hawley}, J.~F. 1991, \apj, 376, 214

\bibitem[{B\'ethune {et~al.}(2017)B\'ethune, Lesur, \&
  Ferreira}]{bethune_global_2017}
B\'ethune, W., Lesur, G., \& Ferreira, J. 2017, A\&A, 600, A75

\bibitem[{Bi \& Fung(2022)}]{bi_2022}
Bi, J. \& Fung, J. 2022, ApJ, 928, 74

\bibitem[{Blaes \& Balbus(1994)}]{blaes_local_1994}
Blaes, O.~M. \& Balbus, S.~A. 1994, ApJ, 421, 163

\bibitem[{Blandford \& Payne(1982)}]{blandford_hydromagnetic_1982}
Blandford, R.~D. \& Payne, D.~G. 1982, MNRAS, 199, 883

\bibitem[{Bouvier {et~al.}(2020{\natexlab{a}})Bouvier, Alecian, Alencar, Sousa,
  Donati, Perraut, Bayo, Rebull, Dougados, Duvert, Berger, Benisty, Pouilly,
  Folsom, \& Moutou}]{bouvier_investigating_2020}
Bouvier, J., Alecian, E., Alencar, S. H.~P., {et~al.} 2020{\natexlab{a}}, A\&A,
  643, A99

\bibitem[{Bouvier {et~al.}(2007)Bouvier, Alencar, Boutelier, Dougados, Balog,
  Grankin, Hodgkin, Ibrahimov, Kun, Magakian, \&
  Pinte}]{bouvier_magnetospheric_2007}
Bouvier, J., Alencar, S. H.~P., Boutelier, T., {et~al.} 2007, A\&A, 463, 1017

\bibitem[{Bouvier {et~al.}(2020{\natexlab{b}})Bouvier, Perraut, Bouquin,
  Duvert, Dougados, Brandner, Benisty, Berger, \&
  Al\'ecian}]{bouvier_probing_2020}
Bouvier, J., Perraut, K., Bouquin, J.-B.~L., {et~al.} 2020{\natexlab{b}}, A\&A,
  636, A108

\bibitem[{Carmona {et~al.}(2014)Carmona, Pinte, Thi, Benisty, M{\'e}nard,
  Grady, Kamp, Woitke, Olofsson, Roberge, Brittain, D{\^u}chene, Meeus,
  Martin-Za{\"\i}di, Dent, Bouquin, \& Berger}]{carmona_constraining_2014}
Carmona, A., Pinte, C., Thi, W.~F., {et~al.} 2014, A\&A, 567, A51

\bibitem[{{Carmona} {et~al.}(2017){Carmona}, {Thi}, {Kamp}, {Baruteau},
  {Matter}, {van den Ancker}, {Pinte}, {K{\'o}sp{\'a}l}, {Audard}, {Liebhart},
  {Sicilia-Aguilar}, {Pinilla}, {Reg{\'a}ly}, {G{\"u}del}, {Henning}, {Cieza},
  {Baldovin-Saavedra}, {Meeus}, \& {Eiroa}}]{Carmona17}
{Carmona}, A., {Thi}, W.~F., {Kamp}, I., {et~al.} 2017, \aap, 598, A118

\bibitem[{Casassus {et~al.}(2015)Casassus, Marino, P{\'{e}}rez, Roman, Dunhill,
  Armitage, Cuadra, Wootten, van~der Plas, Cieza, Moral, Christiaens, \&
  Montesinos}]{Casassus_2015}
Casassus, S., Marino, S., P{\'{e}}rez, S., {et~al.} 2015, ApJ, 811, 92

\bibitem[{{\v C}emelji{\'c} {et~al.}(2013){\v C}emelji{\'c}, Shang, \&
  Chiang}]{cemeljic_magnetospheric_2013}
{\v C}emelji{\'c}, M., Shang, H., \& Chiang, T.-Y. 2013, ApJ, 768, 5

\bibitem[{{Clarke} {et~al.}(2001){Clarke}, {Gendrin}, \&
  {Sotomayor}}]{clarke01}
{Clarke}, C.~J., {Gendrin}, A., \& {Sotomayor}, M. 2001, \mnras, 328, 485

\bibitem[{{Combet} \& {Ferreira}(2008)}]{Combet08}
{Combet}, C. \& {Ferreira}, J. 2008, \aap, 479, 481

\bibitem[{Combet {et~al.}(2010)Combet, Ferreira, \& Casse}]{combet_dead_2010}
Combet, C., Ferreira, J., \& Casse, F. 2010, A\&A, 519, A108

\bibitem[{Cuello {et~al.}(2016)Cuello, Gonzalez, \& Pignatale}]{cuello_2016}
Cuello, N., Gonzalez, J.-F., \& Pignatale, F.~C. 2016, 458, 2140

\bibitem[{Cui \& Bai(2021)}]{cui_global_2021}
Cui, C. \& Bai, X.-N. 2021, MNRAS, 507, 1106

\bibitem[{{Dong} \& {Dawson}(2016)}]{DongDawson16}
{Dong}, R. \& {Dawson}, R. 2016, \apj, 825, 77

\bibitem[{{Dullemond} \& {Dominik}(2005)}]{DullemondDominik05}
{Dullemond}, C.~P. \& {Dominik}, C. 2005, \aap, 434, 971

\bibitem[{{Espaillat} {et~al.}(2014){Espaillat}, {Muzerolle}, {Najita},
  {Andrews}, {Zhu}, {Calvet}, {Kraus}, {Hashimoto}, {Kraus}, \&
  {D'Alessio}}]{espaillat_ppvi}
{Espaillat}, C., {Muzerolle}, J., {Najita}, J., {et~al.} 2014, in Protostars
  and Planets VI, ed. H.~{Beuther}, R.~S. {Klessen}, C.~P. {Dullemond}, \&
  T.~{Henning}, 497

\bibitem[{Evans \& Hawley(1988)}]{evans_simulation_1988}
Evans, C.~R. \& Hawley, J.~F. 1988, ApJ, 332, 659

\bibitem[{{Fang} {et~al.}(2013){Fang}, {Kim}, {van Boekel}, {Sicilia-Aguilar},
  {Henning}, \& {Flaherty}}]{Fang13}
{Fang}, M., {Kim}, J.~S., {van Boekel}, R., {et~al.} 2013, \apjs, 207, 5

\bibitem[{{Ferreira}(1997)}]{Ferreira97}
{Ferreira}, J. 1997, \aap, 319, 340

\bibitem[{Fromang {et~al.}(2002)Fromang, Terquem, \&
  Balbus}]{fromang_ionization_2002}
Fromang, S., Terquem, C., \& Balbus, S.~A. 2002, MNRAS, 329, 18

\bibitem[{{Gammie}(1996)}]{Gammie96}
{Gammie}, C.~F. 1996, \apj, 457, 355

\bibitem[{G\'arate {et~al.}(2021)G\'arate, Delage, Stadler, Pinilla, Birnstiel,
  Stammler, Picogna, Ercolano, Franz, \& Lenz}]{garate_large_2021}
G\'arate, M., Delage, T.~N., Stadler, J., {et~al.} 2021, A\&A, 655, A18

\bibitem[{Gressel {et~al.}(2020)Gressel, Ramsey, Brinch, Nelson, Turner, \&
  Bruderer}]{gressel_global_2020}
Gressel, O., Ramsey, J.~P., Brinch, C., {et~al.} 2020, ApJ, 896, 126

\bibitem[{Guilet \& Ogilvie(2014)}]{guilet_global_2014}
Guilet, J. \& Ogilvie, G. 2014, MNRAS, 441

\bibitem[{Harris {et~al.}(2020)Harris, Millman, van~der Walt, Gommers,
  Virtanen, Cournapeau, Wieser, Taylor, Berg, Smith, Kern, Picus, Hoyer, van
  Kerkwijk, Brett, Haldane, del R{\'{i}}o, Wiebe, Peterson,
  G{\'{e}}rard-Marchant, Sheppard, Reddy, Weckesser, Abbasi, Gohlke, \&
  Oliphant}]{Numpy}
Harris, C.~R., Millman, K.~J., van~der Walt, S.~J., {et~al.} 2020, Nature, 585,
  357

\bibitem[{Hunter(2007)}]{Matplotlib}
Hunter, J.~D. 2007, Computing in science \& engineering, 9, 90

\bibitem[{Igea \& Glassgold(1999)}]{igea_x-ray_1999}
Igea, J. \& Glassgold, A.~E. 1999, ApJ, 518, 848, publisher: IOP Publishing

\bibitem[{Jacquemin-Ide {et~al.}(2021)Jacquemin-Ide, Lesur, \&
  Ferreira}]{jacquemin-ide_magnetic_2020}
Jacquemin-Ide, J., Lesur, G., \& Ferreira, J. 2021, A\&A, 647, A192

\bibitem[{{Kane Yee}(1966)}]{kane_yee_numerical_1966}
{Kane Yee}. 1966, IEEE Transactions on Antennas and Propagation, 14, 302

\bibitem[{Lesur {et~al.}(2014)Lesur, Kunz, \& Fromang}]{lesur_thanatology_2014}
Lesur, G., Kunz, M.~W., \& Fromang, S. 2014, A\&A, 566, A56

\bibitem[{Lesur(2021{\natexlab{a}})}]{lesur_magnetohydrodynamics_2021}
Lesur, G. R.~J. 2021{\natexlab{a}}, Journal of Plasma Physics, 87, publisher:
  Cambridge University Press

\bibitem[{Lesur(2021{\natexlab{b}})}]{lesur_systematic_2021}
Lesur, G. R.~J. 2021{\natexlab{b}}, A\&A, 650, A35

\bibitem[{Li {et~al.}(2000)Li, Finn, Lovelace, \& Colgate}]{li_rossby_2000}
Li, H., Finn, J.~M., Lovelace, R. V.~E., \& Colgate, S.~A. 2000, ApJ, 533,
  1023, publisher: IOP Publishing

\bibitem[{Liska {et~al.}(2022)Liska, Musoke, Tchekhovskoy, Porth, \&
  Beloborodov}]{liska2022formation}
Liska, M. T.~P., Musoke, G., Tchekhovskoy, A., Porth, O., \& Beloborodov, A.~M.
  2022 [\eprint[arXiv]{2201.03526}]

\bibitem[{Lovelace {et~al.}(1999)Lovelace, Li, Colgate, \&
  Nelson}]{lovelace_rossby_1999}
Lovelace, R. V.~E., Li, H., Colgate, S.~A., \& Nelson, A.~F. 1999, ApJ, 513,
  805, publisher: IOP Publishing

\bibitem[{{Manara} {et~al.}(2014){Manara}, {Testi}, {Natta}, {Rosotti},
  {Benisty}, {Ercolano}, \& {Ricci}}]{Manara14}
{Manara}, C.~F., {Testi}, L., {Natta}, A., {et~al.} 2014, \aap, 568, A18

\bibitem[{{Marsh} \& {Mahoney}(1992)}]{Marsh92}
{Marsh}, K.~A. \& {Mahoney}, M.~J. 1992, \apjl, 395, L115

\bibitem[{McKinney {et~al.}(2012)McKinney, Tchekhovskoy, \&
  Blandford}]{mckinney_general_2012}
McKinney, J.~C., Tchekhovskoy, A., \& Blandford, R.~D. 2012, MNRAS, 423, 3083

\bibitem[{Mignone {et~al.}(2007)Mignone, Bodo, Massaglia, Matsakos, Tesileanu,
  Zanni, \& Ferrari}]{mignone_pluto_2007}
Mignone, A., Bodo, G., Massaglia, S., {et~al.} 2007, ApJS, 170, 228, publisher:
  American Astronomical Society

\bibitem[{Mishra {et~al.}(2020)Mishra, Begelman, Armitage, \&
  Simon}]{mishra_strongly_2020}
Mishra, B., Begelman, M.~C., Armitage, P.~J., \& Simon, J.~B. 2020, MNRAS, 492,
  1855

\bibitem[{{Morishima}(2012)}]{Morishima12}
{Morishima}, R. 2012, \mnras, 420, 2851

\bibitem[{{Najita} {et~al.}(2007){Najita}, {Strom}, \& {Muzerolle}}]{Najita07}
{Najita}, J.~R., {Strom}, S.~E., \& {Muzerolle}, J. 2007, \mnras, 378, 369

\bibitem[{Narayan {et~al.}(2003)Narayan, Igumenshchev, \&
  Abramowicz}]{narayan_magnetically_2003}
Narayan, R., Igumenshchev, I.~V., \& Abramowicz, M.~A. 2003, Publications of
  the Astronomical Society of Japan, 55, L69

\bibitem[{Nelson {et~al.}(2013)Nelson, Gressel, \&
  Umurhan}]{nelson_linear_2013}
Nelson, R.~P., Gressel, O., \& Umurhan, O.~M. 2013, MNRAS, 435, 2610

\bibitem[{Perez-Becker \& Chiang(2011)}]{perez-becker_surface_2011}
Perez-Becker, D. \& Chiang, E. 2011, ApJ, 735, 8, publisher: American
  Astronomical Society

\bibitem[{Pouilly {et~al.}(2020)Pouilly, Bouvier, Alecian, Alencar, Cody,
  Donati, Grankin, Hussain, Rebull, \& Folsom}]{pouilly_magnetospheric_2020}
Pouilly, K., Bouvier, J., Alecian, E., {et~al.} 2020, A\&A, 642, A99

\bibitem[{Riols \& Lesur(2018)}]{riols_dust_2018}
Riols, A. \& Lesur, G. 2018, A\&A, 617, A117

\bibitem[{Riols \& Lesur(2019)}]{riols_spontaneous_2019}
Riols, A. \& Lesur, G. 2019, A\&A, 625, A108

\bibitem[{Riols {et~al.}(2020)Riols, Lesur, \& Menard}]{riols_ring_2020}
Riols, A., Lesur, G., \& Menard, F. 2020, A\&A, 639, A95

\bibitem[{Ripperda {et~al.}(2022)Ripperda, Liska, Chatterjee, Musoke,
  Philippov, Markoff, Tchekhovskoy, \& Younsi}]{ripperda_black_2021}
Ripperda, B., Liska, M., Chatterjee, K., {et~al.} 2022, ApJL, 924, L32

\bibitem[{{Rosenfeld} {et~al.}(2014){Rosenfeld}, {Chiang}, \&
  {Andrews}}]{Rosenfeld14}
{Rosenfeld}, K.~A., {Chiang}, E., \& {Andrews}, S.~M. 2014, \apj, 782, 62

\bibitem[{Shakura \& Sunyaev(1973)}]{shakura_black_1973}
Shakura, N.~I. \& Sunyaev, R.~A. 1973, A\&A, 24, 337

\bibitem[{{Simon} {et~al.}(2013){Simon}, {Bai}, {Stone}, {Armitage}, \&
  {Beckwith}}]{Simon13}
{Simon}, J.~B., {Bai}, X.-N., {Stone}, J.~M., {Armitage}, P.~J., \& {Beckwith},
  K. 2013, \apj, 764, 66

\bibitem[{Simon {et~al.}(2015)Simon, Lesur, Kunz, \&
  Armitage}]{simon_magnetically_2015}
Simon, J.~B., Lesur, G., Kunz, M.~W., \& Armitage, P.~J. 2015, MNRAS, 454, 1117

\bibitem[{Spruit {et~al.}(1995)Spruit, Stehle, \&
  Papaloizou}]{spruit_interchange_1995}
Spruit, H.~C., Stehle, R., \& Papaloizou, J. C.~B. 1995, MNRAS, 275, 1223,
  publisher: Oxford Academic

\bibitem[{Spruit \& Taam(1990)}]{spruit_mass_1990}
Spruit, H.~C. \& Taam, R.~E. 1990, A\&A, 229, 475

\bibitem[{Stehle \& Spruit(2001)}]{stehle_stability_2001}
Stehle, R. \& Spruit, H.~C. 2001, MNRAS, 323, 587, publisher: Oxford Academic

\bibitem[{Suriano {et~al.}(2019)Suriano, Li, Krasnopolsky, Suzuki, \&
  Shang}]{suriano_formation_2019}
Suriano, S.~S., Li, Z.-Y., Krasnopolsky, R., Suzuki, T.~K., \& Shang, H. 2019,
  MNRAS, 484, 107

\bibitem[{{Suzuki} {et~al.}(2016){Suzuki}, {Ogihara}, {Morbidelli}, {Crida}, \&
  {Guillot}}]{Suzuki16}
{Suzuki}, T.~K., {Ogihara}, M., {Morbidelli}, A., {Crida}, A., \& {Guillot}, T.
  2016, \aap, 596, A74

\bibitem[{Thi {et~al.}(2019)Thi, Lesur, Woitke, Kamp, Rab, \&
  Carmona}]{thi_radiation_2019}
Thi, W.~F., Lesur, G., Woitke, P., {et~al.} 2019, A\&A, 632, A44

\bibitem[{Umebayashi \& Nakano(1980)}]{umebayashi_recombination_1980}
Umebayashi, T. \& Nakano, T. 1980, Publications of the Astronomical Society of
  Japan, 32, 405

\bibitem[{Umebayashi \& Nakano(2008)}]{umebayashi_effects_2008}
Umebayashi, T. \& Nakano, T. 2008, ApJ, 690, 69, publisher: American
  Astronomical Society

\bibitem[{{van der Marel} {et~al.}(2016){van der Marel}, {van Dishoeck},
  {Bruderer}, {Andrews}, {Pontoppidan}, {Herczeg}, {van Kempen}, \&
  {Miotello}}]{Vdm16}
{van der Marel}, N., {van Dishoeck}, E.~F., {Bruderer}, S., {et~al.} 2016,
  \aap, 585, A58

\bibitem[{{van der Marel} {et~al.}(2015){van der Marel}, {van Dishoeck},
  {Bruderer}, {P{\'e}rez}, \& {Isella}}]{Vdm15}
{van der Marel}, N., {van Dishoeck}, E.~F., {Bruderer}, S., {P{\'e}rez}, L., \&
  {Isella}, A. 2015, \aap, 579, A106

\bibitem[{Van~Rossum(2020)}]{Pickle}
Van~Rossum, G. 2020, The Python Library Reference, release 3.8.2 (Python
  Software Foundation)

\bibitem[{Virtanen {et~al.}(2020)Virtanen, Gommers, Oliphant, Haberland, Reddy,
  Cournapeau, Burovski, Peterson, Weckesser, Bright, {van der Walt}, Brett,
  Wilson, Millman, Mayorov, Nelson, Jones, Kern, Larson, Carey, Polat, Feng,
  Moore, {VanderPlas}, Laxalde, Perktold, Cimrman, Henriksen, Quintero, Harris,
  Archibald, Ribeiro, Pedregosa, {van Mulbregt}, \& {SciPy 1.0
  Contributors}}]{Scipy}
Virtanen, P., Gommers, R., Oliphant, T.~E., {et~al.} 2020, Nature Methods, 17,
  261

\bibitem[{Vlemmings {et~al.}(2019)Vlemmings, Lankhaar, Cazzoletti, Ceccobello,
  Dall’Olio, van Dishoeck, Facchini, Humphreys, Persson, Testi, \&
  Williams}]{vlemmings_stringent_2019}
Vlemmings, W. H.~T., Lankhaar, B., Cazzoletti, P., {et~al.} 2019, A\&A, 624, L7

\bibitem[{Wang \& Goodman(2017)}]{wang_wind-driven_2017}
Wang, L. \& Goodman, J.~J. 2017, ApJ, 835, 59

\bibitem[{Wardle(2007)}]{wardle_magnetic_2007}
Wardle, M. 2007, Astrophys Space Sci, 311, 35

\bibitem[{Wardle \& Koenigl(1993)}]{wardle_structure_1993}
Wardle, M. \& Koenigl, A. 1993, ApJ, 410, 218

\bibitem[{Zanni \& Ferreira(2013)}]{zanni_mhd_2013}
Zanni, C. \& Ferreira, J. 2013, Astronomy and Astrophysics, 550, A99

\bibitem[{{Zhang} {et~al.}(2014){Zhang}, {Isella}, {Carpenter}, \&
  {Blake}}]{Zhang14}
{Zhang}, K., {Isella}, A., {Carpenter}, J.~M., \& {Blake}, G.~A. 2014, \apj,
  791, 42

\bibitem[{{Zhu} {et~al.}(2011){Zhu}, {Nelson}, {Hartmann}, {Espaillat}, \&
  {Calvet}}]{Zhu11}
{Zhu}, Z., {Nelson}, R.~P., {Hartmann}, L., {Espaillat}, C., \& {Calvet}, N.
  2011, \apj, 729, 47

\bibitem[{Zhu \& Stone(2018)}]{zhu_global_2018}
Zhu, Z. \& Stone, J.~M. 2018, ApJ, 857, 34

\end{thebibliography}

\appendix
\section{Ambipolar diffusivity for a transition disc: a simple model}
\label{app_lambda}

The aim of this appendix is to model the ambipolar diffusivity spatial dependence in both a transition disc and a standard protoplanetary disc (i.e. without cavity).
The general procedure to reach such a result follows and adapts the main calculation steps that are presented in \cite{combet_dead_2010}.
As assumed in Eq.~\ref{ohm}, only the ambipolar diffusivity does appear in the MHD equations, which we assume is the dominant non-ideal effect in the regime of discs we use at $R\geq10$ \citep{riols_ring_2020,simon_magnetically_2015}.
Therefore, the only momentum exchange that occurs between particles happens only between ions and neutrals.
In a plasma made of molecular ions, electrons and neutrals, the ambipolar diffusivity is given by \citep{wardle_magnetic_2007}
\begin{equation}
\eta_\text{A}=\frac{\boldsymbol{B}^2}{4\pi\,\gamma_\text{in}\,\rho_\text{n}\,\rho_\text{i}},
\label{eta_A}
\end{equation}
where $\rho_\text{n}$ and $\rho_\text{i}$ are respectively the density of the neutrals (the gas so $\rho_\text{n}=\rho$) and of the ions and $\gamma_\text{in}=\langle\sigma v\rangle_\text{in}/(m_\text{n}+m_\text{i})$ with $\langle\sigma v\rangle_\text{in}$ the ion-neutral collision rate whose value is \citep{bai_magnetorotational_2011}
\begin{equation}
\langle\sigma v\rangle_\text{in} = \num{2.0e-9}\,\left(\frac{m_\text{H}}{\mu}\right)^{1/2}\si{\centi\metre^3\second^{-1}},
\end{equation}
with $m_\text{H}$ the atomic mass and $\mu=\num{2.34}\,m_\text{H}$ is the mean molecular weight. Introducing the ionisation fraction $\xi=\rho_\text{i}/\rho_\text{n}$, one gets
\begin{equation}
\eta_\text{A}=\num{1.6e16}\,\left(\frac{\xi}{\num{1e-13}}\right)^{-1}\,\left(\frac{\boldsymbol{B}}{\num{1}\,\mathrm{G}}\right)^2\,\left(\frac{\rho}{\num{1e14}\,\si{\centi\metre}^{-1}}\right)^{-2}\,\si{\centi\metre^2\second^{-1}}.
\end{equation}
Ambipolar diffusion is usually evaluated with the dimensionless ambipolar Elsasser number $\Lambda_\text{A}$ defined in Eq.~\ref{lambdaA}.
To get this number, we have to evaluate the ionisation fraction. Let us consider a simple chemical lattice with no metals nor grains,
\begin{align}
\text{m}\;+\;\text{ionising radiation}\;&\longrightarrow\;\text{m}^+\;+\;\text{e}^-&\quad\zeta_\text{i}\\
\text{m}^+\;+\;\text{e}^-\;&\longrightarrow\;\text{m}\quad&\delta,
\end{align}
with $\zeta_\text{i}$ the ionisation rate and $\delta$ the dissociative recombination rate. Following \cite{fromang_ionization_2002}, we take  
\begin{equation}
\delta=\num{3e-6}\,T^{-1/2}\,\si{\centi\metre^3\second^{-1}}.
\label{delta}
\end{equation}
In this toy model we then have \citep{lesur_thanatology_2014}
\begin{equation}
\xi=\sqrt{\frac{\zeta_\text{i}}{\delta\,\rho}} + \xi_\text{FUV},
\label{xi}
\end{equation}
where $\xi_\text{FUV}$ accounts for the far UV photons contribution that we model following \cite{perez-becker_surface_2011} as
\begin{equation}
    \xi_\text{FUV} = 2\times 10 ^{-5}\,\exp{\left[-\left(\Sigma_\star / \num{0.03}\,\si{\gram\,\centi\metre^{-2}}\right)^4\right]},
\end{equation}
with $\Sigma_\star$ the column density computed from the star to the point of interest.

\noindent To calculate $\zeta_\text{i}$, we add the ionisation sources listed below
\begin{itemize}[label=\textbullet]%, font=\LARGE \color{cornflowerblue}]
\item X-ray ionisation from the protostar modelled by two bremsstrahlung-emitting corona (following \citealt{bai_heat_2009} and \citealt{igea_x-ray_1999})
\begin{multline}
\zeta_\text{X}=L_{\text{X},\,29}\,\left(\frac{R}{1\,\mathrm{a.u.}}\right)^{-\num{2.2}}\,\left[\zeta_1\left(\e^{-\left(N_{\text{H}1}/N_1\right)^\alpha}+\e^{-\left(N_{\text{H}2}/N_1\right)^\alpha}\right)+\right.\\
\left.\zeta_2\left(\e^{-\left(N_{\text{H}1}/N_2\right)^\beta}+\e^{-\left(N_{\text{H}2}/N_2\right)^\beta}\right)\right],
\label{Xray}
\end{multline}
with $L_{\text{X},29}\equiv L_\text{X}/\num{e29}\text{erg}\,\si{\second^{-1}}$ and $L_\text{X}$, $\zeta_1$, $\zeta_2$, $\alpha$, $\beta$, $N_1$, $N_2$ are the numerical values defined in \cite{bai_heat_2009} while $N_{\text{H}1}$ and $N_{\text{H}2}$ are the columns density of hydrogen vertically computed above and below the calculation point.
\item Cosmic-ray ionisation following \citep{umebayashi_recombination_1980}  
\begin{equation}
\zeta_\text{CR}=\zeta_{\text{CR},0}\,\e^{-\Sigma_\text{col.}/\num{96}\,\si{\gram\,\centi\metre^{-2}}}
\,\si{\second^{-1}},
\label{CR}
\end{equation}
where $\zeta_{\text{CR}, 0}=\num{e-17}\,\si{\second^{-1}}$ and $\Sigma_\text{col.}$ is the matter column density above and below the point of interest.
\item Radioactive decay is assumed constant \citep{umebayashi_effects_2008}
\begin{equation}
\zeta_\text{rad.} = \num{e-19}\,\si{\second^{-1}}.
\label{rad}
\end{equation}
\end{itemize}
Combining the equations \ref{Xray}, \ref{CR} and \ref{rad}, we obtain $\zeta_\text{i}=\zeta_\text{X}+\zeta_\text{CR} + \zeta_\text{rad.}$, paving the way to finally get $\Lambda_\text{A}$ using equations \ref{lambdaA}, \ref{delta} and \ref{xi}.
Note that due to the dependency of $\eta_\text{A}$ and $v_\text{A}$ on the norm of the magnetic field, this latter cancels and does not need to be computed to get $\Lambda_\text{A}$.
The previous calculations can be performed either for a standard protoplanetary disc or for a transition disc.
The only thing that needs to be changed to account for such discs is the surface density profile, where Eq.~\ref{sigmaf} allows to consider or not the effects of the cavity.

\begin{figure}
\begin{center}
\includegraphics[width=0.5\textwidth]{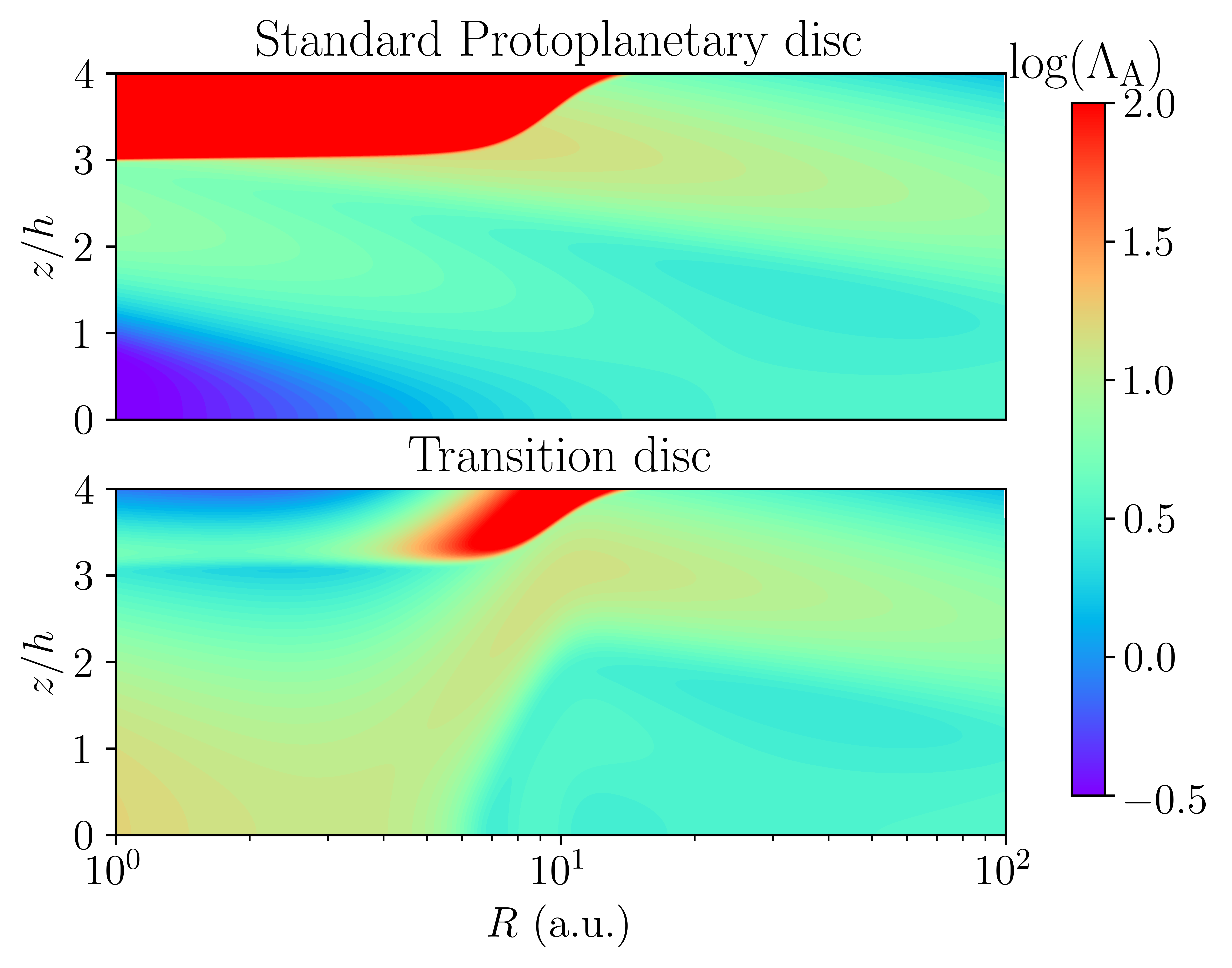}
\caption{\label{LambdaPPD_TD_FUV}Ambipolar Elsasser number $\Lambda_\text{A}$ in a standard protoplanetary disc (top panel) and in a transition disc (bottom panel). In spite of these $2$~profiles being slightly different, no major changes occur from one kind of disc to another around the midplane.}
\end{center}
\end{figure}

The results of such calculations are displayed in Fig.~\ref{LambdaPPD_TD_FUV} that represents the spatial dependency of $\Lambda_\text{A}$ in both a standard protoplanetary disc and a transition disc. Though these $2$~profiles look different at first glance, a deeper investigation reveals that the values taken by $\Lambda_\text{A}$ in the discs remain pretty much close to unity in both cases, while the general trend of $\Lambda_\text{A}$ in a standard protoplanetary disc is recovered even in the case of a transition disc \citep{thi_radiation_2019}. 
Moreover, $\Lambda_\text{A}$ remains fairly below the critical value $\Lambda_\text{A, crit.}=10^2$ with or without a cavity.
$\Lambda_\text{A}$ must stay below $\Lambda_\text{A, crit.}$ so that the MRI effects are negligible \citep{blaes_local_1994, bai_magnetorotational_2011}. 
Therefore, assuming a characteristic value of $\Lambda_{\text{A},\,0}=1$ captures within a reasonable accuracy the physics of ambipolar diffusion and the cavity does not alter the ambipolar Elsasser number profile.
The results we get from this simple toy model are to be compared to the more detailed work of \cite{wang_wind-driven_2017} where many chemical species are taken into account to compute the ambipolar Elsasser number inside the cavity of a wind-driven transition disc.

Following \cite{lesur_systematic_2021} and \cite{thi_radiation_2019}, we implement the profile of $\Lambda_\text{A}$ so that
\begin{equation}
\Lambda_\text{A}(z, R) = \Lambda_{\text{A},\,0}\,\exp{\left(\frac{z}{\lambda\,h}\right)^4},
\end{equation}
where $\lambda$ is a parameter that controls the height where a transition between non-ideal and ideal MHD occurs (the non-ideal MHD part being the inside of the disc) and is chosen constant and equal to~$3\,h$. $\Lambda_{\text{A},\,0}$ remains a free parameter (see \ref{SimuList} for more details).
Additionally, a cutoff is used for the $\eta_\text{A}$ profile so that if $\eta_\text{A}>\eta_{\text{A},\,\text{max}}$, the value of $\eta_\text{A}$ is replaced by $\eta_{\text{A},\,\text{max}}\equiv 10\,\varepsilon^2$ in code units, such a choice being reflected on the $\Lambda_\text{A}$ profile with Eq.~\ref{lambdaA}.

\section{Poloidal velocity relaxation and inner boundary condition}
\label{relaxation}
 
\begin{center}
\begin{figure*}
\includegraphics[width=1.\textwidth]{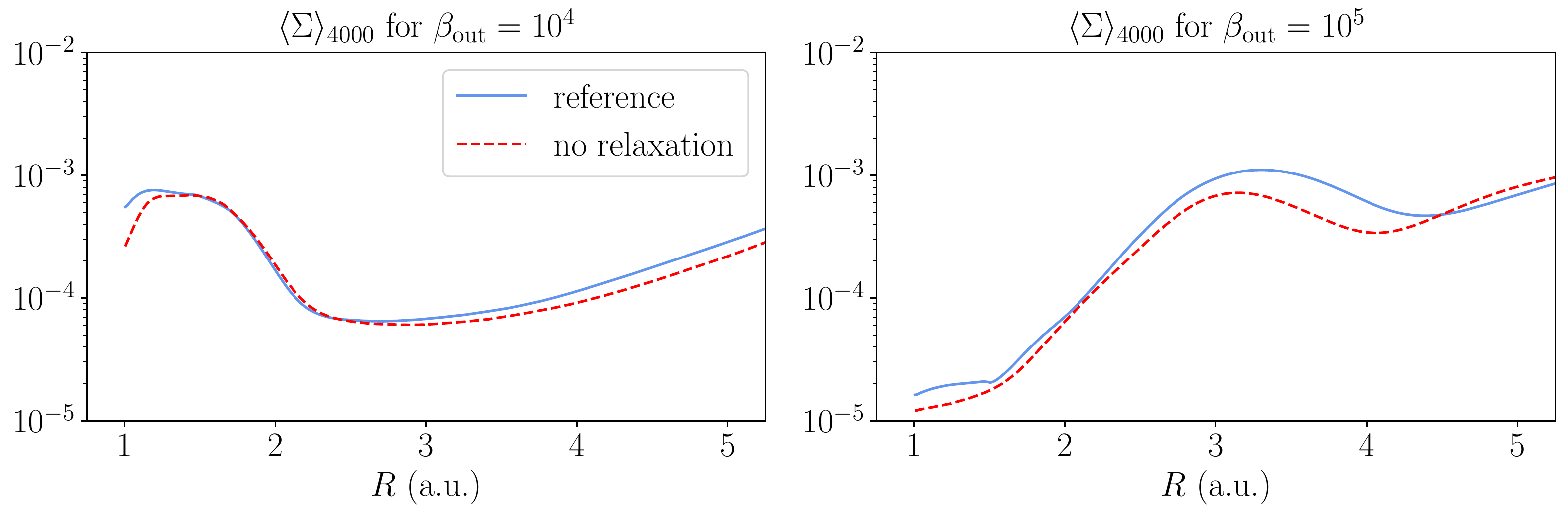}
\caption{\label{vpol_relaxation_CL}Surface density time-averaged on the first 4000~orbits at $R_\mathrm{int}$. The blue lines are the reference runs (left panel: fiducial run, right panel: B5Bin0Am0) and the red-dashed mines are the corresponding runs without the relaxation.}
\end{figure*}
\end{center}

We aim to address the influence of the poloidal velocity relaxation on our results to test our control on the inner boundary condition.
Two additional simulations are conducted respectively with the same setup as B4Bin0Am0 (fiducial run) and B5Bin0Am0, but without the relaxation procedure.
The results are given in Fig.~\ref{vpol_relaxation_CL}, where we show the surface density $\langle\Sigma\rangle_{4000}$ time-averaged on the first 4000 orbits at the internal radius (when the differences are enhanced), with a focus on the innermost radii.
We highlight that these differences do not rise up for $t>4000~$orbits at $R_\mathrm{int}$.
For B5Bin0Am0, the right panel of Fig.\ref{vpol_relaxation_CL} suggests that the relaxation procedure influences how the initial burst is evacuated since we detect differences between the surface density profiles at $R>1.5$.
However, releasing this inner constrain reduces the inner peak of the profile of $\Sigma$, but does not prevent the initial accumulation of matter from appearing.
In particular, the bursts of matter seen in Fig.~\ref{sig_beta_avg_2D_beta} are not due to this condition (and are probably due to the inner boundary condition, see the next paragraph).
For the fiducial simulation, we estimate differences of $15\%$ until $R=2$, $7\%$ until $R=10$ and less than $2\%$ until $R=50$ and conclude that the slight accumulation described in the section~\ref{evolution} is due to this procedure contrary to the occurrence of bursts as seen in Fig.~\ref{sig_beta_avg_2D_beta}. 

\begin{figure*}
\begin{center}
\includegraphics[width=1.\linewidth]{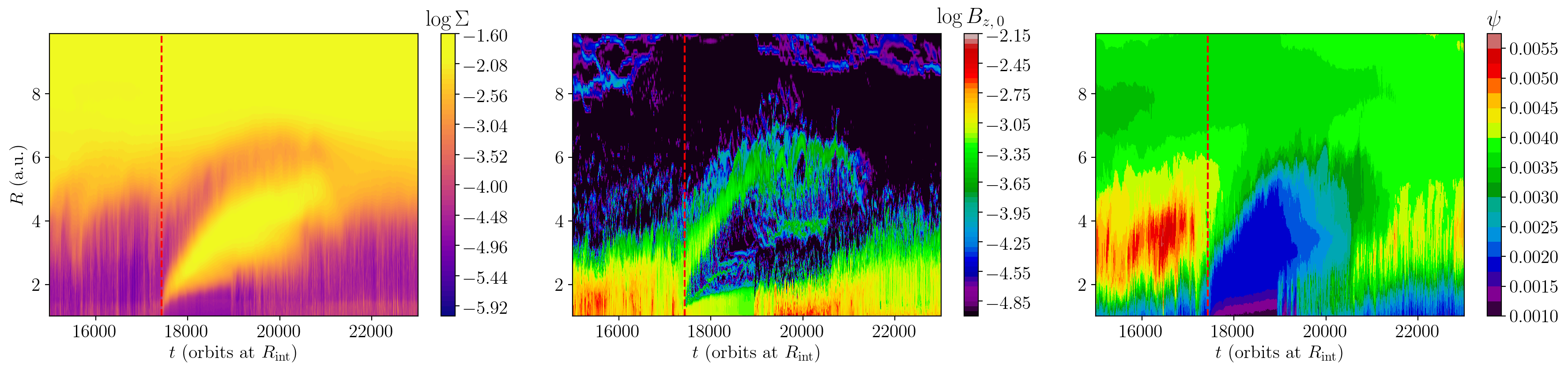}
\caption{\label{burst_S2DB5Bin0Am0}Spatio-temporal diagrams of $\Sigma$ (first panel), $B_{z,\,0}$ the vertical magnetic field at the midplane and $\psi$ the flux function defined in Eq.~\ref{psi}, for simulation B5Bin0Am0. These profiles focus on the second burst detected in the left panels of Fig.~\ref{sig_beta_avg_2D_beta}. The red dashed line marks the beginning of the burst when detected using $\Sigma$.}
\end{center}
\end{figure*}

Regarding the bursts of B5Bin0Am0 (see Fig.~\ref{sig_beta_avg_2D_beta}), we focus on one of them in Fig.~\ref{burst_S2DB5Bin0Am0}.
The first panel displays the spatio-temporal diagram of the surface density on which the burst is clearly detected at $17435~$orbits at $R_\mathrm{int}$ and localised by the red dashed line.
The accumulation of matter is correlated with a decrease of the vertical magnetic field at the midplane (second panel of Fig.~\ref{burst_S2DB5Bin0Am0}).
This magnetic field is not lost but is expelled outwards, as is evident from  the magnetic flux function (third panels of Fig.~\ref{burst_S2DB5Bin0Am0}).
Such a shortage of magnetic field leads to an increase of $\beta$ and blocks accretion (we recall that the accretion speed is $v_\mathrm{acc.}\propto \beta^{-\sigma}$ with $\sigma>0$).
As a result $\dot{M}$ falls from $\num{0.25}$ down to $\num{0.1}\times10^{-7}~M_\odot.\text{yrs}^{-1}$ in the region between the inner radial boundary and the burst, and matter piles up in the cavity.
This episode ends when the magnetic flux is eventually re-accreted, leading to an increase of the mass accretion rate and the disappearance of the density excess in the cavity.
At some point, the magnetic flux is advected back onto the seed up until it saturates so that $B_z$ can accumulate again close to the inner boundary condition before accretion is enhanced back to normal.
The reason why such magnetic flux evacuates from the seed from time to time remains unclear and these occurrences close to the inner boundary suggest that these might be a boundary condition artefact.
However, we mention that the total magnetisation of the seed eventually saturates with a roughly constant value, so that a sharp increase of magnetic field (as it is the case for this burst, see the middle panel of Fig.~\ref{burst_S2DB5Bin0Am0}, a few orbits before the location of the red dashed line) could force the seed to lose magnetic flux to ensure its conservation.
We end up by adding that these bursts are only detected for the weakly magnetised simulations (the ones with $\beta_\text{out}=10^5$).

Regarding the inner radial boundary condition for the magnetic field, we tried several configuration (outflow conditions which is the one we eventually chose and perfect conductor).
Both of these conditions lead to the same steady-states.

We also ran a simulation with a stronger magnetic field close to the inner boundary condition, but no significant changes were noticed.
The additional magnetic field was chosen so that the magnetisation of the seed is set close to its saturation value in the fiducial run.
However and in any case, the same transient state occurs and leaves the stage to a similar steady state (the magnetisation of the seed reaches the same saturation value and the same stripes are observed in the spatio-temporal diagram of $\psi$).

Therefore, we conclude that our setup is robust regarding the initial state and the boundary conditions.
The inner boundary still plays a role because of its magnetisation and the fact that only a given amount of magnetic field can be advected.
This probably leads to the burst events seen in simulation B5Bin0Am0.

\section{Interchange instability criterion calculations}
\label{Interchange_appendix}
We express the instability criterion for the interchange instability (or RTI) calculated in \cite{spruit_interchange_1995} (equation 59) in terms of the plasma parameter.
This criterion reads
\begin{equation}
    g_\text{m}\,\partial_R\ln\frac{\Sigma}{B_z} > 2\,\left(r\frac{\text{d}\Omega}{\text{d}r}\right)^2\equiv 2\,S^2,
\end{equation}
where $S$ is the shear that we approximate with $S^2 = 9/4\,\Omega^2$ and $g_\text{m}$ is
\begin{equation}
    g_\mathrm{m}\equiv \frac{B_R^{\,+}\,B_z}{2\pi\,\Sigma}.
\end{equation}
$B_R^{\,+}$ is the radial component of the magnetic field at the disc surface.
Let us rewrite the previous expression in terms of $\beta$, $q$ (defined with $B_R^{\,+}=q\,B_z$) and $\delta$ (defined as $\delta=-\text{d}\ln\Sigma/\text{d}\ln R$).

\begin{align}
\frac{B_R^{\,+}\,B_z}{2\pi\,\Sigma}\partial_R\ln\frac{\Sigma}{B_z}
		       &=\frac{B_R^{\,+}\,B_z}{2\pi\,\Sigma}\frac{\Sigma'}{\Sigma}-\frac{B_R^{\,+}\,B_z}{2\pi\,\Sigma}\frac{B_z'}{B_z}\\
		       &=\frac{q\,B_z^{\;2}}{2\pi\,\Sigma}\frac{-\delta}{R}-\frac{q}{4\pi\,\Sigma}\left(B_z^{\;2}\right)',
\end{align}
where $X'$ denotes the derivative of X with respect to $R$.
With $P=c_\text{s}^2\,\rho=(h\,\Omega_\text{K})^2\,\Sigma/(\sqrt{2\pi}\,h)$, we get
\begin{equation}
\beta =\frac{4\sqrt{2\pi}\, R\,\varepsilon\,\Omega_\text{K}^{\;2}\,\Sigma}{{B_z}^2}.
\end{equation}
Therefore, the instability criterion becomes
\begin{equation}
S^{2}<-\frac{4\,\varepsilon\,\Omega_\text{K}^{\;2}\,q\,\delta}{\sqrt{2\pi}\,\beta}-\frac{q}{4\pi\,\Sigma}\partial_R\left(\frac{4\sqrt{2\pi}\,R\,\varepsilon\,\Omega_\text{K}^{\;2}\,\Sigma}{\beta}\right),
\end{equation}
$\varepsilon$ being constant in the disc as well as $\beta$ inside the cavity. $\Omega_\text{K}$ varies as $R^{-3/2}$ and $\Sigma$ as $R^{-\delta}$ so that
 \begin{equation}
 S^2<\frac{4\,\varepsilon\,\Omega_\text{K}^{\;2}}{\sqrt{2\pi}\,\beta}\,q\,(-\delta+1+\delta/2)\label{crit1}
 \end{equation}
By taking $S^{2}/\Omega_\text{K}^{\;2}=9/4$, the RTI can be triggered when
\begin{equation}
    \beta < \frac{16\,\varepsilon}{9\,\sqrt{2\pi}}\,q\,\left(1-\frac{\delta}{2}\right).
    \label{full_crit}
\end{equation}
If we now assume that $\delta=q=1$ for simplicity, we finally get
\begin{equation}
    \beta < \frac{8\,\varepsilon}{9\,\sqrt{2\pi}}\approx 0.355\,\varepsilon = 0.0355 \equiv \beta_\text{crit.},
    \label{crit2}
\end{equation}
where $\varepsilon=0.1$.

\begin{figure}
    \centering
    \includegraphics[width=0.5\textwidth]{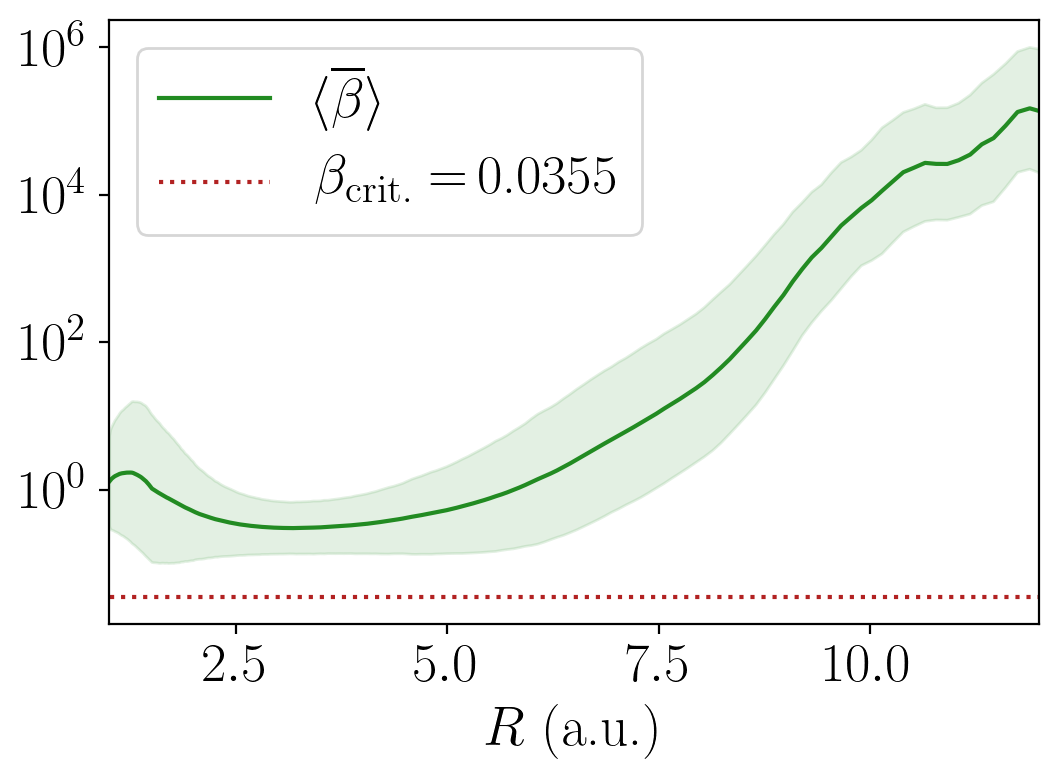}
    \caption{Interchange instability criteria. In red dotted line is shown the critical value of $\beta_\text{crit.}$ while the blue dashed line is obtained with Eq.~\ref{full_crit}.}
    \label{beta_RTI_cor_fid2D}
\end{figure}

Figure \ref{beta_RTI_cor_fid2D} compares the time-averaged values of $\overline{\beta}$ with the criterion given in Eq.~\ref{crit2}.
The value of $\beta_\text{crit.}$ is anyhow below the time-averaged values of $\overline{\beta}$.
Though this simple analysis makes it difficult to be definitive on this subject, it seems that the interchange instability is not triggered inside the cavity.

\section{Laminar transport coefficients}
\label{Laminar_coef}

\begin{figure}
    \centering
    \includegraphics[width=0.5\textwidth]{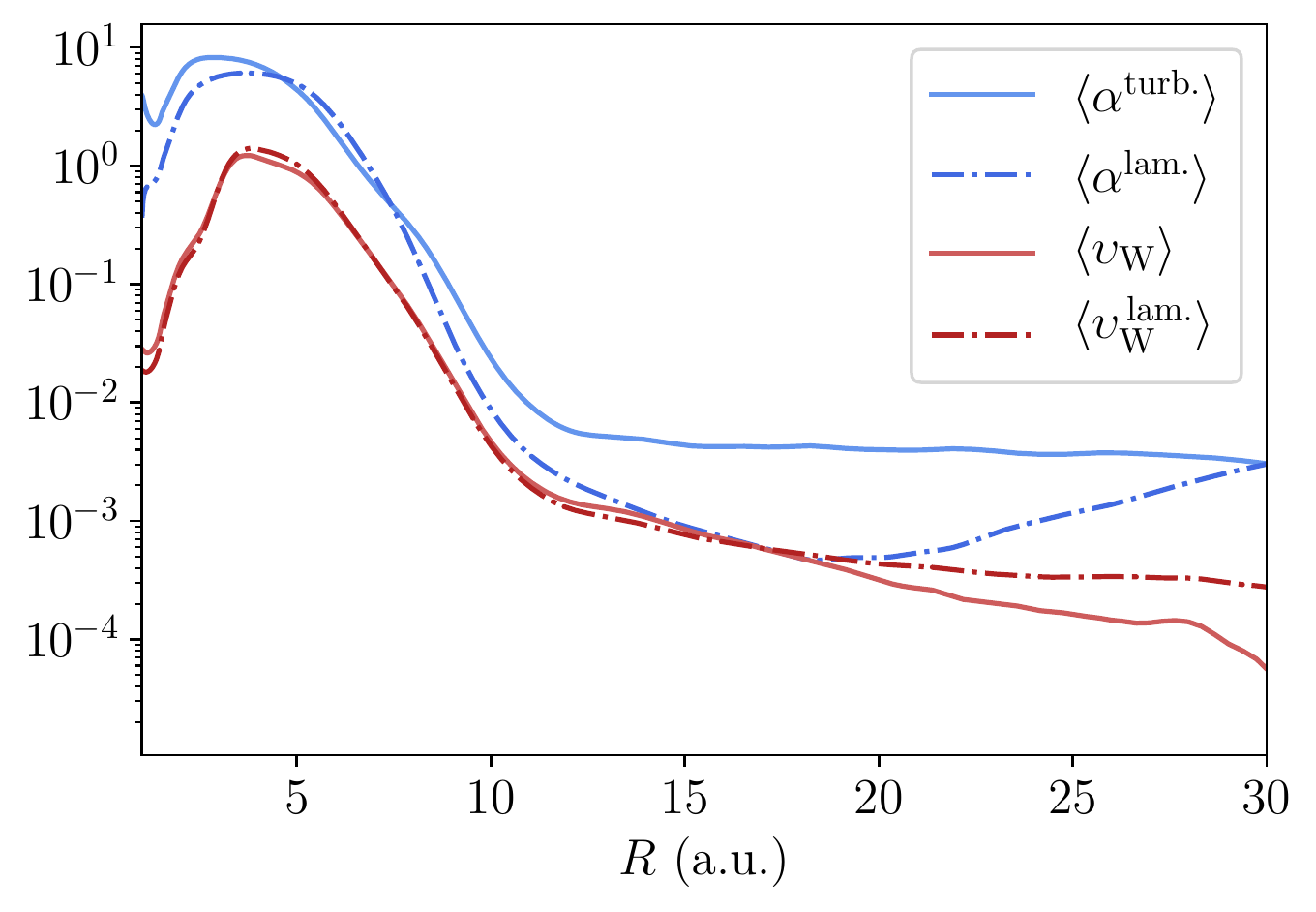}
    \caption{Time-averaged transport coefficients and their laminar and turbulent contributions. We give the laminar and turbulent contributions for $\langle\alpha\rangle$ and the total profile with its laminar contribution for $\langle\upsilon_\mathrm{W}\rangle$.}
    \label{alpha_upsilon_turb_lam_tot_fid2D}
\end{figure}

In order to discuss the role of the MRI, we must highlight the impact of the laminar stress and its contribution to the transport coefficients.
In this article, we focus on the total stresses, defined in Eq.~\ref{stresses}.
To compare the turbulent effects, we decompose the stresses with a turbulent and a laminar part.
In this prospect, we introduce the deviation to the temporal mean such that
\begin{equation}
    \delta X \equiv X - \langle X \rangle.
\end{equation}
Focusing on $W_{r\varphi}$, we expand the magnetic term as
\begin{equation}
    \label{decomposition}
    \overline{\langle B_r\,B_\varphi\rangle} = \underbrace{\overline{\langle B_r\rangle\,\langle B_\varphi\rangle}}_{\text{laminar}} + \underbrace{\overline{\langle\delta B_r\,\delta B_\varphi\rangle}}_{\text{turbulent}}.
\end{equation}
Concerning the turbulent stresses, we refer to \cite{jacquemin-ide_magnetic_2020} (see their appendix~A) as we only compute the laminar ones and compare the laminar transport coefficients to the ones studied in the article.
Therefore, we adopt the following definition for the laminar radial stress
\begin{equation}
    \langle W_{r\varphi}^\text{\,lam.} \rangle \equiv -\frac{1}{4\pi}\,\overline{\sin\theta\,\langle B_r\rangle\,\langle B_\varphi \rangle},
\end{equation}
and for the laminar   surface stress
\begin{equation}
    \langle W_{\theta\varphi}^\text{\,lam.}\rangle\equiv -r\,\left[\sin^2\theta\,\frac{\langle B_\theta\rangle\,\langle B_\varphi\rangle}{4\pi}\right]^{\theta_-}_{\theta_+}.
\end{equation}
These definitions are coherent with previous works \citep{bethune_global_2017, mishra_strongly_2020, jacquemin-ide_magnetic_2020}.
Hence, the laminar transport coefficients are given by
\begin{equation*}
    \left \{
    \begin{aligned}
    \langle\alpha^\text{lam.}\rangle &\equiv \frac{\langle W_{r\varphi}^\text{\,lam.}\rangle}{\langle \overline{P}\rangle} \\
    \langle \upsilon_\mathrm{W}^\text{lam.}\rangle &\equiv \frac{\langle W_{\theta\varphi}^\text{\,lam.}\rangle}{r\,\langle P_0\rangle}
    \end{aligned} \right. ,
\end{equation*}
  while we define their turbulent counterparts as
\begin{equation*}
    \left \{
    \begin{aligned}
    \langle\alpha^\text{turb.}\rangle &\equiv \langle\alpha\rangle - \langle\alpha^\mathrm{lam.}\rangle\\
    \langle \upsilon_\mathrm{W}^{\,\text{turb.}}\rangle &\equiv \langle\upsilon_\mathrm{W}\rangle - \langle\upsilon_\mathrm{W}^{\,\mathrm{lam.}}\rangle
    \end{aligned} \right. .
\end{equation*}
The results are shown in Fig.~\ref{alpha_upsilon_turb_lam_tot_fid2D}.
The laminar contribution is the major one for $\langle\upsilon_\mathrm{W}\rangle$ in the whole disc so that we only show its laminar contribution with respect to the full coefficient, as they take essentially the same values.
Nevertheless, despite the laminar term being high for $\langle\alpha\rangle$, a strong turbulent term is at stake, especially in the external part of the disc where it is dominant.
Inside the cavity, $\langle\alpha\rangle$ is fairly distributed between the laminar and turbulent contributions. 
However, we recall that the wind may act on the turbulent component of $\langle\alpha\rangle$ too since the magnetic field also appears in Eq.~\ref{decomposition}.

We finally conclude that the MRI is probably acting on the disc outer parts in the $\langle\alpha\rangle$ coefficient, while the surface stress embodied by $\langle\upsilon_\mathrm{W}\rangle$ is definitely dominated by its the laminar part and due to the wind.

\end{document}